\documentclass[aps,showkeys,twocolumn]{revtex4-2}
\usepackage{mathrsfs}
\usepackage{amssymb}
\usepackage{amsmath}
\usepackage{bm}
\usepackage{steinmetz}
\usepackage{graphics}
\usepackage{natbib}
\usepackage{color}

\newcommand\degree{^\circ}

\begin{document}

\title{Optically Driven Orbital Hall Transport in Floquet Odd-Parity Collinear Altermagnets with High Chern Numbers}

\author{Yuping Tian$^1 $}
\thanks{Y. Tian and C.-H. Zhao contributed equally to this work.}
\author{Chen-Hao Zhao $^1$ }
\thanks{Y. Tian and C.-H. Zhao contributed equally to this work.}
\author{Chao-Bo Wang $^1$ }
\author{Binyuan Zhang$^1$ }
\author{Xiangru Kong$^1$}\email{kongxiangru@mail.neu.edu.cn}

%\author{}
\author{Wei-Jiang Gong$^1$}\email{gwj@mail.neu.edu.cn}

\affiliation{ College of Sciences, Northeastern University, Shenyang 110819, China}

\date{\today}

\begin{abstract}
Recent studies have attracted increasing interest in nonrelativistic odd-parity magnetism and its associated topology in collinear altermagnets.
Here, based on symmetry analysis and an effective model, we demonstrate that Floquet engineering can induce $f$-wave odd-parity altermagnetism in two-dimensional collinear antiferromagnetic multilayers via the coupling between circularly polarized light (CPL) and layer degrees of freedom. 
Furthermore, modifying the CPL induces nonequilibrium quantum anomalous Hall effect (QAHE) with tunable Chern numbers up to $C=\pm8$, arising from layer- and valley-dependent band inversions.
The induced topological phase transitions provide an efficient means to manipulate the orbital Hall effect (OHE) by redistributing orbital angular momentum.
First-principles calculations reveal that experimentally accessible VSi$_2$N$_4$ serves as a viable platform for topological phase diagram of the QAHE and OHE, featuring pronounced trigonal warping.
Our findings establish a versatile route toward optically controllable topological phenomena, opening new opportunities for future developments in topological spintronics and orbitronics.

\end{abstract}


\maketitle

\bigskip
\textit{Introduction}-Altermagnets provide a unique and significant platform for spin-polarized phenomena \cite{PhysRevB.104.024401,Ma2021Multifunctional,Reichlova2024Observation,PhysRevLett.130.046401,PhysRevLett.133.166701,PhysRevLett.133.206702,PhysRevLett.134.166701,PhysRevLett.131.076003,PhysRevLett.134.106802,zm5y-vy41,PhysRevLett.134.106801,Gao2025AIaccelerated,Fukaya_2025,v3fg-6smc}, distinguished by their nonrelativistic spin splitting in 
\textit{\textbf{k}}-space protected by crystalline symmetries \cite{PhysRevB.102.144441,PhysRevLett.128.197202,PhysRevX.12.040002,PhysRevB.75.115103,PhysRevLett.130.036702,doi:10.7566/JPSJ.88.123702,PhysRevLett.132.176702,Krempasky2024Altermagnetic,Song2025Altermagnets,https://doi.org/10.1002/adfm.202409327,Cheong2025Altermagnetism,Zhang2025Crystal,PhysRevX.15.021083}.
Typically, altermagnets exhibit even-parity spin splitting, giving rise to Fermi surfaces with \textit{d}-, \textit{g}-, and \textit{i}-wave symmetries \cite{PhysRevX.12.040501,PhysRevX.12.031042}.
In contrast, magnets hosting unconventional odd-parity spin splitting (e.g., 
\textit{p}- and \textit{f}-wave types) are particularly significant due to their intimate connection with unconventional superfluidity \cite{RevModPhys.76.323,Koo2020Rashba,Manchon2015New}.
However, odd-parity spin splitting has been largely limited to noncollinear magnets \cite{Song2025Electrical,Hellenes2023Pwave,zk69-k6b2,PhysRevB.101.220403,PhysRevLett.133.236703}, whose complex magnetic textures are often fragile against external perturbations, restricting their practical applicability. 
Realizing odd-parity altermagnetism (AM) in collinear magnetic systems is therefore highly desirable, yet remains a fundamental challenge.
Floquet engineering via periodic driving provides a versatile paradigm for dynamically reshaping electronic landscapes and inducing exotic nonequilibrium quantum phases \cite{Bukov04032015,Zhan2024Floquet,10.1038/s41586-021-04051-8,Bielinski_2025_FloquetBloch,McIver_2020_LightInducedAHEGraphene,Merboldt_2025_FloquetGraphene,Choi_2025_FloquetBlochGraphene,PhysRevLett.134.146401,Bielinski_2025_FloquetBloch,D4TC02438A,Fragkos_2025_FloquetBlochValleytronics,Kong_2022,10.1063/5.0006446,doi:10.1021/acs.nanolett.2c04651,https://doi.org/10.1002/adfm.202501934,doi:10.1021/acsnano.5c10277,D4MH00237G,PhysRevLett.133.246606}.
While periodic light can effectively modify magnetic symmetries, the dynamical manipulation of spin-space-group symmetries to engineer higher-order odd-parity AM remains a pivotal and formidable frontier \cite{9346-9jpf,li2025floquetspinsplittingspin,7ywb-ml2q,wnqs-3djt,zhuang2025oddparityaltermagnetismoriginatedorbital,zhu2026lightinducedevenparityunidirectionalspin,lkf9-jgv6,li2026robusttunablefloquetaltermagnets,lin2026oddparityaltermagnetismsublatticecurrents}.

In this Letter, we demonstrate that Floquet engineering induces 
\textit{f}-wave odd-parity AM in two-dimensional (2D) collinear antiferromagnetic (AFM) multilayers based on symmetry arguments and effective model analysis.
In the parity-time ($\mathcal{P}\mathcal{T}$)-symmetric bilayer characterized by layer-locked spin polarization, circularly polarized light (CPL) couples to the hidden layer- and valley-dependent orbital textures.
Through spin-layer locking, this coupling converts the orbital Floquet response into an odd-parity AM state. 
Furthermore, the formation of photon-dressed Floquet-Bloch states drives layer-selective band inversion, enabling distinct quantum anomalous Hall (QAH) phases with Chern numbers tunable by the CPL intensity or frequency. 
The accompanying variation of orbital angular momentum (OAM) plays a key role in controlling the magnitude of the orbital Hall effect (OHE).
The OHE which is the orbital counterpart to the spin Hall effect, enables the electrical generation of transverse OAM flow \cite{PhysRevLett.95.066601,Go2021Orbitronics,PhysRevLett.121.086602,PhysRevB.98.214405}. 
Recent experimental milestones in light metals \cite{Choi2023Observation,PhysRevLett.131.156702,PhysRevLett.131.156703} have solidified the OHE as a pivotal orbitronic mechanism, offering a robust pathway to control orbital degrees of freedom in magnetic multilayers \cite{PhysRevLett.125.177201,PhysRevResearch.4.033037,PhysRevLett.134.036304,PhysRevLett.100.096601,PhysRevLett.134.136201,Kumar2023Ultrafast,Mishra2024Active}.

Building on the link between orbital textures and the OHE \cite{PhysRevLett.125.216404,PhysRevB.101.161409,doi:10.1021/acs.nanolett.9b02802,Chen2020Strong,PhysRevLett.132.186302,PhysRevB.94.121114}, we reveal a light-driven mechanism for sub-picosecond OAM modulation.
While the driving frequency selects resonant photo-hybridizations, the field strength dictates band renormalization and symmetry breaking, which bridges static topological invariants with Floquet-engineered states and enables deterministic OAM manipulation in the petahertz regime.
To substantiate these findings, we construct a $d$-orbital tight-binding (TB) model that reproduces the light-induced 
$f$-wave AM and the topological phase diagram with topology-engineered OHE.
Remarkably, high Chern numbers of $C=\pm8$ are achieved, facilitated by triple warping effects.
Finally, we propose VSi$_2$N$_4$ as a realistic platform to host the predicted ultrafast phenomena.
Our approach, rooted in CPL-induced $\mathcal{T}$-symmetry breaking, is broadly applicable to both $\mathcal{PT}$-symmetric and non-$\mathcal{PT}$-symmetric magnetic multilayers.

\begin{figure}[htp]
\begin{center}\scalebox{1.13}{\includegraphics{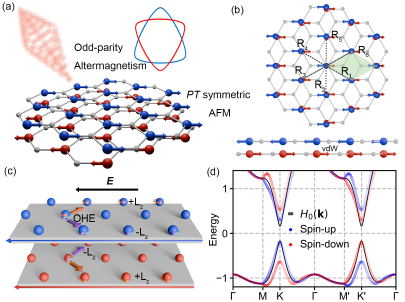}}
\caption{(a) and (b) Illustration of light-induced odd-parity AM in vdW AFM bilayes composed of FM monolayers with in-plane magnetization. The two sublattices with opposite spin polarization are depicted by antiparallel red and blue spin arrows. (c) Schematic illustration of the light-induced QAHE and light-modulated OHE. Red and blue spheres: sublattices with opposite spins. Orbiting orange and purple spheres: distinct OAM. Red and blue lines: direction and spin polarization of edge currents. (d) Spin-degenerate $H_{0}(\textbf{\textit{k}})$ (black lines) and spin-resolved $H_{\rm{eff}}(\textbf{\textit{k}})$ (red and blue circles) band structures. }
\end{center}
\end{figure}

\textit{Symmetry}-
We consider a 2D collinear AFM bilayer where the magnetic sublattices are related by $\mathcal{PT}$ symmetry. 
Due to negligible spin-orbital coupling (SOC), the system is governed by spin-group symmetries $[R_i||R_j]$, where $R_i$ and $R_j$ act in spin and real spaces, respectively \cite{PhysRevX.12.031042,PhysRevX.12.040501,PhysRevX.14.031039,PhysRevX.14.031038,PhysRevX.14.031037}.
The spin structure is primarily dictated by two operations: $[C_2||\mathcal{P}]$, where a spin-space twofold rotation $C_2$ is coupled with spatial inversion $\mathcal{P}$ exchanging the sublattices, and $[C_2\mathcal{T}||E]$, a generic symmetry for collinear magnets that combines time reversal $\mathcal{T}$ with $C_2$ \cite{PhysRevX.14.031038}. 
These symmetries enforce the constraints $\varepsilon(s,\textbf{\textit{k}})=\varepsilon(-s,-\textbf{\textit{k}})$ and $\varepsilon(s,\textbf{\textit{k}})=\varepsilon(s,-\textbf{\textit{k}})$ (where $s=\uparrow,\downarrow$), respectively, thereby ensuring global spin degeneracy.
Crucially, CPL breaks the $\mathcal{T}$-related symmetry $[C_2\mathcal{T}||E]$ while preserving the spin-rotation-related inversion $[C_2||\mathcal{P}]$, as the light field does not directly couple to spin degrees of freedom in the absence of SOC.
This selective symmetry breaking lifts the spin degeneracy, leaving only the odd-parity constraint $\varepsilon(s,\textbf{\textit{k}})=\varepsilon(-s,-\textbf{\textit{k}})$, which characterizes an odd-parity Fermi surface.
When an additional $C_3$ symmetry is introduced, an $f$-wave AM state can emerge [Fig.1(a)], protected by the $[C_2||S_{6z}]$ symmetry (whose cubic operation corresponds to $[C_2||\mathcal{P}]$).

Building on the $f$-wave AM state, on-resonant CPL driving with photon energies spanning half to the full band gap reshapes the electronic landscape.
Through the optical Stark effect, photon-dressed Floquet-Bloch bands are shifted toward the Fermi level \cite{PhysRevB.97.045307,Sie2015Valleyselective,doi:10.1126/science.aal2241,DeGiovannini2016Monitoring}, lifting spin degeneracies and triggering spin-layer-resolved topological phase transitions (TPTs) characterized by band gap closing and reopening. 
The associated inversion of orbital characters provides an efficient means to redistribute the OAM and thereby modulate the OHE \cite{doi:10.1021/acs.nanolett.3c05129}.
The OAM current generated by the OHE exerts an orbital torque upon injection into a ferromagnet.
As illustrated in Fig.1(c), the Floquet AM manifests spin-layer locking, which generates layer-polarized torques and consequently drives counter-propagating electronic OAM currents under an external electric field.
When the system is in a topologically nontrivial phase, the resulting dissipationless edge currents exhibit layer-locked orbital torque characteristics.

\textit{Model}-We consider a hexagonal AFM bilayer composed of $d$ orbitals to demonstrate the emergence of light-induced $f$-wave AM. 
The lattice geometry is illustrated in Fig.1(b), and the corresponding TB Hamiltonian is given by \cite{PhysRevB.111.L140404,https://doi.org/10.1002/adfm.202501506,D5MH00242G}:
\begin{equation}
	\begin{aligned}
		H_0(\textbf{\textit{k}})\! \!=\! \!&\sum_{\left \langle i,j \right \rangle }\sum_{m,n}\Lambda_{ij}(\textbf{\textit{k}})\sigma_0 t_{ij}^{mn} c_{m}^{\dagger} c_{n} \\ 
		&+\sum_{m}\tau_z \boldsymbol{\sigma} \cdot \textbf{\textit{m}}_{\xi} c_{m}^{\dagger} c_{m}
		+t_\perp \tau_x \sigma_0  I_3 
	\end{aligned}
\end{equation}
where $\Lambda_{ij}(\textbf{\textit{k}})\! \!=\! \!\sum_{L=1,2}P_L e^{i(-1)^{L-1}\textbf{\textit{k}}\cdot (\textbf{\textit{R}}_i \!- \!\textbf{\textit{R}}_j)}$ is the layer-dependent phase operator incorporating the 180$\degree$ relative rotation via phase conjugation in the lower layer.
The layer projection operators are $P_{1,2} \!\!= \!\!\frac{1\pm\tau_z}{2}$,
where $\tau_i$ and $\sigma_i~(i \!= \! x,y,z)$ denote Pauli matrices acting on the layer and spin degrees of freedom, respectively. 
The Hilbert space is spanned by the $d$-orbital basis $\Psi \!= \!(d_{z^2},d_{xy},d_{x^2-y^2})$.
Here, $ t_{ij}^{mn}$ denotes the nearest-neighbor hopping amplitudes between orbital $n$ at site $j$ and orbital $m$ at site $i$, with $c_{m}^\dagger$ ($c_{n}$) being the electron creation (annihilation) operators.
The AFM exchange interaction is characterized by an in-plane magnetization $\textit{\textbf{m}}_{\xi} \! \!= \! \! m_{\xi}(\mathrm{sin}\theta,0,\mathrm{cos}\theta)$ with $\theta  \!\!= \! \!90\degree$.
Due to the weak vdW interaction, the interlayer coupling $t_\perp$ is significantly smaller than the intralayer hopping amplitudes $t_{ij}^{mn}$.
At equilibrium, the preserved $\mathcal{PT}$ symmetry of $H_0(\textbf{\textit{k}})$ enforces spin degeneracy throughout the Brillouin zone [Fig.1(d)].
The valence band maximum and conduction band minimum are dominated by $d_{xy}/d_{x^2-y^2}$ and $d_{z^2}$ orbitals, respectively [Fig.S1] \cite{SuppMat}, where inter-orbital hybridization encodes a nontrivial momentum-space orbital texture. 
This texture generates a finite orbital Berry curvature (OBC), underpinning an intrinsic OHE that persists even in the absence of SOC.

To quantify the intrinsic orbital response, we evaluate the orbital Hall conductivity (OHC) by employing the Fermi-sea term of the Kubo formula within linear response theory \cite{PhysRevLett.123.236403,PhysRevB.104.155146}
$\sigma_{xy}^{L_z}\!=\!-e \! \int_{BZ}\frac{d^2\textbf{\textit{k}}}{(2\pi)^2}\sum_{\nu} \! f_\nu(\textbf{\textit{k}})\Omega_{\nu,xy}^{O,L_z}\!(\textbf{\textit{k}})$
where $\Omega_{\nu,xy}^{O,L_z}\!(\textbf{\textit{k}})$ is the OBC, given by $\Omega_{\nu,xy}^{O,L_z}\!(\textbf{\textit{k}})\!=\!2\hbar \mathrm{Im}\sum_{\mu \neq \nu}\!\frac{\left\langle u_{\textbf{\textit{k}}}^\nu\right| \mathcal{J}^{L_z}_{o,x}\left | u_{\textbf{\textit{k}}}^\mu  \right \rangle
    \left\langle u_{\textbf{\textit{k}}}^\mu\right| \hat{v}_y\left | u_{\textbf{\textit{k}}}^\nu  \right \rangle}{(
    \varepsilon_{\textbf{\textit{k}}}^\nu-\varepsilon_{\textbf{\textit{k}}}^\mu)^2}$,
with $u_{\textbf{\textit{k}}}^\nu$ representing the periodic part of the Bloch state associated with the energy $\varepsilon_{\textbf{\textit{k}}}^\nu$. $f_\nu(\textbf{\textit{k}})$ is the equilibrium Fermi distribution function and the velocity operator is denoted as $\hat{v}_y \!=\!\frac{1}{\hbar}\frac{\partial H_{\textbf{\textit{k}}}}{\partial \textbf{\textit{k}}_y}$.
The OAM current operator is expressed as $\mathcal{J}^{L_z}_{o,x}\!=\!\frac{\left \{\hat{v}_x, \hat{L}_z  \right \}}{2} $, and for the Berry curvature (BC) calculations, this operator is substituted by the velocity operator $\hat{v}_x$.
The OBC represents the geometric response of the momentum-space texture of OAM $\left \langle \textbf{\textit{L}}_z(\textbf{\textit{k}}) \right \rangle\!=\! \left \langle u_{\textbf{\textit{k}}}\right| \!\textbf{\textit{L}}_z(\textbf{\textit{k}}) \!\left | u_{\textbf{\textit{k}}} \right \rangle$.
In the equilibrium bilayer, the OHC exhibits a plateau of $-3.94~e/2\pi$ within the band gap [Fig.2(c)], with the OBC showing positive peaks at the $K$ and $K'$ valleys. 
Under the coexistence of inversion $\mathcal{P}$ and time-reversal $\mathcal{T}$ symmetries, the OAM vanishes at the $K$ and $K'$ points [Figs.S10-11] \cite{SuppMat}.
Furthermore, the emergence of floating edge states, in-gap modes, and localized charge distributions evidences the second-order topological insulator nature of the equilibrium system [Fig.S1] \cite{PhysRevLett.124.036803,PhysRevB.109.L201109,Choi_2020_HigherOrderTopologyWTe2,PhysRevB.108.205410,SuppMat}.

We now explore the dynamical response of the AFM bilayer under normally incident CPL.
The driving field, with frequency $\omega$ and chirality $\eta\!=\!\pm1$, is described by the vector potential $A(t)\!=\! A(\eta \mathrm{sin}(\omega t),\mathrm{cos}(\omega t),0)$.
The time-dependent hoppings are governed by the Peierls substitution $t_{ij}^{mn}(t)\!=\! t_{ij}^{mn}e^{i\frac{e}{\hbar}\textit{\textbf{A}}(t)\cdot\textbf{\textit{d}}_{ij}^{mn}}$ \cite{PhysRevA.27.72,PhysRevLett.110.200403}, where $\textbf{\textit{d}}_{ij}^{mn}$ is the displacement vector between orbitals.
The light-driven operator can be expressed as $c_{m}(t)\!=\!\sum_{\alpha=-\infty}^{\infty}\!c_{\alpha m}e^{i\alpha\omega t}$, where $c_{\alpha m}$ denotes the Floquet component.
According to the Floquet–Bloch theory \cite{Kong_2022,doi:10.1021/acs.nanolett.2c04651,Zhan2024Floquet}, an effective static Hamiltonian $H_F(\textit{\textbf{k}},\omega)$ is constructed in the extended Hilbert space spanned by multiphoton replicas [Section I \cite{SuppMat}].
To qualitatively characterize the spin splitting in this periodically driven AFM bilayer, we employ the Magnus expansion \cite{Bukov04032015} to derive the effective Floquet Hamiltonian from $H_F(\textit{\textbf{k}},\omega)$. 
While rigorous only in the high-frequency limit, this expansion remains a valuable tool for identifying symmetry origins of odd-parity AM.
The effective Floquet Hamiltonian reads
\begin{equation}
	\begin{aligned}
		H_{\rm{eff}}(\textbf{\textit{k}})\!&=\! H_0(\textbf{\textit{k}}, J_0(Aa)t_{ij}^{mn} )\!+\!\sum_{n\geq1}\!\! \frac{[H_{-1},H_1]}{n \hbar \omega}\!+\!\mathcal{O}(\frac{1}{\omega^2})\\
		&=\!\sum_{\left \langle i,j \right \rangle }\!\sum_{m,n}\Lambda_{ij}(\textbf{\textit{k}})\sigma_0 \tilde{t}_{ij}^{mn}c_m^{\dagger}c_n\\
		&~~+ \!\!\!\sum_{\left \langle \left \langle i,j \right \rangle  \right \rangle }\!\sum_{m,n}\Lambda_{ij}(\textbf{\textit{k}})\sigma_0 T_{ij}^{mn} c_m^{\dagger}c_n
	\end{aligned}
\end{equation}
where $H_{\pm1}=\frac{1}{T}\int_o^TH(t) e^{\pm i\omega t}dt$ is the first Fourier component in the frequency space.
The commutator term manifests as a high-frequency correction that lifts the band degeneracy [Fig.1(d)]. 
Here, the effective intralayer hopping amplitudes $\tilde{t}_{ij}^{mn}$ and $T_{ij}^{mn}$ acquire complex phases, directly encoding the CPL-imparted chirality and the consequent breaking of $\mathcal{T}$ symmetry.
Critically, these second-order virtual processes expand the TB connectivity beyond the nearest-neighbor limit of $H_0$, generating effective next-nearest-neighbor transitions via photon-assisted pathways.
These emergent couplings act as synthetic Peierls phases, effectively threading the lattice with a non-zero magnetic flux. 
This mechanism enables the ultrafast reconfiguration of the band structure and provides a viable route to engineering nonequilibrium topological states.
The renormalization of the band edges under CPL can be analytically captured by the effective eigenvalues at the $K$ point [Eqs.S9-12] \cite{SuppMat}:
\begin{equation}
	\begin{aligned}
		&E_v^F \!\!= \!  E_v \! \pm\! \frac{\eta J_1^2(Aa)[3(t_{11}\!-\! t_{22})^2\!\!-\!2(t_{1}\!+\! t_2)^2]}{\hbar \omega}, \\ 
		&E_c^F  \!\!=\!  E_c \! \pm \! 8\eta J_1^2(Aa)t_1 t_2/\hbar \omega,
	\end{aligned}
\end{equation}
where the positive (negative) sign distinguishes the lower (upper) layer, and $E_{c/v}$ are the eigenvalues of $H_0(\textbf{\textit{k}}, J_0(Aa)t_{ij}^{mn} )$. 
Crucially, these splitting signs invert at the $K'$ valley ($\textbf{\textit{k}} \! \! \rightarrow   \! \!-\textbf{\textit{k}}$) , resulting in an odd-parity layer splitting [Figs.S2-S3] \cite{SuppMat}.
Interplaying with the intrinsic spin-layer locking, this effect induces an odd parity spin-splitting texture satisfying $E_s(\textbf{\textit{k}}) \! =  \! E_{-s}(-\textbf{\textit{k}})$.
The emerging $f$-wave odd-parity AM is governed by the symmetry constraint $E_s(\textbf{\textit{k}})  \!\!=  \!\! E_{-s}(\mathcal{O}^{-1}\textbf{\textit{k}})$, where $\mathcal{O}$ represents the spatial component of the $[C_2||S_{6z}]^2$.
Fig.2(a) shows the spin-resolved 3D band structure of $ H_{\rm{eff}}(\textbf{\textit{k}})$, which demonstrates this light-induced higher-order odd-parity AM.
Importantly, since AM originates entirely from light-induced symmetry breaking, time-periodic driving provides a general route to realize and engineer AM states.

\begin{figure}[htp]
	\begin{center}\scalebox{1.05}{\includegraphics{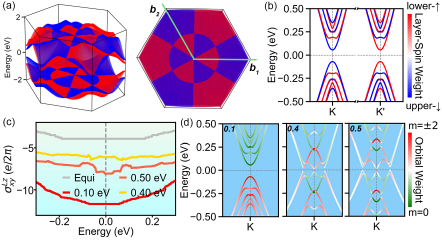}}
		\caption{ (a) Spin-resolved 3D band structures for the $f$-wave odd-parity AM state described by $H_{\rm{eff}}(\textbf{\textit{k}})$. 
			(b) Spin-layer-resolved band structure near the $K$ and $K'$ valley under CPL with $\hbar\omega = 0.10$ eV at light intensity $eA/\hbar = 0.20~\mathrm{\AA}^{-1}$.(c) OHC as a function of energy for the equilibrium system and the nonequilibrium system driven by CPL with different photon energies $\hbar\omega$ at $eA/\hbar = 0.20~\mathrm{\AA}^{-1}$.
			(d) The orbital-resolved band structures near the $K$ valley under different photon energies $\hbar \omega=0.1, 0.4, 0.5$ eV at $eA/\hbar = 0.20~\mathrm{\AA}^{-1}$. 
		}
	\end{center}
\end{figure}

Notably, the photon-mediated processes encapsulated in $H_F(\textit{\textbf{k}},\omega)$
induce Floquet-Bloch sidebands that replicate the equilibrium electronic structure. 
Increasing the photon energy $\hbar\omega$ shifts these sidebands toward the Fermi level [Fig.S4] \cite{SuppMat}, providing a spectral signature for experimental detection via time-resolved angle-resolved photoemission spectroscopy (Tr-ARPES) \cite{Zhou_2023_PseudospinFloquetBlackPhosphorus}.
In the sub-resonant regime where the driving energy remains well below the band-touching threshold (e.g., $\hbar\omega = 0.10$ eV), the OHC is markedly amplified to nearly triple its equilibrium value, forming a robust plateau of $-12.00~e/2\pi$.
This surge originates from the coherent dressing of states, which triples the positive OBC peaks at the $K$ and $K'$ valleys without necessitating a band gap closing [Fig.S10] \cite{SuppMat}.
Meanwhile, a zero anomalous Hall conductivity (AHC) plateau appears near the Fermi level [Figs.3(a) and (d)], determined by the momentum-space integration of the BC.
Crucially, with increasing light intensity $eA/\hbar$, the spin degeneracy is further lifted, resulting in a significantly enhanced odd-parity spin splitting at the $K$ and $K'$ valleys [Fig.2(b)].
The massive, field-tunable enhancement of orbital Hall transport and odd-parity spin splitting via coherent Floquet-dressing establishes altermagnets as a reliable foundation for petahertz orbitronics.

\begin{figure}[htp]
	\begin{center}\scalebox{1}{\includegraphics{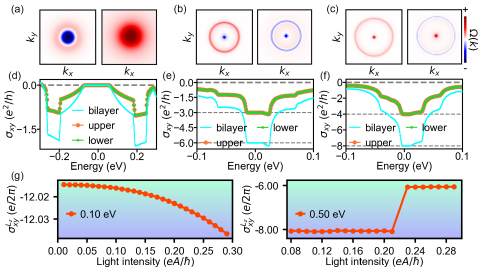}}
		\caption{ BC (left panel) and OBC (right panel) of the TB model in the $K$ valley under the CPL with photon energy $\hbar\omega $ (a) $= 0.10$ eV, (b) $= 0.40$ eV and (c) $= 0.50$ eV. 
			Total and layer-resolved AHC under the CPL with photon energy $\hbar\omega $ (d) $= 0.10$ eV, (e) $= 0.40$ eV and (f) $= 0.50$ eV. 
			The light intensity is $eA/\hbar = 0.20~\mathrm{\AA^{-1}}$.
			(g) OHC of the TB model under CPL with $\hbar\omega =$ $ 0.10$ eV and 0.50 eV as a function of light intensity $eA/\hbar$.}
	\end{center}
\end{figure}

As the photon energy advances through the half-gap resonance, the system enters a topological regime where the emergence and subsequent gapping of spin-degenerate type-I nodal rings trigger a nontrivial orbital inversion [Fig.2(d)].
This evolution initiates a sequence of intricate TPTs that fundamentally reconfigure the electronic landscape [Figs.S4–S11] \cite{SuppMat}.
Specifically, at $\hbar \omega \ge 0.20$ eV, orbital character exchange between the $E_v^{1}$ and $E_c^{-1}$ bands coincides with the first reopening of the global gap. 
This doubly degenerate band inversion generates positive OBC peaks at the $K$ and $K'$ valleys, encircled by negative contributions, whereas the BC exhibits positive valley-centered peaks surrounded by positive rings [Fig.3(c)].
The resulting TPT establishes a quantized AHC plateau with a Chern number $C=-8$ and an OHC of $-8.05~e/2\pi$ [Figs.2(c) and 3(f)], with each layer contributing an equal $-4~e^2/h$ to the anomalous Hall transport \cite{PhysRevB.111.155404}.
Significantly, the light-induced $f$-wave AM determines a valley-contrasting band inversion as light intensity further increases.
This secondary inversion involving spin-polarized states (e.g., spin-up in the lower layer at $K$ valleys) reverses the valley-centered positive OBC peaks while leaving the surrounding negative rings intact [Fig.3(b)]. 
This evolution modifies the OHC to $-6.07~e/2\pi$ and reconfigures the topological invariant to $C=-6$.
The quantized AHC reemerges with symmetric layer contributions of $-3~e^2/h$, hallmarked by a BC landscape where negative valley peaks are enclosed by positive spectral rings [Fig.3(e)].
Collectively, light-induced AM underpins the co-evolution of discrete topological invariants and the orbital Hall response.

Evidently, the topology-related band inversion exert modulation on orbital transport.
From Eq.(3), it is the phase perturbation of the manifold induced by the OAM that gives rise to the nontrivial behavior of the OBC describing geometric rotations. 
In the topologically trivial phase, where $\mathcal{T}$ symmetry is broken while $\mathcal{P}$ is preserved, the OAM is no longer pinned to zero at $K$ and $K'$ but instead acquires identical positive values at both valleys [Fig.S11] \cite{SuppMat}.
The OHC remains relatively stable in this regime, with a marginal increase driven by light-induced band-gap narrowing [Fig.3(g)].
However, traversing the TPT from $C=-8$ to $C=-6$ triggers a pronounced drop in the OHC. 
This transition is marked by a reconfiguration of the OAM texture: two sign reversals from the valley center to the outer region for $C=-8$ phase reduce to a single reversal for QAH phase with $C=-6$ phase.
This correspondence establishes the OHC plateau as a macroscopic fingerprint of the microscopic topological restructuring of orbital degrees of freedom.

\textit{Material Realization}-To validate the theoretical analysis presented above, we identify bilayer VSi$_2$N$_4$ as a promising platform to host the predicted ultrafast phenomena \cite{doi:10.1126/science.abb7023,Wang_2021_MA2Z4TopologicalMagneticSuperconducting}.
Our first-principles calculations reveal that the bilayer favors a van der Waals AFM ground state with opposite magnetic moments on the two V sublattices, characterized by an in-plane magnetization and a magnetocrystalline anisotropy energy (MAE = $E_x - E_z$) of -70 $\mu$eV.
Following our theoretical model, the bilayer is constructed by rotating the upper layer by $180^{\circ}$ around the $C_{2z}$ axis relative to the lower layer [Figs.4(a) and S13-S14] \cite{Liu2024Tailoring,doi:10.1021/acs.nanolett.4c00597,doi:10.1021/acs.nanolett.3c04597,D2MH00906D,SuppMat}.
Near the Fermi level, the equilibrium electronic structure features degenerate spin-up (lower layer) and spin-down (upper layer) bands, with a direct gap of 0.3544 eV.
The bilayer OHC reaches $2.34~e/2\pi$, nearly twice the monolayer value ($1.16~e/2\pi$), which confirms that weak interlayer coupling allows the system to act as a coherent superposition of individual layers, consistent with observations in transition-metal dichalcogenide bilayers \cite{PhysRevLett.126.056601}.
Belonging to the $p\bar{3}m1$ space group, the bilayer preserves the $[C_2||\bar{3}_{001}]$ spin-group symmetry while breaking $[C_2||M_z]$.
This specific symmetry configuration dictates that the light-induced odd-parity AM exhibits a characteristic $f$-wave angular dependence [Fig.4(b)].

\begin{figure}[htp]
	\begin{center}\scalebox{1.1}{\includegraphics{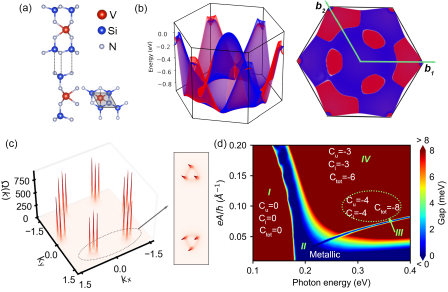}}
		\caption{(a) Crystal structure of the VSi$_2$N$_4$ bilayer. The gray region denotes the primitive cell. (b) The spin-resolved 3D band structures under CPL.
			(c) Distribution of BC corresponding to the $C=-8$ QAH phase.
			(d) Phase diagram of band gap as functions of photon energy $\hbar\omega$ and light intensity $eA/\hbar$. }
	\end{center}
\end{figure}

Guided by symmetry considerations, we reveal that the interplay between photon energy and light intensity drives the VSi$_2$N$_4$ bilayer through a rich sequence of TPTs [Figs.4(d) and S15-S19] \cite{SuppMat}.
The phase diagram, mapped by the band gap in the ($\hbar\omega$, $eA/\hbar$) parameter space, exhibits four distinct regimes: a trivial insulator (I, $C_l=0$, $C_u=0$, $C_{\mathrm{tot}}=0$), a globally gapless phase (II), and two QAH phases with high Chern numbers (III, $C_l=-4$, $C_u=-4$, $C_{\mathrm{tot}}=-8$; IV, $C_l=-3$, $C_u=-3$, $C_{\mathrm{tot}}=-6$) \cite{PhysRevLett.127.236402}, where $C_{l(u)}$ is the layer-resolved Chern number for the lower (upper) layer.
In the trivial region I, the OHC is amplified to $7.01~e/2\pi$, approximately triple its equilibrium value. 
Upon entering the topological regime III, the OHC drops to $3.15~e/2\pi$, and further diminishes to $1.1~e/2\pi$ in region IV as valley-contrasting band inversions reshape the OBC distribution.
Remarkably, the large topological invariants in these phases are underpinned by trigonal warping at the $K$ and $K'$ valleys \cite{PhysRevLett.98.176806,PhysRevB.82.113405,PhysRevB.95.045424,PhysRevB.104.195427}, which manifests as three distinct BC peaks in the valence bands [Fig.4(c)].
Notably, this warping-induced proliferation of BC hot spots provides a geometric mechanism to elevate the Chern numbers. 
While $\mathcal{P}$ symmetry ensures mirror-symmetric BC distributions at $K$ and $K'$, the light-driven warping expands the Floquet QAH phase space, facilitating the engineering of high-Chern-number states.
The above mechanism can be observed in both trilayer TB models and VSi$_2$N$_4$ trilayer, where CPL can induce QAH phases with Chern numbers $C=\pm12$ and $C=\pm9$, and the OHC can reach up to $10.5~e/2\pi$ in VSi$_2$N$_4$ [Figs.S12 and S21-22] \cite{SuppMat}. 
Moreover, reversing the CPL handedness flips the chirality of QAH phase while leaving the overall trend and sign of the OHC unchanged.

\textit{Conclusion}-In this work, we have demonstrated that Floquet engineering can induce odd-parity AM in collinear antiferromagnets by coupling CPL and layer degree of freedom. 
This mechanism transcends conventional paradigms, enabling the sub-picosecond manipulation of the OHE and the realization of nonequilibrium topological states with high Chern numbers through the hybridization of Floquet-Bloch bands. 
First-principles calculations substantiate VSi$_2$N$_4$ as a robust platform for hosting light-induced QAH phases and topology-engineered orbital transport, characterized by $m_z=\pm2$.
Given its crystal-field origin and resilience to realistic perturbations, the OHE offers a scalable and energy-efficient alternative to the spin Hall effect, particularly for orbitronic applications based on abundant light elements. 
Considering that the required light parameters are already accessible in current experimental setups \cite{Bielinski_2025_FloquetBloch,Choi_2025_FloquetBlochGraphene,McIver_2020_LightInducedAHEGraphene,Merboldt_2025_FloquetGraphene} and that several MSi$_2$Z$_4$ monolayers have been successfully synthesized, our findings provide a timely encouragement for the experimental exploration of light-driven topological phenomena and petahertz-scale orbitronics.

\section*{Acknowledgements}

This work was financially supported by the LiaoNing Revitalization Talents Program (Grant No. XLYC1907033) and the Natural Science Foundation of Liaoning province (Grant No. 2023-MS-072).
X. K. acknowledges the start up funding from Northeastern University, China. This work is supported by the Fundamental Research Funds for the Central Universities (No. N25LPY025).

\section{Reference}
\bibliography{reference}

\begin{thebibliography}{132}%
\makeatletter
\providecommand \@ifxundefined [1]{%
 \@ifx{#1\undefined}
}%
\providecommand \@ifnum [1]{%
 \ifnum #1\expandafter \@firstoftwo
 \else \expandafter \@secondoftwo
 \fi
}%
\providecommand \@ifx [1]{%
 \ifx #1\expandafter \@firstoftwo
 \else \expandafter \@secondoftwo
 \fi
}%
\providecommand \natexlab [1]{#1}%
\providecommand \enquote  [1]{``#1''}%
\providecommand \bibnamefont  [1]{#1}%
\providecommand \bibfnamefont [1]{#1}%
\providecommand \citenamefont [1]{#1}%
\providecommand \href@noop [0]{\@secondoftwo}%
\providecommand \href [0]{\begingroup \@sanitize@url \@href}%
\providecommand \@href[1]{\@@startlink{#1}\@@href}%
\providecommand \@@href[1]{\endgroup#1\@@endlink}%
\providecommand \@sanitize@url [0]{\catcode `\\12\catcode `\$12\catcode
  `\&12\catcode `\#12\catcode `\^12\catcode `\_12\catcode `\%12\relax}%
\providecommand \@@startlink[1]{}%
\providecommand \@@endlink[0]{}%
\providecommand \url  [0]{\begingroup\@sanitize@url \@url }%
\providecommand \@url [1]{\endgroup\@href {#1}{\urlprefix }}%
\providecommand \urlprefix  [0]{URL }%
\providecommand \Eprint [0]{\href }%
\providecommand \doibase [0]{https://doi.org/}%
\providecommand \selectlanguage [0]{\@gobble}%
\providecommand \bibinfo  [0]{\@secondoftwo}%
\providecommand \bibfield  [0]{\@secondoftwo}%
\providecommand \translation [1]{[#1]}%
\providecommand \BibitemOpen [0]{}%
\providecommand \bibitemStop [0]{}%
\providecommand \bibitemNoStop [0]{.\EOS\space}%
\providecommand \EOS [0]{\spacefactor3000\relax}%
\providecommand \BibitemShut  [1]{\csname bibitem#1\endcsname}%
\let\auto@bib@innerbib\@empty
%</preamble>
\bibitem [{\citenamefont {Zhou}\ \emph {et~al.}(2021)\citenamefont {Zhou},
  \citenamefont {Feng}, \citenamefont {Yang}, \citenamefont {Guo},\ and\
  \citenamefont {Yao}}]{PhysRevB.104.024401}%
  \BibitemOpen
  \bibfield  {author} {\bibinfo {author} {\bibfnamefont {X.}~\bibnamefont
  {Zhou}}, \bibinfo {author} {\bibfnamefont {W.}~\bibnamefont {Feng}}, \bibinfo
  {author} {\bibfnamefont {X.}~\bibnamefont {Yang}}, \bibinfo {author}
  {\bibfnamefont {G.-Y.}\ \bibnamefont {Guo}},\ and\ \bibinfo {author}
  {\bibfnamefont {Y.}~\bibnamefont {Yao}},\ }\bibinfo {title} {Crystal
  chirality magneto-optical effects in collinear antiferromagnets},\ \href
  {https://doi.org/10.1103/PhysRevB.104.024401} {\bibfield  {journal} {\bibinfo
   {journal} {Phys. Rev. B}\ }\textbf {\bibinfo {volume} {104}},\ \bibinfo
  {pages} {024401} (\bibinfo {year} {2021})}\BibitemShut {NoStop}%
\bibitem [{\citenamefont {Ma}\ \emph {et~al.}(2021)\citenamefont {Ma},
  \citenamefont {Hu}, \citenamefont {Li}, \citenamefont {Liu}, \citenamefont
  {Yao}, \citenamefont {Jia},\ and\ \citenamefont
  {Liu}}]{Ma2021Multifunctional}%
  \BibitemOpen
  \bibfield  {author} {\bibinfo {author} {\bibfnamefont {H.-Y.}\ \bibnamefont
  {Ma}}, \bibinfo {author} {\bibfnamefont {M.}~\bibnamefont {Hu}}, \bibinfo
  {author} {\bibfnamefont {N.}~\bibnamefont {Li}}, \bibinfo {author}
  {\bibfnamefont {J.}~\bibnamefont {Liu}}, \bibinfo {author} {\bibfnamefont
  {W.}~\bibnamefont {Yao}}, \bibinfo {author} {\bibfnamefont {J.-F.}\
  \bibnamefont {Jia}},\ and\ \bibinfo {author} {\bibfnamefont {J.}~\bibnamefont
  {Liu}},\ }\bibinfo {title} {Multifunctional antiferromagnetic materials with
  giant piezomagnetism and noncollinear spin current},\ \bibfield  {journal}
  {\bibinfo  {journal} {Nature Communications}\ }\textbf {\bibinfo {volume}
  {12}},\ \href {https://doi.org/10.1038/s41467-021-23127-7}
  {10.1038/s41467-021-23127-7} (\bibinfo {year} {2021})\BibitemShut {NoStop}%
\bibitem [{\citenamefont {Reichlova}\ \emph {et~al.}(2024)\citenamefont
  {Reichlova}, \citenamefont {Lopes~Seeger}, \citenamefont
  {González-Hernández}, \citenamefont {Kounta}, \citenamefont {Schlitz},
  \citenamefont {Kriegner}, \citenamefont {Ritzinger}, \citenamefont {Lammel},
  \citenamefont {Leiviskä}, \citenamefont {Birk~Hellenes}, \citenamefont
  {Olejník}, \citenamefont {Petřiček}, \citenamefont {Doležal},
  \citenamefont {Horak}, \citenamefont {Schmoranzerova}, \citenamefont
  {Badura}, \citenamefont {Bertaina}, \citenamefont {Thomas}, \citenamefont
  {Baltz}, \citenamefont {Michez}, \citenamefont {Sinova}, \citenamefont
  {Goennenwein}, \citenamefont {Jungwirth},\ and\ \citenamefont
  {Šmejkal}}]{Reichlova2024Observation}%
  \BibitemOpen
  \bibfield  {author} {\bibinfo {author} {\bibfnamefont {H.}~\bibnamefont
  {Reichlova}}, \bibinfo {author} {\bibfnamefont {R.}~\bibnamefont
  {Lopes~Seeger}}, \bibinfo {author} {\bibfnamefont {R.}~\bibnamefont
  {González-Hernández}}, \bibinfo {author} {\bibfnamefont {I.}~\bibnamefont
  {Kounta}}, \bibinfo {author} {\bibfnamefont {R.}~\bibnamefont {Schlitz}},
  \bibinfo {author} {\bibfnamefont {D.}~\bibnamefont {Kriegner}}, \bibinfo
  {author} {\bibfnamefont {P.}~\bibnamefont {Ritzinger}}, \bibinfo {author}
  {\bibfnamefont {M.}~\bibnamefont {Lammel}}, \bibinfo {author} {\bibfnamefont
  {M.}~\bibnamefont {Leiviskä}}, \bibinfo {author} {\bibfnamefont
  {A.}~\bibnamefont {Birk~Hellenes}}, \bibinfo {author} {\bibfnamefont
  {K.}~\bibnamefont {Olejník}}, \bibinfo {author} {\bibfnamefont
  {V.}~\bibnamefont {Petřiček}}, \bibinfo {author} {\bibfnamefont
  {P.}~\bibnamefont {Doležal}}, \bibinfo {author} {\bibfnamefont
  {L.}~\bibnamefont {Horak}}, \bibinfo {author} {\bibfnamefont
  {E.}~\bibnamefont {Schmoranzerova}}, \bibinfo {author} {\bibfnamefont
  {A.}~\bibnamefont {Badura}}, \bibinfo {author} {\bibfnamefont
  {S.}~\bibnamefont {Bertaina}}, \bibinfo {author} {\bibfnamefont
  {A.}~\bibnamefont {Thomas}}, \bibinfo {author} {\bibfnamefont
  {V.}~\bibnamefont {Baltz}}, \bibinfo {author} {\bibfnamefont
  {L.}~\bibnamefont {Michez}}, \bibinfo {author} {\bibfnamefont
  {J.}~\bibnamefont {Sinova}}, \bibinfo {author} {\bibfnamefont {S.~T.~B.}\
  \bibnamefont {Goennenwein}}, \bibinfo {author} {\bibfnamefont
  {T.}~\bibnamefont {Jungwirth}},\ and\ \bibinfo {author} {\bibfnamefont
  {L.}~\bibnamefont {Šmejkal}},\ }\bibinfo {title} {Observation of a
  spontaneous anomalous Hall response in the Mn$_5$Si$_3$ d-wave altermagnet},\
  \href {https://doi.org/10.1038/s41467-024-48493-w} {\bibfield  {journal}
  {\bibinfo  {journal} {Nature Communications}\ }\textbf {\bibinfo {volume}
  {15}} (\bibinfo {year} {2024})}\BibitemShut {NoStop}%
\bibitem [{\citenamefont {He}\ \emph {et~al.}(2023)\citenamefont {He},
  \citenamefont {Wang}, \citenamefont {Luo}, \citenamefont {Zeng},
  \citenamefont {Chen},\ and\ \citenamefont {Tang}}]{PhysRevLett.130.046401}%
  \BibitemOpen
  \bibfield  {author} {\bibinfo {author} {\bibfnamefont {R.}~\bibnamefont
  {He}}, \bibinfo {author} {\bibfnamefont {D.}~\bibnamefont {Wang}}, \bibinfo
  {author} {\bibfnamefont {N.}~\bibnamefont {Luo}}, \bibinfo {author}
  {\bibfnamefont {J.}~\bibnamefont {Zeng}}, \bibinfo {author} {\bibfnamefont
  {K.-Q.}\ \bibnamefont {Chen}},\ and\ \bibinfo {author} {\bibfnamefont
  {L.-M.}\ \bibnamefont {Tang}},\ }\bibinfo {title} {Nonrelativistic
  Spin-Momentum Coupling in Antiferromagnetic Twisted Bilayers},\ \href
  {https://doi.org/10.1103/PhysRevLett.130.046401} {\bibfield  {journal}
  {\bibinfo  {journal} {Phys. Rev. Lett.}\ }\textbf {\bibinfo {volume} {130}},\
  \bibinfo {pages} {046401} (\bibinfo {year} {2023})}\BibitemShut {NoStop}%
\bibitem [{\citenamefont {Pan}\ \emph {et~al.}(2024)\citenamefont {Pan},
  \citenamefont {Zhou}, \citenamefont {Lyu}, \citenamefont {Xiao},
  \citenamefont {Yang},\ and\ \citenamefont {Sun}}]{PhysRevLett.133.166701}%
  \BibitemOpen
  \bibfield  {author} {\bibinfo {author} {\bibfnamefont {B.}~\bibnamefont
  {Pan}}, \bibinfo {author} {\bibfnamefont {P.}~\bibnamefont {Zhou}}, \bibinfo
  {author} {\bibfnamefont {P.}~\bibnamefont {Lyu}}, \bibinfo {author}
  {\bibfnamefont {H.}~\bibnamefont {Xiao}}, \bibinfo {author} {\bibfnamefont
  {X.}~\bibnamefont {Yang}},\ and\ \bibinfo {author} {\bibfnamefont
  {L.}~\bibnamefont {Sun}},\ }\bibinfo {title} {General Stacking Theory for
  Altermagnetism in Bilayer Systems},\ \href
  {https://doi.org/10.1103/PhysRevLett.133.166701} {\bibfield  {journal}
  {\bibinfo  {journal} {Phys. Rev. Lett.}\ }\textbf {\bibinfo {volume} {133}},\
  \bibinfo {pages} {166701} (\bibinfo {year} {2024})}\BibitemShut {NoStop}%
\bibitem [{\citenamefont {Liu}\ \emph {et~al.}(2024)\citenamefont {Liu},
  \citenamefont {Yu},\ and\ \citenamefont {Liu}}]{PhysRevLett.133.206702}%
  \BibitemOpen
  \bibfield  {author} {\bibinfo {author} {\bibfnamefont {Y.}~\bibnamefont
  {Liu}}, \bibinfo {author} {\bibfnamefont {J.}~\bibnamefont {Yu}},\ and\
  \bibinfo {author} {\bibfnamefont {C.-C.}\ \bibnamefont {Liu}},\ }\bibinfo
  {title} {Twisted Magnetic Van der Waals Bilayers: An Ideal Platform for
  Altermagnetism},\ \href {https://doi.org/10.1103/PhysRevLett.133.206702}
  {\bibfield  {journal} {\bibinfo  {journal} {Phys. Rev. Lett.}\ }\textbf
  {\bibinfo {volume} {133}},\ \bibinfo {pages} {206702} (\bibinfo {year}
  {2024})}\BibitemShut {NoStop}%
\bibitem [{\citenamefont {Zhu}\ \emph {et~al.}(2025)\citenamefont {Zhu},
  \citenamefont {Huo}, \citenamefont {Feng}, \citenamefont {Zhang},
  \citenamefont {Yang},\ and\ \citenamefont {Guo}}]{PhysRevLett.134.166701}%
  \BibitemOpen
  \bibfield  {author} {\bibinfo {author} {\bibfnamefont {X.}~\bibnamefont
  {Zhu}}, \bibinfo {author} {\bibfnamefont {X.}~\bibnamefont {Huo}}, \bibinfo
  {author} {\bibfnamefont {S.}~\bibnamefont {Feng}}, \bibinfo {author}
  {\bibfnamefont {S.-B.}\ \bibnamefont {Zhang}}, \bibinfo {author}
  {\bibfnamefont {S.~A.}\ \bibnamefont {Yang}},\ and\ \bibinfo {author}
  {\bibfnamefont {H.}~\bibnamefont {Guo}},\ }\bibinfo {title} {Design of
  Altermagnetic Models from Spin Clusters},\ \href
  {https://doi.org/10.1103/PhysRevLett.134.166701} {\bibfield  {journal}
  {\bibinfo  {journal} {Phys. Rev. Lett.}\ }\textbf {\bibinfo {volume} {134}},\
  \bibinfo {pages} {166701} (\bibinfo {year} {2025})}\BibitemShut {NoStop}%
\bibitem [{\citenamefont {Ouassou}\ \emph {et~al.}(2023)\citenamefont
  {Ouassou}, \citenamefont {Brataas},\ and\ \citenamefont
  {Linder}}]{PhysRevLett.131.076003}%
  \BibitemOpen
  \bibfield  {author} {\bibinfo {author} {\bibfnamefont {J.~A.}\ \bibnamefont
  {Ouassou}}, \bibinfo {author} {\bibfnamefont {A.}~\bibnamefont {Brataas}},\
  and\ \bibinfo {author} {\bibfnamefont {J.}~\bibnamefont {Linder}},\ }\bibinfo
  {title} {dc Josephson Effect in Altermagnets},\ \href
  {https://doi.org/10.1103/PhysRevLett.131.076003} {\bibfield  {journal}
  {\bibinfo  {journal} {Phys. Rev. Lett.}\ }\textbf {\bibinfo {volume} {131}},\
  \bibinfo {pages} {076003} (\bibinfo {year} {2023})}\BibitemShut {NoStop}%
\bibitem [{\citenamefont {Gu}\ \emph {et~al.}(2025)\citenamefont {Gu},
  \citenamefont {Liu}, \citenamefont {Zhu}, \citenamefont {Yananose},
  \citenamefont {Chen}, \citenamefont {Hu}, \citenamefont {Stroppa},\ and\
  \citenamefont {Liu}}]{PhysRevLett.134.106802}%
  \BibitemOpen
  \bibfield  {author} {\bibinfo {author} {\bibfnamefont {M.}~\bibnamefont
  {Gu}}, \bibinfo {author} {\bibfnamefont {Y.}~\bibnamefont {Liu}}, \bibinfo
  {author} {\bibfnamefont {H.}~\bibnamefont {Zhu}}, \bibinfo {author}
  {\bibfnamefont {K.}~\bibnamefont {Yananose}}, \bibinfo {author}
  {\bibfnamefont {X.}~\bibnamefont {Chen}}, \bibinfo {author} {\bibfnamefont
  {Y.}~\bibnamefont {Hu}}, \bibinfo {author} {\bibfnamefont {A.}~\bibnamefont
  {Stroppa}},\ and\ \bibinfo {author} {\bibfnamefont {Q.}~\bibnamefont {Liu}},\
  }\bibinfo {title} {Ferroelectric Switchable Altermagnetism},\ \href
  {https://doi.org/10.1103/PhysRevLett.134.106802} {\bibfield  {journal}
  {\bibinfo  {journal} {Phys. Rev. Lett.}\ }\textbf {\bibinfo {volume} {134}},\
  \bibinfo {pages} {106802} (\bibinfo {year} {2025})}\BibitemShut {NoStop}%
\bibitem [{\citenamefont {Chen}\ \emph {et~al.}(2025)\citenamefont {Chen},
  \citenamefont {Liu}, \citenamefont {Lu},\ and\ \citenamefont
  {Xie}}]{zm5y-vy41}%
  \BibitemOpen
  \bibfield  {author} {\bibinfo {author} {\bibfnamefont {Y.}~\bibnamefont
  {Chen}}, \bibinfo {author} {\bibfnamefont {X.}~\bibnamefont {Liu}}, \bibinfo
  {author} {\bibfnamefont {H.-Z.}\ \bibnamefont {Lu}},\ and\ \bibinfo {author}
  {\bibfnamefont {X.~C.}\ \bibnamefont {Xie}},\ }\bibinfo {title} {Electrical
  Switching of Altermagnetism},\ \href {https://doi.org/10.1103/zm5y-vy41}
  {\bibfield  {journal} {\bibinfo  {journal} {Phys. Rev. Lett.}\ }\textbf
  {\bibinfo {volume} {135}},\ \bibinfo {pages} {016701} (\bibinfo {year}
  {2025})}\BibitemShut {NoStop}%
\bibitem [{\citenamefont {Duan}\ \emph {et~al.}(2025)\citenamefont {Duan},
  \citenamefont {Zhang}, \citenamefont {Zhu}, \citenamefont {Liu},
  \citenamefont {Zhang}, \citenamefont {\ifmmode \check{Z}\else
  \v{Z}\fi{}uti\ifmmode~\acute{c}\else \'{c}\fi{}},\ and\ \citenamefont
  {Zhou}}]{PhysRevLett.134.106801}%
  \BibitemOpen
  \bibfield  {author} {\bibinfo {author} {\bibfnamefont {X.}~\bibnamefont
  {Duan}}, \bibinfo {author} {\bibfnamefont {J.}~\bibnamefont {Zhang}},
  \bibinfo {author} {\bibfnamefont {Z.}~\bibnamefont {Zhu}}, \bibinfo {author}
  {\bibfnamefont {Y.}~\bibnamefont {Liu}}, \bibinfo {author} {\bibfnamefont
  {Z.}~\bibnamefont {Zhang}}, \bibinfo {author} {\bibfnamefont
  {I.}~\bibnamefont {\ifmmode \check{Z}\else
  \v{Z}\fi{}uti\ifmmode~\acute{c}\else \'{c}\fi{}}},\ and\ \bibinfo {author}
  {\bibfnamefont {T.}~\bibnamefont {Zhou}},\ }\bibinfo {title}
  {Antiferroelectric Altermagnets: Antiferroelectricity Alters Magnets},\ \href
  {https://doi.org/10.1103/PhysRevLett.134.106801} {\bibfield  {journal}
  {\bibinfo  {journal} {Phys. Rev. Lett.}\ }\textbf {\bibinfo {volume} {134}},\
  \bibinfo {pages} {106801} (\bibinfo {year} {2025})}\BibitemShut {NoStop}%
\bibitem [{\citenamefont {Gao}\ \emph {et~al.}(2025)\citenamefont {Gao},
  \citenamefont {Qu}, \citenamefont {Zeng}, \citenamefont {Liu}, \citenamefont
  {Wen}, \citenamefont {Sun}, \citenamefont {Guo},\ and\ \citenamefont
  {Lu}}]{Gao2025AIaccelerated}%
  \BibitemOpen
  \bibfield  {author} {\bibinfo {author} {\bibfnamefont {Z.-F.}\ \bibnamefont
  {Gao}}, \bibinfo {author} {\bibfnamefont {S.}~\bibnamefont {Qu}}, \bibinfo
  {author} {\bibfnamefont {B.}~\bibnamefont {Zeng}}, \bibinfo {author}
  {\bibfnamefont {Y.}~\bibnamefont {Liu}}, \bibinfo {author} {\bibfnamefont
  {J.-R.}\ \bibnamefont {Wen}}, \bibinfo {author} {\bibfnamefont
  {H.}~\bibnamefont {Sun}}, \bibinfo {author} {\bibfnamefont {P.-J.}\
  \bibnamefont {Guo}},\ and\ \bibinfo {author} {\bibfnamefont {Z.-Y.}\
  \bibnamefont {Lu}},\ }\bibinfo {title} {AI-accelerated discovery of
  altermagnetic materials},\ \href {https://doi.org/10.1093/nsr/nwaf066}
  {\bibfield  {journal} {\bibinfo  {journal} {National Science Review}\
  }\textbf {\bibinfo {volume} {12}},\ \bibinfo {pages} {nwaf066} (\bibinfo
  {year} {2025})}\BibitemShut {NoStop}%
\bibitem [{\citenamefont {Fukaya}\ \emph {et~al.}(2025)\citenamefont {Fukaya},
  \citenamefont {Lu}, \citenamefont {Yada}, \citenamefont {Tanaka},\ and\
  \citenamefont {Cayao}}]{Fukaya_2025}%
  \BibitemOpen
  \bibfield  {author} {\bibinfo {author} {\bibfnamefont {Y.}~\bibnamefont
  {Fukaya}}, \bibinfo {author} {\bibfnamefont {B.}~\bibnamefont {Lu}}, \bibinfo
  {author} {\bibfnamefont {K.}~\bibnamefont {Yada}}, \bibinfo {author}
  {\bibfnamefont {Y.}~\bibnamefont {Tanaka}},\ and\ \bibinfo {author}
  {\bibfnamefont {J.}~\bibnamefont {Cayao}},\ }\bibinfo {title}
  {Superconducting phenomena in systems with unconventional magnets},\ \href
  {https://doi.org/10.1088/1361-648X/adf1cf} {\bibfield  {journal} {\bibinfo
  {journal} {Journal of Physics: Condensed Matter}\ }\textbf {\bibinfo {volume}
  {37}},\ \bibinfo {pages} {313003} (\bibinfo {year} {2025})}\BibitemShut
  {NoStop}%
\bibitem [{\citenamefont {Urru}\ \emph {et~al.}(2025)\citenamefont {Urru},
  \citenamefont {Seleznev}, \citenamefont {Teng}, \citenamefont {Park},
  \citenamefont {Reyes-Lillo},\ and\ \citenamefont {Rabe}}]{v3fg-6smc}%
  \BibitemOpen
  \bibfield  {author} {\bibinfo {author} {\bibfnamefont {A.}~\bibnamefont
  {Urru}}, \bibinfo {author} {\bibfnamefont {D.}~\bibnamefont {Seleznev}},
  \bibinfo {author} {\bibfnamefont {Y.}~\bibnamefont {Teng}}, \bibinfo {author}
  {\bibfnamefont {S.~Y.}\ \bibnamefont {Park}}, \bibinfo {author}
  {\bibfnamefont {S.~E.}\ \bibnamefont {Reyes-Lillo}},\ and\ \bibinfo {author}
  {\bibfnamefont {K.~M.}\ \bibnamefont {Rabe}},\ }\bibinfo {title} {$G$-type
  antiferromagnetic ${\mathrm{BiFeO}}_{3}$ is a multiferroic $g$-wave
  altermagnet},\ \href {https://doi.org/10.1103/v3fg-6smc} {\bibfield
  {journal} {\bibinfo  {journal} {Phys. Rev. B}\ }\textbf {\bibinfo {volume}
  {112}},\ \bibinfo {pages} {104411} (\bibinfo {year} {2025})}\BibitemShut
  {NoStop}%
\bibitem [{\citenamefont {Hayami}\ \emph {et~al.}(2020)\citenamefont {Hayami},
  \citenamefont {Yanagi},\ and\ \citenamefont
  {Kusunose}}]{PhysRevB.102.144441}%
  \BibitemOpen
  \bibfield  {author} {\bibinfo {author} {\bibfnamefont {S.}~\bibnamefont
  {Hayami}}, \bibinfo {author} {\bibfnamefont {Y.}~\bibnamefont {Yanagi}},\
  and\ \bibinfo {author} {\bibfnamefont {H.}~\bibnamefont {Kusunose}},\
  }\bibinfo {title} {Bottom-up design of spin-split and reshaped electronic
  band structures in antiferromagnets without spin-orbit coupling: Procedure on
  the basis of augmented multipoles},\ \href
  {https://doi.org/10.1103/PhysRevB.102.144441} {\bibfield  {journal} {\bibinfo
   {journal} {Phys. Rev. B}\ }\textbf {\bibinfo {volume} {102}},\ \bibinfo
  {pages} {144441} (\bibinfo {year} {2020})}\BibitemShut {NoStop}%
\bibitem [{\citenamefont {Bai}\ \emph {et~al.}(2022)\citenamefont {Bai},
  \citenamefont {Han}, \citenamefont {Feng}, \citenamefont {Zhou},
  \citenamefont {Su}, \citenamefont {Wang}, \citenamefont {Liao}, \citenamefont
  {Zhu}, \citenamefont {Chen}, \citenamefont {Pan}, \citenamefont {Fan},\ and\
  \citenamefont {Song}}]{PhysRevLett.128.197202}%
  \BibitemOpen
  \bibfield  {author} {\bibinfo {author} {\bibfnamefont {H.}~\bibnamefont
  {Bai}}, \bibinfo {author} {\bibfnamefont {L.}~\bibnamefont {Han}}, \bibinfo
  {author} {\bibfnamefont {X.~Y.}\ \bibnamefont {Feng}}, \bibinfo {author}
  {\bibfnamefont {Y.~J.}\ \bibnamefont {Zhou}}, \bibinfo {author}
  {\bibfnamefont {R.~X.}\ \bibnamefont {Su}}, \bibinfo {author} {\bibfnamefont
  {Q.}~\bibnamefont {Wang}}, \bibinfo {author} {\bibfnamefont {L.~Y.}\
  \bibnamefont {Liao}}, \bibinfo {author} {\bibfnamefont {W.~X.}\ \bibnamefont
  {Zhu}}, \bibinfo {author} {\bibfnamefont {X.~Z.}\ \bibnamefont {Chen}},
  \bibinfo {author} {\bibfnamefont {F.}~\bibnamefont {Pan}}, \bibinfo {author}
  {\bibfnamefont {X.~L.}\ \bibnamefont {Fan}},\ and\ \bibinfo {author}
  {\bibfnamefont {C.}~\bibnamefont {Song}},\ }\bibinfo {title} {Observation of
  Spin Splitting Torque in a Collinear Antiferromagnet ${\mathrm{RuO}}_{2}$},\
  \href {https://doi.org/10.1103/PhysRevLett.128.197202} {\bibfield  {journal}
  {\bibinfo  {journal} {Phys. Rev. Lett.}\ }\textbf {\bibinfo {volume} {128}},\
  \bibinfo {pages} {197202} (\bibinfo {year} {2022})}\BibitemShut {NoStop}%
\bibitem [{\citenamefont {Mazin}(2022)}]{PhysRevX.12.040002}%
  \BibitemOpen
  \bibfield  {author} {\bibinfo {author} {\bibfnamefont {I.}~\bibnamefont
  {Mazin}} (\bibinfo {collaboration} {The PRX Editors}),\ }\bibinfo {title}
  {Editorial: Altermagnetism---A New Punch Line of Fundamental Magnetism},\
  \href {https://doi.org/10.1103/PhysRevX.12.040002} {\bibfield  {journal}
  {\bibinfo  {journal} {Phys. Rev. X}\ }\textbf {\bibinfo {volume} {12}},\
  \bibinfo {pages} {040002} (\bibinfo {year} {2022})}\BibitemShut {NoStop}%
\bibitem [{\citenamefont {Wu}\ \emph {et~al.}(2007)\citenamefont {Wu},
  \citenamefont {Sun}, \citenamefont {Fradkin},\ and\ \citenamefont
  {Zhang}}]{PhysRevB.75.115103}%
  \BibitemOpen
  \bibfield  {author} {\bibinfo {author} {\bibfnamefont {C.}~\bibnamefont
  {Wu}}, \bibinfo {author} {\bibfnamefont {K.}~\bibnamefont {Sun}}, \bibinfo
  {author} {\bibfnamefont {E.}~\bibnamefont {Fradkin}},\ and\ \bibinfo {author}
  {\bibfnamefont {S.-C.}\ \bibnamefont {Zhang}},\ }\bibinfo {title} {Fermi
  liquid instabilities in the spin channel},\ \href
  {https://doi.org/10.1103/PhysRevB.75.115103} {\bibfield  {journal} {\bibinfo
  {journal} {Phys. Rev. B}\ }\textbf {\bibinfo {volume} {75}},\ \bibinfo
  {pages} {115103} (\bibinfo {year} {2007})}\BibitemShut {NoStop}%
\bibitem [{\citenamefont {Gonzalez~Betancourt}\ \emph
  {et~al.}(2023)\citenamefont {Gonzalez~Betancourt}, \citenamefont
  {Zub\'a\ifmmode~\check{c}\else \v{c}\fi{}}, \citenamefont
  {Gonzalez-Hernandez}, \citenamefont {Geishendorf}, \citenamefont {\ifmmode
  \check{S}\else \v{S}\fi{}ob\'a\ifmmode~\check{n}\else \v{n}\fi{}},
  \citenamefont {Springholz}, \citenamefont {Olejn\'{\i}k}, \citenamefont
  {\ifmmode~\check{S}\else \v{S}\fi{}mejkal}, \citenamefont {Sinova},
  \citenamefont {Jungwirth}, \citenamefont {Goennenwein}, \citenamefont
  {Thomas}, \citenamefont {Reichlov\'a}, \citenamefont {\ifmmode~\check{Z}\else
  \v{Z}\fi{}elezn\'y},\ and\ \citenamefont
  {Kriegner}}]{PhysRevLett.130.036702}%
  \BibitemOpen
  \bibfield  {author} {\bibinfo {author} {\bibfnamefont {R.~D.}\ \bibnamefont
  {Gonzalez~Betancourt}}, \bibinfo {author} {\bibfnamefont {J.}~\bibnamefont
  {Zub\'a\ifmmode~\check{c}\else \v{c}\fi{}}}, \bibinfo {author} {\bibfnamefont
  {R.}~\bibnamefont {Gonzalez-Hernandez}}, \bibinfo {author} {\bibfnamefont
  {K.}~\bibnamefont {Geishendorf}}, \bibinfo {author} {\bibfnamefont
  {Z.}~\bibnamefont {\ifmmode \check{S}\else
  \v{S}\fi{}ob\'a\ifmmode~\check{n}\else \v{n}\fi{}}}, \bibinfo {author}
  {\bibfnamefont {G.}~\bibnamefont {Springholz}}, \bibinfo {author}
  {\bibfnamefont {K.}~\bibnamefont {Olejn\'{\i}k}}, \bibinfo {author}
  {\bibfnamefont {L.}~\bibnamefont {\ifmmode~\check{S}\else \v{S}\fi{}mejkal}},
  \bibinfo {author} {\bibfnamefont {J.}~\bibnamefont {Sinova}}, \bibinfo
  {author} {\bibfnamefont {T.}~\bibnamefont {Jungwirth}}, \bibinfo {author}
  {\bibfnamefont {S.~T.~B.}\ \bibnamefont {Goennenwein}}, \bibinfo {author}
  {\bibfnamefont {A.}~\bibnamefont {Thomas}}, \bibinfo {author} {\bibfnamefont
  {H.}~\bibnamefont {Reichlov\'a}}, \bibinfo {author} {\bibfnamefont
  {J.}~\bibnamefont {\ifmmode~\check{Z}\else \v{Z}\fi{}elezn\'y}},\ and\
  \bibinfo {author} {\bibfnamefont {D.}~\bibnamefont {Kriegner}},\ }\bibinfo
  {title} {Spontaneous Anomalous Hall Effect Arising from an Unconventional
  Compensated Magnetic Phase in a Semiconductor},\ \href
  {https://doi.org/10.1103/PhysRevLett.130.036702} {\bibfield  {journal}
  {\bibinfo  {journal} {Phys. Rev. Lett.}\ }\textbf {\bibinfo {volume} {130}},\
  \bibinfo {pages} {036702} (\bibinfo {year} {2023})}\BibitemShut {NoStop}%
\bibitem [{\citenamefont {Hayami}\ \emph {et~al.}(2019)\citenamefont {Hayami},
  \citenamefont {Yanagi},\ and\ \citenamefont
  {Kusunose}}]{doi:10.7566/JPSJ.88.123702}%
  \BibitemOpen
  \bibfield  {author} {\bibinfo {author} {\bibfnamefont {S.}~\bibnamefont
  {Hayami}}, \bibinfo {author} {\bibfnamefont {Y.}~\bibnamefont {Yanagi}},\
  and\ \bibinfo {author} {\bibfnamefont {H.}~\bibnamefont {Kusunose}},\
  }\bibinfo {title} {Momentum-Dependent Spin Splitting by Collinear
  Antiferromagnetic Ordering},\ \href {https://doi.org/10.7566/JPSJ.88.123702}
  {\bibfield  {journal} {\bibinfo  {journal} {Journal of the Physical Society
  of Japan}\ }\textbf {\bibinfo {volume} {88}},\ \bibinfo {pages} {123702}
  (\bibinfo {year} {2019})}\BibitemShut {NoStop}%
\bibitem [{\citenamefont {McClarty}\ and\ \citenamefont
  {Rau}(2024)}]{PhysRevLett.132.176702}%
  \BibitemOpen
  \bibfield  {author} {\bibinfo {author} {\bibfnamefont {P.~A.}\ \bibnamefont
  {McClarty}}\ and\ \bibinfo {author} {\bibfnamefont {J.~G.}\ \bibnamefont
  {Rau}},\ }\bibinfo {title} {Landau Theory of Altermagnetism},\ \href
  {https://doi.org/10.1103/PhysRevLett.132.176702} {\bibfield  {journal}
  {\bibinfo  {journal} {Phys. Rev. Lett.}\ }\textbf {\bibinfo {volume} {132}},\
  \bibinfo {pages} {176702} (\bibinfo {year} {2024})}\BibitemShut {NoStop}%
\bibitem [{\citenamefont {Krempaský}\ \emph {et~al.}(2024)\citenamefont
  {Krempaský}, \citenamefont {Šmejkal}, \citenamefont {D'Souza},
  \citenamefont {Hajlaoui}, \citenamefont {Springholz}, \citenamefont
  {Uhlířová}, \citenamefont {Alarab}, \citenamefont {Constantinou},
  \citenamefont {Strocov}, \citenamefont {Usanov}, \citenamefont {Pudelko},
  \citenamefont {González-Hernández}, \citenamefont {Hellenes}, \citenamefont
  {Jansa}, \citenamefont {Reichlová}, \citenamefont {Šobáň}, \citenamefont
  {Gonzalez~Betancourt}, \citenamefont {Wadley}, \citenamefont {Sinova},
  \citenamefont {Kriegner}, \citenamefont {Minár}, \citenamefont {Dil},\ and\
  \citenamefont {Jungwirth}}]{Krempasky2024Altermagnetic}%
  \BibitemOpen
  \bibfield  {author} {\bibinfo {author} {\bibfnamefont {J.}~\bibnamefont
  {Krempaský}}, \bibinfo {author} {\bibfnamefont {L.}~\bibnamefont
  {Šmejkal}}, \bibinfo {author} {\bibfnamefont {S.~W.}\ \bibnamefont
  {D'Souza}}, \bibinfo {author} {\bibfnamefont {M.}~\bibnamefont {Hajlaoui}},
  \bibinfo {author} {\bibfnamefont {G.}~\bibnamefont {Springholz}}, \bibinfo
  {author} {\bibfnamefont {K.}~\bibnamefont {Uhlířová}}, \bibinfo {author}
  {\bibfnamefont {F.}~\bibnamefont {Alarab}}, \bibinfo {author} {\bibfnamefont
  {P.~C.}\ \bibnamefont {Constantinou}}, \bibinfo {author} {\bibfnamefont
  {V.}~\bibnamefont {Strocov}}, \bibinfo {author} {\bibfnamefont
  {D.}~\bibnamefont {Usanov}}, \bibinfo {author} {\bibfnamefont {W.~R.}\
  \bibnamefont {Pudelko}}, \bibinfo {author} {\bibfnamefont {R.}~\bibnamefont
  {González-Hernández}}, \bibinfo {author} {\bibfnamefont {A.~B.}\
  \bibnamefont {Hellenes}}, \bibinfo {author} {\bibfnamefont {Z.}~\bibnamefont
  {Jansa}}, \bibinfo {author} {\bibfnamefont {H.}~\bibnamefont {Reichlová}},
  \bibinfo {author} {\bibfnamefont {Z.}~\bibnamefont {Šobáň}}, \bibinfo
  {author} {\bibfnamefont {R.~D.}\ \bibnamefont {Gonzalez~Betancourt}},
  \bibinfo {author} {\bibfnamefont {P.}~\bibnamefont {Wadley}}, \bibinfo
  {author} {\bibfnamefont {J.}~\bibnamefont {Sinova}}, \bibinfo {author}
  {\bibfnamefont {D.}~\bibnamefont {Kriegner}}, \bibinfo {author}
  {\bibfnamefont {J.}~\bibnamefont {Minár}}, \bibinfo {author} {\bibfnamefont
  {J.~H.}\ \bibnamefont {Dil}},\ and\ \bibinfo {author} {\bibfnamefont
  {T.}~\bibnamefont {Jungwirth}},\ }\bibinfo {title} {Altermagnetic lifting of
  Kramers spin degeneracy},\ \href {https://doi.org/10.1038/s41586-023-06907-7}
  {\bibfield  {journal} {\bibinfo  {journal} {Nature}\ }\textbf {\bibinfo
  {volume} {626}},\ \bibinfo {pages} {517} (\bibinfo {year}
  {2024})}\BibitemShut {NoStop}%
\bibitem [{\citenamefont {Song}\ \emph {et~al.}(2025)\citenamefont {Song},
  \citenamefont {Bai}, \citenamefont {Zhou}, \citenamefont {Han}, \citenamefont
  {Reichlova}, \citenamefont {Dil}, \citenamefont {Liu}, \citenamefont {Chen},\
  and\ \citenamefont {Pan}}]{Song2025Altermagnets}%
  \BibitemOpen
  \bibfield  {author} {\bibinfo {author} {\bibfnamefont {C.}~\bibnamefont
  {Song}}, \bibinfo {author} {\bibfnamefont {H.}~\bibnamefont {Bai}}, \bibinfo
  {author} {\bibfnamefont {Z.}~\bibnamefont {Zhou}}, \bibinfo {author}
  {\bibfnamefont {L.}~\bibnamefont {Han}}, \bibinfo {author} {\bibfnamefont
  {H.}~\bibnamefont {Reichlova}}, \bibinfo {author} {\bibfnamefont {J.~H.}\
  \bibnamefont {Dil}}, \bibinfo {author} {\bibfnamefont {J.}~\bibnamefont
  {Liu}}, \bibinfo {author} {\bibfnamefont {X.}~\bibnamefont {Chen}},\ and\
  \bibinfo {author} {\bibfnamefont {F.}~\bibnamefont {Pan}},\ }\bibinfo {title}
  {Altermagnets as a new class of functional materials},\ \href
  {https://doi.org/10.1038/s41578-025-00779-1} {\bibfield  {journal} {\bibinfo
  {journal} {Nature Reviews Materials}\ }\textbf {\bibinfo {volume} {10}},\
  \bibinfo {pages} {473} (\bibinfo {year} {2025})}\BibitemShut {NoStop}%
\bibitem [{\citenamefont {Bai}\ \emph {et~al.}(2024)\citenamefont {Bai},
  \citenamefont {Feng}, \citenamefont {Liu}, \citenamefont {Šmejkal},
  \citenamefont {Mokrousov},\ and\ \citenamefont
  {Yao}}]{https://doi.org/10.1002/adfm.202409327}%
  \BibitemOpen
  \bibfield  {author} {\bibinfo {author} {\bibfnamefont {L.}~\bibnamefont
  {Bai}}, \bibinfo {author} {\bibfnamefont {W.}~\bibnamefont {Feng}}, \bibinfo
  {author} {\bibfnamefont {S.}~\bibnamefont {Liu}}, \bibinfo {author}
  {\bibfnamefont {L.}~\bibnamefont {Šmejkal}}, \bibinfo {author}
  {\bibfnamefont {Y.}~\bibnamefont {Mokrousov}},\ and\ \bibinfo {author}
  {\bibfnamefont {Y.}~\bibnamefont {Yao}},\ }\bibinfo {title} {Altermagnetism:
  Exploring New Frontiers in Magnetism and Spintronics},\ \href
  {https://doi.org/https://doi.org/10.1002/adfm.202409327} {\bibfield
  {journal} {\bibinfo  {journal} {Advanced Functional Materials}\ }\textbf
  {\bibinfo {volume} {34}},\ \bibinfo {pages} {2409327} (\bibinfo {year}
  {2024})}\BibitemShut {NoStop}%
\bibitem [{\citenamefont {Cheong}\ and\ \citenamefont
  {Huang}(2025)}]{Cheong2025Altermagnetism}%
  \BibitemOpen
  \bibfield  {author} {\bibinfo {author} {\bibfnamefont {S.-W.}\ \bibnamefont
  {Cheong}}\ and\ \bibinfo {author} {\bibfnamefont {F.-T.}\ \bibnamefont
  {Huang}},\ }\bibinfo {title} {Altermagnetism classification},\ \href
  {https://doi.org/10.1038/s41535-025-00756-5} {\bibfield  {journal} {\bibinfo
  {journal} {npj Quantum Materials}\ }\textbf {\bibinfo {volume} {10}}
  (\bibinfo {year} {2025})}\BibitemShut {NoStop}%
\bibitem [{\citenamefont {Zhang}\ \emph {et~al.}(2025)\citenamefont {Zhang},
  \citenamefont {Cheng}, \citenamefont {Yin}, \citenamefont {Liu},
  \citenamefont {Deng}, \citenamefont {Qiao}, \citenamefont {Shi},
  \citenamefont {Zhang}, \citenamefont {Lin}, \citenamefont {Liu},
  \citenamefont {Ye}, \citenamefont {Huang}, \citenamefont {Meng},
  \citenamefont {Zhang}, \citenamefont {Okuda}, \citenamefont {Shimada},
  \citenamefont {Cui}, \citenamefont {Zhao}, \citenamefont {Cao}, \citenamefont
  {Qiao}, \citenamefont {Liu},\ and\ \citenamefont {Chen}}]{Zhang2025Crystal}%
  \BibitemOpen
  \bibfield  {author} {\bibinfo {author} {\bibfnamefont {F.}~\bibnamefont
  {Zhang}}, \bibinfo {author} {\bibfnamefont {X.}~\bibnamefont {Cheng}},
  \bibinfo {author} {\bibfnamefont {Z.}~\bibnamefont {Yin}}, \bibinfo {author}
  {\bibfnamefont {C.}~\bibnamefont {Liu}}, \bibinfo {author} {\bibfnamefont
  {L.}~\bibnamefont {Deng}}, \bibinfo {author} {\bibfnamefont {Y.}~\bibnamefont
  {Qiao}}, \bibinfo {author} {\bibfnamefont {Z.}~\bibnamefont {Shi}}, \bibinfo
  {author} {\bibfnamefont {S.}~\bibnamefont {Zhang}}, \bibinfo {author}
  {\bibfnamefont {J.}~\bibnamefont {Lin}}, \bibinfo {author} {\bibfnamefont
  {Z.}~\bibnamefont {Liu}}, \bibinfo {author} {\bibfnamefont {M.}~\bibnamefont
  {Ye}}, \bibinfo {author} {\bibfnamefont {Y.}~\bibnamefont {Huang}}, \bibinfo
  {author} {\bibfnamefont {X.}~\bibnamefont {Meng}}, \bibinfo {author}
  {\bibfnamefont {C.}~\bibnamefont {Zhang}}, \bibinfo {author} {\bibfnamefont
  {T.}~\bibnamefont {Okuda}}, \bibinfo {author} {\bibfnamefont
  {K.}~\bibnamefont {Shimada}}, \bibinfo {author} {\bibfnamefont
  {S.}~\bibnamefont {Cui}}, \bibinfo {author} {\bibfnamefont {Y.}~\bibnamefont
  {Zhao}}, \bibinfo {author} {\bibfnamefont {G.-H.}\ \bibnamefont {Cao}},
  \bibinfo {author} {\bibfnamefont {S.}~\bibnamefont {Qiao}}, \bibinfo {author}
  {\bibfnamefont {J.}~\bibnamefont {Liu}},\ and\ \bibinfo {author}
  {\bibfnamefont {C.}~\bibnamefont {Chen}},\ }\bibinfo {title}
  {Crystal-symmetry-paired spin–valley locking in a layered room-temperature
  metallic altermagnet candidate},\ \href
  {https://doi.org/10.1038/s41567-025-02864-2} {\bibfield  {journal} {\bibinfo
  {journal} {Nature Physics}\ }\textbf {\bibinfo {volume} {21}},\ \bibinfo
  {pages} {760} (\bibinfo {year} {2025})}\BibitemShut {NoStop}%
\bibitem [{\citenamefont {Hu}\ \emph {et~al.}(2025)\citenamefont {Hu},
  \citenamefont {Cheng}, \citenamefont {Huang},\ and\ \citenamefont
  {Liu}}]{PhysRevX.15.021083}%
  \BibitemOpen
  \bibfield  {author} {\bibinfo {author} {\bibfnamefont {M.}~\bibnamefont
  {Hu}}, \bibinfo {author} {\bibfnamefont {X.}~\bibnamefont {Cheng}}, \bibinfo
  {author} {\bibfnamefont {Z.}~\bibnamefont {Huang}},\ and\ \bibinfo {author}
  {\bibfnamefont {J.}~\bibnamefont {Liu}},\ }\bibinfo {title} {Catalog of
  $C$-Paired Spin-Momentum Locking in Antiferromagnetic Systems},\ \href
  {https://doi.org/10.1103/PhysRevX.15.021083} {\bibfield  {journal} {\bibinfo
  {journal} {Phys. Rev. X}\ }\textbf {\bibinfo {volume} {15}},\ \bibinfo
  {pages} {021083} (\bibinfo {year} {2025})}\BibitemShut {NoStop}%
\bibitem [{\citenamefont {\ifmmode~\check{S}\else \v{S}\fi{}mejkal}\ \emph
  {et~al.}(2022{\natexlab{a}})\citenamefont {\ifmmode~\check{S}\else
  \v{S}\fi{}mejkal}, \citenamefont {Sinova},\ and\ \citenamefont
  {Jungwirth}}]{PhysRevX.12.040501}%
  \BibitemOpen
  \bibfield  {author} {\bibinfo {author} {\bibfnamefont {L.}~\bibnamefont
  {\ifmmode~\check{S}\else \v{S}\fi{}mejkal}}, \bibinfo {author} {\bibfnamefont
  {J.}~\bibnamefont {Sinova}},\ and\ \bibinfo {author} {\bibfnamefont
  {T.}~\bibnamefont {Jungwirth}},\ }\bibinfo {title} {Emerging Research
  Landscape of Altermagnetism},\ \href
  {https://doi.org/10.1103/PhysRevX.12.040501} {\bibfield  {journal} {\bibinfo
  {journal} {Phys. Rev. X}\ }\textbf {\bibinfo {volume} {12}},\ \bibinfo
  {pages} {040501} (\bibinfo {year} {2022}{\natexlab{a}})}\BibitemShut
  {NoStop}%
\bibitem [{\citenamefont {\ifmmode~\check{S}\else \v{S}\fi{}mejkal}\ \emph
  {et~al.}(2022{\natexlab{b}})\citenamefont {\ifmmode~\check{S}\else
  \v{S}\fi{}mejkal}, \citenamefont {Sinova},\ and\ \citenamefont
  {Jungwirth}}]{PhysRevX.12.031042}%
  \BibitemOpen
  \bibfield  {author} {\bibinfo {author} {\bibfnamefont {L.}~\bibnamefont
  {\ifmmode~\check{S}\else \v{S}\fi{}mejkal}}, \bibinfo {author} {\bibfnamefont
  {J.}~\bibnamefont {Sinova}},\ and\ \bibinfo {author} {\bibfnamefont
  {T.}~\bibnamefont {Jungwirth}},\ }\bibinfo {title} {Beyond Conventional
  Ferromagnetism and Antiferromagnetism: A Phase with Nonrelativistic Spin and
  Crystal Rotation Symmetry},\ \href
  {https://doi.org/10.1103/PhysRevX.12.031042} {\bibfield  {journal} {\bibinfo
  {journal} {Phys. Rev. X}\ }\textbf {\bibinfo {volume} {12}},\ \bibinfo
  {pages} {031042} (\bibinfo {year} {2022}{\natexlab{b}})}\BibitemShut
  {NoStop}%
\bibitem [{\citenamefont {\ifmmode \check{Z}\else
  \v{Z}\fi{}uti\ifmmode~\acute{c}\else \'{c}\fi{}}\ \emph
  {et~al.}(2004)\citenamefont {\ifmmode \check{Z}\else
  \v{Z}\fi{}uti\ifmmode~\acute{c}\else \'{c}\fi{}}, \citenamefont {Fabian},\
  and\ \citenamefont {Das~Sarma}}]{RevModPhys.76.323}%
  \BibitemOpen
  \bibfield  {author} {\bibinfo {author} {\bibfnamefont {I.}~\bibnamefont
  {\ifmmode \check{Z}\else \v{Z}\fi{}uti\ifmmode~\acute{c}\else \'{c}\fi{}}},
  \bibinfo {author} {\bibfnamefont {J.}~\bibnamefont {Fabian}},\ and\ \bibinfo
  {author} {\bibfnamefont {S.}~\bibnamefont {Das~Sarma}},\ }\bibinfo {title}
  {Spintronics: Fundamentals and applications},\ \href
  {https://doi.org/10.1103/RevModPhys.76.323} {\bibfield  {journal} {\bibinfo
  {journal} {Rev. Mod. Phys.}\ }\textbf {\bibinfo {volume} {76}},\ \bibinfo
  {pages} {323} (\bibinfo {year} {2004})}\BibitemShut {NoStop}%
\bibitem [{\citenamefont {Koo}\ \emph {et~al.}(2020)\citenamefont {Koo},
  \citenamefont {Kim}, \citenamefont {Kim}, \citenamefont {Park}, \citenamefont
  {Choi}, \citenamefont {Kim}, \citenamefont {Go}, \citenamefont {Oh},
  \citenamefont {Lee}, \citenamefont {Park}, \citenamefont {Hong},\ and\
  \citenamefont {Lee}}]{Koo2020Rashba}%
  \BibitemOpen
  \bibfield  {author} {\bibinfo {author} {\bibfnamefont {H.~C.}\ \bibnamefont
  {Koo}}, \bibinfo {author} {\bibfnamefont {S.~B.}\ \bibnamefont {Kim}},
  \bibinfo {author} {\bibfnamefont {H.}~\bibnamefont {Kim}}, \bibinfo {author}
  {\bibfnamefont {T.-E.}\ \bibnamefont {Park}}, \bibinfo {author}
  {\bibfnamefont {J.~W.}\ \bibnamefont {Choi}}, \bibinfo {author}
  {\bibfnamefont {K.-W.}\ \bibnamefont {Kim}}, \bibinfo {author} {\bibfnamefont
  {G.}~\bibnamefont {Go}}, \bibinfo {author} {\bibfnamefont {J.~H.}\
  \bibnamefont {Oh}}, \bibinfo {author} {\bibfnamefont {D.-K.}\ \bibnamefont
  {Lee}}, \bibinfo {author} {\bibfnamefont {E.-S.}\ \bibnamefont {Park}},
  \bibinfo {author} {\bibfnamefont {I.-S.}\ \bibnamefont {Hong}},\ and\
  \bibinfo {author} {\bibfnamefont {K.-J.}\ \bibnamefont {Lee}},\ }\bibinfo
  {title} {Rashba Effect in Functional Spintronic Devices},\ \href
  {https://doi.org/10.1002/adma.202002117} {\bibfield  {journal} {\bibinfo
  {journal} {Advanced Materials}\ }\textbf {\bibinfo {volume} {32}},\ \bibinfo
  {pages} {2002117} (\bibinfo {year} {2020})}\BibitemShut {NoStop}%
\bibitem [{\citenamefont {Manchon}\ \emph {et~al.}(2015)\citenamefont
  {Manchon}, \citenamefont {Koo}, \citenamefont {Nitta}, \citenamefont
  {Frolov},\ and\ \citenamefont {Duine}}]{Manchon2015New}%
  \BibitemOpen
  \bibfield  {author} {\bibinfo {author} {\bibfnamefont {A.}~\bibnamefont
  {Manchon}}, \bibinfo {author} {\bibfnamefont {H.~C.}\ \bibnamefont {Koo}},
  \bibinfo {author} {\bibfnamefont {J.}~\bibnamefont {Nitta}}, \bibinfo
  {author} {\bibfnamefont {S.~M.}\ \bibnamefont {Frolov}},\ and\ \bibinfo
  {author} {\bibfnamefont {R.~A.}\ \bibnamefont {Duine}},\ }\bibinfo {title}
  {New perspectives for Rashba spin–orbit coupling},\ \href
  {https://doi.org/10.1038/nmat4360} {\bibfield  {journal} {\bibinfo  {journal}
  {Nature Materials}\ }\textbf {\bibinfo {volume} {14}},\ \bibinfo {pages}
  {871} (\bibinfo {year} {2015})}\BibitemShut {NoStop}%
\bibitem [{\citenamefont {Song}\ \emph {et~al.}(2025)\citenamefont {Song},
  \citenamefont {Stavrić}, \citenamefont {Barone}, \citenamefont {Droghetti},
  \citenamefont {Antonenko}, \citenamefont {Venderbos}, \citenamefont
  {Occhialini}, \citenamefont {Ilyas}, \citenamefont {Ergeçen}, \citenamefont
  {Gedik}, \citenamefont {Cheong}, \citenamefont {Fernandes}, \citenamefont
  {Picozzi},\ and\ \citenamefont {Comin}}]{Song2025Electrical}%
  \BibitemOpen
  \bibfield  {author} {\bibinfo {author} {\bibfnamefont {Q.}~\bibnamefont
  {Song}}, \bibinfo {author} {\bibfnamefont {S.}~\bibnamefont {Stavrić}},
  \bibinfo {author} {\bibfnamefont {P.}~\bibnamefont {Barone}}, \bibinfo
  {author} {\bibfnamefont {A.}~\bibnamefont {Droghetti}}, \bibinfo {author}
  {\bibfnamefont {D.~S.}\ \bibnamefont {Antonenko}}, \bibinfo {author}
  {\bibfnamefont {J.~W.~F.}\ \bibnamefont {Venderbos}}, \bibinfo {author}
  {\bibfnamefont {C.~A.}\ \bibnamefont {Occhialini}}, \bibinfo {author}
  {\bibfnamefont {B.}~\bibnamefont {Ilyas}}, \bibinfo {author} {\bibfnamefont
  {E.}~\bibnamefont {Ergeçen}}, \bibinfo {author} {\bibfnamefont
  {N.}~\bibnamefont {Gedik}}, \bibinfo {author} {\bibfnamefont {S.-W.}\
  \bibnamefont {Cheong}}, \bibinfo {author} {\bibfnamefont {R.~M.}\
  \bibnamefont {Fernandes}}, \bibinfo {author} {\bibfnamefont {S.}~\bibnamefont
  {Picozzi}},\ and\ \bibinfo {author} {\bibfnamefont {R.}~\bibnamefont
  {Comin}},\ }\bibinfo {title} {Electrical switching of a p-wave magnet},\
  \href {https://doi.org/10.1038/s41586-025-09034-7} {\bibfield  {journal}
  {\bibinfo  {journal} {Nature}\ }\textbf {\bibinfo {volume} {642}},\ \bibinfo
  {pages} {64} (\bibinfo {year} {2025})}\BibitemShut {NoStop}%
\bibitem [{\citenamefont {Hellenes}\ \emph {et~al.}(2023)\citenamefont
  {Hellenes}, \citenamefont {Jungwirth}, \citenamefont {Jaeschke-Ubiergo},
  \citenamefont {Chakraborty}, \citenamefont {Sinova},\ and\ \citenamefont
  {Šmejkal}}]{Hellenes2023Pwave}%
  \BibitemOpen
  \bibfield  {author} {\bibinfo {author} {\bibfnamefont {A.~B.}\ \bibnamefont
  {Hellenes}}, \bibinfo {author} {\bibfnamefont {T.}~\bibnamefont {Jungwirth}},
  \bibinfo {author} {\bibfnamefont {R.}~\bibnamefont {Jaeschke-Ubiergo}},
  \bibinfo {author} {\bibfnamefont {A.}~\bibnamefont {Chakraborty}}, \bibinfo
  {author} {\bibfnamefont {J.}~\bibnamefont {Sinova}},\ and\ \bibinfo {author}
  {\bibfnamefont {L.}~\bibnamefont {Šmejkal}},\ }\href
  {https://arxiv.org/abs/2309.01607} {\bibinfo {title} {P-wave magnets}}
  (\bibinfo {year} {2023})\BibitemShut {NoStop}%
\bibitem [{\citenamefont {Yu}\ \emph {et~al.}(2025)\citenamefont {Yu},
  \citenamefont {Lyngby}, \citenamefont {Shishidou}, \citenamefont {Roig},
  \citenamefont {Kreisel}, \citenamefont {Weinert}, \citenamefont {Andersen},\
  and\ \citenamefont {Agterberg}}]{zk69-k6b2}%
  \BibitemOpen
  \bibfield  {author} {\bibinfo {author} {\bibfnamefont {Y.}~\bibnamefont
  {Yu}}, \bibinfo {author} {\bibfnamefont {M.~B.}\ \bibnamefont {Lyngby}},
  \bibinfo {author} {\bibfnamefont {T.}~\bibnamefont {Shishidou}}, \bibinfo
  {author} {\bibfnamefont {M.}~\bibnamefont {Roig}}, \bibinfo {author}
  {\bibfnamefont {A.}~\bibnamefont {Kreisel}}, \bibinfo {author} {\bibfnamefont
  {M.}~\bibnamefont {Weinert}}, \bibinfo {author} {\bibfnamefont {B.~M.}\
  \bibnamefont {Andersen}},\ and\ \bibinfo {author} {\bibfnamefont {D.~F.}\
  \bibnamefont {Agterberg}},\ }\bibinfo {title} {Odd-Parity Magnetism Driven by
  Antiferromagnetic Exchange},\ \href {https://doi.org/10.1103/zk69-k6b2}
  {\bibfield  {journal} {\bibinfo  {journal} {Phys. Rev. Lett.}\ }\textbf
  {\bibinfo {volume} {135}},\ \bibinfo {pages} {046701} (\bibinfo {year}
  {2025})}\BibitemShut {NoStop}%
\bibitem [{\citenamefont {Hayami}\ \emph {et~al.}(2020)\citenamefont {Hayami},
  \citenamefont {Yanagi},\ and\ \citenamefont
  {Kusunose}}]{PhysRevB.101.220403}%
  \BibitemOpen
  \bibfield  {author} {\bibinfo {author} {\bibfnamefont {S.}~\bibnamefont
  {Hayami}}, \bibinfo {author} {\bibfnamefont {Y.}~\bibnamefont {Yanagi}},\
  and\ \bibinfo {author} {\bibfnamefont {H.}~\bibnamefont {Kusunose}},\
  }\bibinfo {title} {Spontaneous antisymmetric spin splitting in noncollinear
  antiferromagnets without spin-orbit coupling},\ \href
  {https://doi.org/10.1103/PhysRevB.101.220403} {\bibfield  {journal} {\bibinfo
   {journal} {Phys. Rev. B}\ }\textbf {\bibinfo {volume} {101}},\ \bibinfo
  {pages} {220403} (\bibinfo {year} {2020})}\BibitemShut {NoStop}%
\bibitem [{\citenamefont {Brekke}\ \emph {et~al.}(2024)\citenamefont {Brekke},
  \citenamefont {Sukhachov}, \citenamefont {Giil}, \citenamefont {Brataas},\
  and\ \citenamefont {Linder}}]{PhysRevLett.133.236703}%
  \BibitemOpen
  \bibfield  {author} {\bibinfo {author} {\bibfnamefont {B.}~\bibnamefont
  {Brekke}}, \bibinfo {author} {\bibfnamefont {P.}~\bibnamefont {Sukhachov}},
  \bibinfo {author} {\bibfnamefont {H.~G.}\ \bibnamefont {Giil}}, \bibinfo
  {author} {\bibfnamefont {A.}~\bibnamefont {Brataas}},\ and\ \bibinfo {author}
  {\bibfnamefont {J.}~\bibnamefont {Linder}},\ }\bibinfo {title} {Minimal
  Models and Transport Properties of Unconventional $p$-Wave Magnets},\ \href
  {https://doi.org/10.1103/PhysRevLett.133.236703} {\bibfield  {journal}
  {\bibinfo  {journal} {Phys. Rev. Lett.}\ }\textbf {\bibinfo {volume} {133}},\
  \bibinfo {pages} {236703} (\bibinfo {year} {2024})}\BibitemShut {NoStop}%
\bibitem [{\citenamefont {Bukov}\ \emph {et~al.}(2015)\citenamefont {Bukov},
  \citenamefont {D'Alessio},\ and\ \citenamefont
  {Polkovnikov}}]{Bukov04032015}%
  \BibitemOpen
  \bibfield  {author} {\bibinfo {author} {\bibfnamefont {M.}~\bibnamefont
  {Bukov}}, \bibinfo {author} {\bibfnamefont {L.}~\bibnamefont {D'Alessio}},\
  and\ \bibinfo {author} {\bibfnamefont {A.}~\bibnamefont {Polkovnikov}},\
  }\bibinfo {title} {Universal high-frequency behavior of periodically driven
  systems: from dynamical stabilization to Floquet engineering},\ \href
  {https://doi.org/10.1080/00018732.2015.1055918} {\bibfield  {journal}
  {\bibinfo  {journal} {Advances in Physics}\ }\textbf {\bibinfo {volume}
  {64}},\ \bibinfo {pages} {139} (\bibinfo {year} {2015})}\BibitemShut
  {NoStop}%
\bibitem [{\citenamefont {Zhan}\ \emph {et~al.}(2024)\citenamefont {Zhan},
  \citenamefont {Chen}, \citenamefont {Ning}, \citenamefont {Ma}, \citenamefont
  {Wang}, \citenamefont {Xu},\ and\ \citenamefont {Wang}}]{Zhan2024Floquet}%
  \BibitemOpen
  \bibfield  {author} {\bibinfo {author} {\bibfnamefont {F.}~\bibnamefont
  {Zhan}}, \bibinfo {author} {\bibfnamefont {R.}~\bibnamefont {Chen}}, \bibinfo
  {author} {\bibfnamefont {Z.}~\bibnamefont {Ning}}, \bibinfo {author}
  {\bibfnamefont {D.-S.}\ \bibnamefont {Ma}}, \bibinfo {author} {\bibfnamefont
  {Z.}~\bibnamefont {Wang}}, \bibinfo {author} {\bibfnamefont {D.-H.}\
  \bibnamefont {Xu}},\ and\ \bibinfo {author} {\bibfnamefont {R.}~\bibnamefont
  {Wang}},\ }\bibinfo {title} {Perspective: Floquet engineering topological
  states from effective models towards realistic materials},\ \href
  {https://doi.org/10.1007/s44214-024-00067-z} {\bibfield  {journal} {\bibinfo
  {journal} {Quantum Front}\ }\textbf {\bibinfo {volume} {3}} (\bibinfo {year}
  {2024})}\BibitemShut {NoStop}%
\bibitem [{\citenamefont {Shan}\ \emph {et~al.}(2021)\citenamefont {Shan},
  \citenamefont {M.~Ye}, \citenamefont {Lee}, \citenamefont {Park},
  \citenamefont {Balents},\ and\ \citenamefont
  {Hsieh}}]{10.1038/s41586-021-04051-8}%
  \BibitemOpen
  \bibfield  {author} {\bibinfo {author} {\bibfnamefont {J.-Y.}\ \bibnamefont
  {Shan}}, \bibinfo {author} {\bibfnamefont {H.~C.}\ \bibnamefont {M.~Ye}},
  \bibinfo {author} {\bibfnamefont {S.}~\bibnamefont {Lee}}, \bibinfo {author}
  {\bibfnamefont {J.-G.}\ \bibnamefont {Park}}, \bibinfo {author}
  {\bibfnamefont {L.}~\bibnamefont {Balents}},\ and\ \bibinfo {author}
  {\bibfnamefont {D.}~\bibnamefont {Hsieh}},\ }\bibinfo {title} {Giant
  modulation of optical nonlinearity by Floquet engineering},\ \href
  {https://doi.org/10.1038/s41586-021-04051-8} {\bibfield  {journal} {\bibinfo
  {journal} {Nature}\ }\textbf {\bibinfo {volume} {600}},\ \bibinfo {pages}
  {235–239} (\bibinfo {year} {2021})}\BibitemShut {NoStop}%
\bibitem [{\citenamefont {Bielinski}\ \emph {et~al.}(2025)\citenamefont
  {Bielinski}, \citenamefont {Chari}, \citenamefont {May-Mann}, \citenamefont
  {Kim}, \citenamefont {Zwettler}, \citenamefont {Deng}, \citenamefont
  {Aishwarya}, \citenamefont {Roychowdhury}, \citenamefont {Shekhar},
  \citenamefont {Hashimoto}, \citenamefont {Lu}, \citenamefont {Yan},
  \citenamefont {Felser}, \citenamefont {Madhavan}, \citenamefont {Shen},
  \citenamefont {Hughes},\ and\ \citenamefont
  {Mahmood}}]{Bielinski_2025_FloquetBloch}%
  \BibitemOpen
  \bibfield  {author} {\bibinfo {author} {\bibfnamefont {N.}~\bibnamefont
  {Bielinski}}, \bibinfo {author} {\bibfnamefont {R.}~\bibnamefont {Chari}},
  \bibinfo {author} {\bibfnamefont {J.}~\bibnamefont {May-Mann}}, \bibinfo
  {author} {\bibfnamefont {S.}~\bibnamefont {Kim}}, \bibinfo {author}
  {\bibfnamefont {J.}~\bibnamefont {Zwettler}}, \bibinfo {author}
  {\bibfnamefont {Y.}~\bibnamefont {Deng}}, \bibinfo {author} {\bibfnamefont
  {A.}~\bibnamefont {Aishwarya}}, \bibinfo {author} {\bibfnamefont
  {S.}~\bibnamefont {Roychowdhury}}, \bibinfo {author} {\bibfnamefont
  {C.}~\bibnamefont {Shekhar}}, \bibinfo {author} {\bibfnamefont
  {M.}~\bibnamefont {Hashimoto}}, \bibinfo {author} {\bibfnamefont
  {D.}~\bibnamefont {Lu}}, \bibinfo {author} {\bibfnamefont {J.}~\bibnamefont
  {Yan}}, \bibinfo {author} {\bibfnamefont {C.}~\bibnamefont {Felser}},
  \bibinfo {author} {\bibfnamefont {V.}~\bibnamefont {Madhavan}}, \bibinfo
  {author} {\bibfnamefont {Z.-X.}\ \bibnamefont {Shen}}, \bibinfo {author}
  {\bibfnamefont {T.~L.}\ \bibnamefont {Hughes}},\ and\ \bibinfo {author}
  {\bibfnamefont {F.}~\bibnamefont {Mahmood}},\ }\bibinfo {title}
  {Floquet–Bloch manipulation of the Dirac gap in a topological
  antiferromagnet},\ \href
  {https://doi.org/https://doi.org/10.1038/s41567-024-02769-6} {\bibfield
  {journal} {\bibinfo  {journal} {Nature Physics}\ }\textbf {\bibinfo {volume}
  {21}},\ \bibinfo {pages} {458} (\bibinfo {year} {2025})}\BibitemShut
  {NoStop}%
\bibitem [{\citenamefont {McIver}\ \emph {et~al.}(2020)\citenamefont {McIver},
  \citenamefont {Schulte}, \citenamefont {Stein}, \citenamefont {Matsuyama},
  \citenamefont {Jotzu}, \citenamefont {Meier},\ and\ \citenamefont
  {Cavalleri}}]{McIver_2020_LightInducedAHEGraphene}%
  \BibitemOpen
  \bibfield  {author} {\bibinfo {author} {\bibfnamefont {J.~W.}\ \bibnamefont
  {McIver}}, \bibinfo {author} {\bibfnamefont {B.}~\bibnamefont {Schulte}},
  \bibinfo {author} {\bibfnamefont {F.-U.}\ \bibnamefont {Stein}}, \bibinfo
  {author} {\bibfnamefont {T.}~\bibnamefont {Matsuyama}}, \bibinfo {author}
  {\bibfnamefont {G.}~\bibnamefont {Jotzu}}, \bibinfo {author} {\bibfnamefont
  {G.}~\bibnamefont {Meier}},\ and\ \bibinfo {author} {\bibfnamefont
  {A.}~\bibnamefont {Cavalleri}},\ }\bibinfo {title} {Light-induced anomalous
  Hall effect in graphene},\ \href {https://doi.org/10.1038/s41567-019-0698-y}
  {\bibfield  {journal} {\bibinfo  {journal} {Nature Physics}\ }\textbf
  {\bibinfo {volume} {16}},\ \bibinfo {pages} {38} (\bibinfo {year}
  {2020})}\BibitemShut {NoStop}%
\bibitem [{\citenamefont {Merboldt}\ \emph {et~al.}(2025)\citenamefont
  {Merboldt}, \citenamefont {Schüler}, \citenamefont {Schmitt}, \citenamefont
  {Bange}, \citenamefont {Bennecke}, \citenamefont {Gadge}, \citenamefont
  {Pierz}, \citenamefont {Schumacher}, \citenamefont {Momeni}, \citenamefont
  {Steil}, \citenamefont {Manmana}, \citenamefont {Sentef}, \citenamefont
  {Reutzel},\ and\ \citenamefont {Mathias}}]{Merboldt_2025_FloquetGraphene}%
  \BibitemOpen
  \bibfield  {author} {\bibinfo {author} {\bibfnamefont {M.}~\bibnamefont
  {Merboldt}}, \bibinfo {author} {\bibfnamefont {M.}~\bibnamefont {Schüler}},
  \bibinfo {author} {\bibfnamefont {D.}~\bibnamefont {Schmitt}}, \bibinfo
  {author} {\bibfnamefont {J.~P.}\ \bibnamefont {Bange}}, \bibinfo {author}
  {\bibfnamefont {W.}~\bibnamefont {Bennecke}}, \bibinfo {author}
  {\bibfnamefont {K.}~\bibnamefont {Gadge}}, \bibinfo {author} {\bibfnamefont
  {K.}~\bibnamefont {Pierz}}, \bibinfo {author} {\bibfnamefont {H.~W.}\
  \bibnamefont {Schumacher}}, \bibinfo {author} {\bibfnamefont
  {D.}~\bibnamefont {Momeni}}, \bibinfo {author} {\bibfnamefont
  {D.}~\bibnamefont {Steil}}, \bibinfo {author} {\bibfnamefont {S.~R.}\
  \bibnamefont {Manmana}}, \bibinfo {author} {\bibfnamefont {M.~A.}\
  \bibnamefont {Sentef}}, \bibinfo {author} {\bibfnamefont {M.}~\bibnamefont
  {Reutzel}},\ and\ \bibinfo {author} {\bibfnamefont {S.}~\bibnamefont
  {Mathias}},\ }\bibinfo {title} {Observation of Floquet states in graphene},\
  \href {https://doi.org/https://doi.org/10.1038/s41567-025-02889-7} {\bibfield
   {journal} {\bibinfo  {journal} {Nature Physics}\ }\textbf {\bibinfo {volume}
  {21}},\ \bibinfo {pages} {1093} (\bibinfo {year} {2025})}\BibitemShut
  {NoStop}%
\bibitem [{\citenamefont {Choi}\ \emph {et~al.}(2025)\citenamefont {Choi},
  \citenamefont {Mogi}, \citenamefont {De~Giovannini}, \citenamefont {Azoury},
  \citenamefont {Lv}, \citenamefont {Su}, \citenamefont {Hübener},
  \citenamefont {Rubio},\ and\ \citenamefont
  {Gedik}}]{Choi_2025_FloquetBlochGraphene}%
  \BibitemOpen
  \bibfield  {author} {\bibinfo {author} {\bibfnamefont {D.}~\bibnamefont
  {Choi}}, \bibinfo {author} {\bibfnamefont {M.}~\bibnamefont {Mogi}}, \bibinfo
  {author} {\bibfnamefont {U.}~\bibnamefont {De~Giovannini}}, \bibinfo {author}
  {\bibfnamefont {D.}~\bibnamefont {Azoury}}, \bibinfo {author} {\bibfnamefont
  {B.}~\bibnamefont {Lv}}, \bibinfo {author} {\bibfnamefont {Y.}~\bibnamefont
  {Su}}, \bibinfo {author} {\bibfnamefont {H.}~\bibnamefont {Hübener}},
  \bibinfo {author} {\bibfnamefont {A.}~\bibnamefont {Rubio}},\ and\ \bibinfo
  {author} {\bibfnamefont {N.}~\bibnamefont {Gedik}},\ }\bibinfo {title}
  {Observation of Floquet–Bloch states in monolayer graphene},\ \href
  {https://doi.org/https://doi.org/10.1038/s41567-025-02888-8} {\bibfield
  {journal} {\bibinfo  {journal} {Nature Physics}\ }\textbf {\bibinfo {volume}
  {21}},\ \bibinfo {pages} {1100} (\bibinfo {year} {2025})}\BibitemShut
  {NoStop}%
\bibitem [{\citenamefont {Wang}\ \emph {et~al.}(2025)\citenamefont {Wang},
  \citenamefont {Chen}, \citenamefont {Bao}, \citenamefont {Lin}, \citenamefont
  {Zhong}, \citenamefont {Zhang},\ and\ \citenamefont
  {Zhou}}]{PhysRevLett.134.146401}%
  \BibitemOpen
  \bibfield  {author} {\bibinfo {author} {\bibfnamefont {F.}~\bibnamefont
  {Wang}}, \bibinfo {author} {\bibfnamefont {W.}~\bibnamefont {Chen}}, \bibinfo
  {author} {\bibfnamefont {C.}~\bibnamefont {Bao}}, \bibinfo {author}
  {\bibfnamefont {T.}~\bibnamefont {Lin}}, \bibinfo {author} {\bibfnamefont
  {H.}~\bibnamefont {Zhong}}, \bibinfo {author} {\bibfnamefont
  {H.}~\bibnamefont {Zhang}},\ and\ \bibinfo {author} {\bibfnamefont
  {S.}~\bibnamefont {Zhou}},\ }\bibinfo {title} {Light-Field Dressing of
  Transient Photoexcited States above the Fermi Energy},\ \href
  {https://doi.org/10.1103/PhysRevLett.134.146401} {\bibfield  {journal}
  {\bibinfo  {journal} {Phys. Rev. Lett.}\ }\textbf {\bibinfo {volume} {134}},\
  \bibinfo {pages} {146401} (\bibinfo {year} {2025})}\BibitemShut {NoStop}%
\bibitem [{\citenamefont {Kong}\ \emph {et~al.}(2024)\citenamefont {Kong},
  \citenamefont {Zhang}, \citenamefont {Gong},\ and\ \citenamefont
  {Li}}]{D4TC02438A}%
  \BibitemOpen
  \bibfield  {author} {\bibinfo {author} {\bibfnamefont {X.}~\bibnamefont
  {Kong}}, \bibinfo {author} {\bibfnamefont {B.}~\bibnamefont {Zhang}},
  \bibinfo {author} {\bibfnamefont {W.-j.}\ \bibnamefont {Gong}},\ and\
  \bibinfo {author} {\bibfnamefont {L.}~\bibnamefont {Li}},\ }\bibinfo {title}
  {Tunable light-induced topological edge states in strain engineering of
  bismuthene monolayers},\ \href {https://doi.org/10.1039/D4TC02438A}
  {\bibfield  {journal} {\bibinfo  {journal} {J. Mater. Chem. C}\ }\textbf
  {\bibinfo {volume} {12}},\ \bibinfo {pages} {13325} (\bibinfo {year}
  {2024})}\BibitemShut {NoStop}%
\bibitem [{\citenamefont {Fragkos}\ \emph {et~al.}(2025)\citenamefont
  {Fragkos}, \citenamefont {Fabre}, \citenamefont {Tkach}, \citenamefont
  {Petit}, \citenamefont {Descamps}, \citenamefont {Schönhense}, \citenamefont
  {Mairesse}, \citenamefont {Schüler},\ and\ \citenamefont
  {Beaulieu}}]{Fragkos_2025_FloquetBlochValleytronics}%
  \BibitemOpen
  \bibfield  {author} {\bibinfo {author} {\bibfnamefont {S.}~\bibnamefont
  {Fragkos}}, \bibinfo {author} {\bibfnamefont {B.}~\bibnamefont {Fabre}},
  \bibinfo {author} {\bibfnamefont {O.}~\bibnamefont {Tkach}}, \bibinfo
  {author} {\bibfnamefont {S.}~\bibnamefont {Petit}}, \bibinfo {author}
  {\bibfnamefont {D.}~\bibnamefont {Descamps}}, \bibinfo {author}
  {\bibfnamefont {G.}~\bibnamefont {Schönhense}}, \bibinfo {author}
  {\bibfnamefont {Y.}~\bibnamefont {Mairesse}}, \bibinfo {author}
  {\bibfnamefont {M.}~\bibnamefont {Schüler}},\ and\ \bibinfo {author}
  {\bibfnamefont {S.}~\bibnamefont {Beaulieu}},\ }\bibinfo {title}
  {Floquet-Bloch valleytronics},\ \href@noop {} {\bibfield  {journal} {\bibinfo
   {journal} {Nature Communications}\ }\textbf {\bibinfo {volume} {16}}
  (\bibinfo {year} {2025})}\BibitemShut {NoStop}%
\bibitem [{\citenamefont {Kong}\ \emph {et~al.}(2022)\citenamefont {Kong},
  \citenamefont {Luo}, \citenamefont {Li}, \citenamefont {Yoon}, \citenamefont
  {Berlijn},\ and\ \citenamefont {Liang}}]{Kong_2022}%
  \BibitemOpen
  \bibfield  {author} {\bibinfo {author} {\bibfnamefont {X.}~\bibnamefont
  {Kong}}, \bibinfo {author} {\bibfnamefont {W.}~\bibnamefont {Luo}}, \bibinfo
  {author} {\bibfnamefont {L.}~\bibnamefont {Li}}, \bibinfo {author}
  {\bibfnamefont {M.}~\bibnamefont {Yoon}}, \bibinfo {author} {\bibfnamefont
  {T.}~\bibnamefont {Berlijn}},\ and\ \bibinfo {author} {\bibfnamefont
  {L.}~\bibnamefont {Liang}},\ }\bibinfo {title} {Floquet band engineering and
  topological phase transitions in 1T$’$ transition metal dichalcogenides},\
  \href {https://doi.org/10.1088/2053-1583/ac4957} {\bibfield  {journal}
  {\bibinfo  {journal} {2D Materials}\ }\textbf {\bibinfo {volume} {9}},\
  \bibinfo {pages} {025005} (\bibinfo {year} {2022})}\BibitemShut {NoStop}%
\bibitem [{\citenamefont {Kong}\ \emph {et~al.}(2020)\citenamefont {Kong},
  \citenamefont {Li}, \citenamefont {Liang}, \citenamefont {Peeters},\ and\
  \citenamefont {Liu}}]{10.1063/5.0006446}%
  \BibitemOpen
  \bibfield  {author} {\bibinfo {author} {\bibfnamefont {X.}~\bibnamefont
  {Kong}}, \bibinfo {author} {\bibfnamefont {L.}~\bibnamefont {Li}}, \bibinfo
  {author} {\bibfnamefont {L.}~\bibnamefont {Liang}}, \bibinfo {author}
  {\bibfnamefont {F.~M.}\ \bibnamefont {Peeters}},\ and\ \bibinfo {author}
  {\bibfnamefont {X.-J.}\ \bibnamefont {Liu}},\ }\bibinfo {title} {The
  magnetic, electronic, and light-induced topological properties in
  two-dimensional hexagonal FeX$_2$ (X=Cl, Br, I) monolayers},\ \href
  {https://doi.org/10.1063/5.0006446} {\bibfield  {journal} {\bibinfo
  {journal} {Applied Physics Letters}\ }\textbf {\bibinfo {volume} {116}},\
  \bibinfo {pages} {192404} (\bibinfo {year} {2020})}\BibitemShut {NoStop}%
\bibitem [{\citenamefont {Zhan}\ \emph {et~al.}(2023)\citenamefont {Zhan},
  \citenamefont {Zeng}, \citenamefont {Chen}, \citenamefont {Jin},
  \citenamefont {Fan}, \citenamefont {Chen},\ and\ \citenamefont
  {Wang}}]{doi:10.1021/acs.nanolett.2c04651}%
  \BibitemOpen
  \bibfield  {author} {\bibinfo {author} {\bibfnamefont {F.}~\bibnamefont
  {Zhan}}, \bibinfo {author} {\bibfnamefont {J.}~\bibnamefont {Zeng}}, \bibinfo
  {author} {\bibfnamefont {Z.}~\bibnamefont {Chen}}, \bibinfo {author}
  {\bibfnamefont {X.}~\bibnamefont {Jin}}, \bibinfo {author} {\bibfnamefont
  {J.}~\bibnamefont {Fan}}, \bibinfo {author} {\bibfnamefont {T.}~\bibnamefont
  {Chen}},\ and\ \bibinfo {author} {\bibfnamefont {R.}~\bibnamefont {Wang}},\
  }\bibinfo {title} {Floquet Engineering of Nonequilibrium Valley-Polarized
  Quantum Anomalous Hall Effect with Tunable Chern Number},\ \href
  {https://doi.org/10.1021/acs.nanolett.2c04651} {\bibfield  {journal}
  {\bibinfo  {journal} {Nano Letters}\ }\textbf {\bibinfo {volume} {23}},\
  \bibinfo {pages} {2166} (\bibinfo {year} {2023})}\BibitemShut {NoStop}%
\bibitem [{\citenamefont {Feng}\ \emph {et~al.}(2025)\citenamefont {Feng},
  \citenamefont {Bai}, \citenamefont {Chen}, \citenamefont {Dai}, \citenamefont
  {Huang},\ and\ \citenamefont {Niu}}]{https://doi.org/10.1002/adfm.202501934}%
  \BibitemOpen
  \bibfield  {author} {\bibinfo {author} {\bibfnamefont {X.}~\bibnamefont
  {Feng}}, \bibinfo {author} {\bibfnamefont {Y.}~\bibnamefont {Bai}}, \bibinfo
  {author} {\bibfnamefont {Z.}~\bibnamefont {Chen}}, \bibinfo {author}
  {\bibfnamefont {Y.}~\bibnamefont {Dai}}, \bibinfo {author} {\bibfnamefont
  {B.}~\bibnamefont {Huang}},\ and\ \bibinfo {author} {\bibfnamefont
  {C.}~\bibnamefont {Niu}},\ }\bibinfo {title} {Engineering Quantum Anomalous
  Hall Effect with a High Chern Number in Nonmagnetic Second-Order Topological
  Insulator},\ \href {https://doi.org/https://doi.org/10.1002/adfm.202501934}
  {\bibfield  {journal} {\bibinfo  {journal} {Advanced Functional Materials}\
  }\textbf {\bibinfo {volume} {35}},\ \bibinfo {pages} {2501934} (\bibinfo
  {year} {2025})}\BibitemShut {NoStop}%
\bibitem [{\citenamefont {Zou}\ \emph {et~al.}(2025)\citenamefont {Zou},
  \citenamefont {Feng}, \citenamefont {Dai}, \citenamefont {Huang},\ and\
  \citenamefont {Niu}}]{doi:10.1021/acsnano.5c10277}%
  \BibitemOpen
  \bibfield  {author} {\bibinfo {author} {\bibfnamefont {X.}~\bibnamefont
  {Zou}}, \bibinfo {author} {\bibfnamefont {X.}~\bibnamefont {Feng}}, \bibinfo
  {author} {\bibfnamefont {Y.}~\bibnamefont {Dai}}, \bibinfo {author}
  {\bibfnamefont {B.}~\bibnamefont {Huang}},\ and\ \bibinfo {author}
  {\bibfnamefont {C.}~\bibnamefont {Niu}},\ }\bibinfo {title} {Floquet Quantum
  Anomalous Hall Effect with In-Plane Magnetization in Two-Dimensional
  Altermagnets},\ \href {https://doi.org/10.1021/acsnano.5c10277} {\bibfield
  {journal} {\bibinfo  {journal} {ACS Nano}\ }\textbf {\bibinfo {volume}
  {19}},\ \bibinfo {pages} {35575} (\bibinfo {year} {2025})}\BibitemShut
  {NoStop}%
\bibitem [{\citenamefont {Li}\ \emph {et~al.}(2024)\citenamefont {Li},
  \citenamefont {Zou}, \citenamefont {Chen}, \citenamefont {Feng},
  \citenamefont {Huang}, \citenamefont {Dai},\ and\ \citenamefont
  {Niu}}]{D4MH00237G}%
  \BibitemOpen
  \bibfield  {author} {\bibinfo {author} {\bibfnamefont {R.}~\bibnamefont
  {Li}}, \bibinfo {author} {\bibfnamefont {X.}~\bibnamefont {Zou}}, \bibinfo
  {author} {\bibfnamefont {Z.}~\bibnamefont {Chen}}, \bibinfo {author}
  {\bibfnamefont {X.}~\bibnamefont {Feng}}, \bibinfo {author} {\bibfnamefont
  {B.}~\bibnamefont {Huang}}, \bibinfo {author} {\bibfnamefont
  {Y.}~\bibnamefont {Dai}},\ and\ \bibinfo {author} {\bibfnamefont
  {C.}~\bibnamefont {Niu}},\ }\bibinfo {title} {Floquet engineering of the
  orbital Hall effect and valleytronics in two-dimensional topological
  magnets},\ \href {https://doi.org/10.1039/D4MH00237G} {\bibfield  {journal}
  {\bibinfo  {journal} {Mater. Horiz.}\ }\textbf {\bibinfo {volume} {11}},\
  \bibinfo {pages} {3819} (\bibinfo {year} {2024})}\BibitemShut {NoStop}%
\bibitem [{\citenamefont {Ning}\ \emph {et~al.}(2024)\citenamefont {Ning},
  \citenamefont {Ma}, \citenamefont {Zeng}, \citenamefont {Xu},\ and\
  \citenamefont {Wang}}]{PhysRevLett.133.246606}%
  \BibitemOpen
  \bibfield  {author} {\bibinfo {author} {\bibfnamefont {Z.}~\bibnamefont
  {Ning}}, \bibinfo {author} {\bibfnamefont {D.-S.}\ \bibnamefont {Ma}},
  \bibinfo {author} {\bibfnamefont {J.}~\bibnamefont {Zeng}}, \bibinfo {author}
  {\bibfnamefont {D.-H.}\ \bibnamefont {Xu}},\ and\ \bibinfo {author}
  {\bibfnamefont {R.}~\bibnamefont {Wang}},\ }\bibinfo {title} {Flexible
  Control of Chiral Superconductivity in Optically Driven Nodal Point
  Superconductors with Antiferromagnetism},\ \href
  {https://doi.org/10.1103/PhysRevLett.133.246606} {\bibfield  {journal}
  {\bibinfo  {journal} {Phys. Rev. Lett.}\ }\textbf {\bibinfo {volume} {133}},\
  \bibinfo {pages} {246606} (\bibinfo {year} {2024})}\BibitemShut {NoStop}%
\bibitem [{\citenamefont {Huang}\ \emph {et~al.}(2026)\citenamefont {Huang},
  \citenamefont {Qin}, \citenamefont {Zhan}, \citenamefont {Xu}, \citenamefont
  {Ma},\ and\ \citenamefont {Wang}}]{9346-9jpf}%
  \BibitemOpen
  \bibfield  {author} {\bibinfo {author} {\bibfnamefont {S.}~\bibnamefont
  {Huang}}, \bibinfo {author} {\bibfnamefont {Z.}~\bibnamefont {Qin}}, \bibinfo
  {author} {\bibfnamefont {F.}~\bibnamefont {Zhan}}, \bibinfo {author}
  {\bibfnamefont {D.-H.}\ \bibnamefont {Xu}}, \bibinfo {author} {\bibfnamefont
  {D.-S.}\ \bibnamefont {Ma}},\ and\ \bibinfo {author} {\bibfnamefont
  {R.}~\bibnamefont {Wang}},\ }\bibinfo {title} {Light-induced odd-parity
  magnetism in conventional antiferromagnetism},\ \href
  {https://doi.org/10.1103/9346-9jpf} {\bibfield  {journal} {\bibinfo
  {journal} {Phys. Rev. Lett.}\ ,\ } (\bibinfo {year} {2026})}\BibitemShut
  {NoStop}%
\bibitem [{\citenamefont {Li}\ \emph {et~al.}(2025)\citenamefont {Li},
  \citenamefont {Shao},\ and\ \citenamefont
  {Kovalev}}]{li2025floquetspinsplittingspin}%
  \BibitemOpen
  \bibfield  {author} {\bibinfo {author} {\bibfnamefont {B.}~\bibnamefont
  {Li}}, \bibinfo {author} {\bibfnamefont {D.-F.}\ \bibnamefont {Shao}},\ and\
  \bibinfo {author} {\bibfnamefont {A.~A.}\ \bibnamefont {Kovalev}},\ }\href
  {https://arxiv.org/abs/2507.22884} {\bibinfo {title} {Floquet spin splitting
  and spin generation in antiferromagnets}} (\bibinfo {year} {2025}),\ \Eprint
  {https://arxiv.org/abs/2507.22884} {arXiv:2507.22884 [cond-mat.mes-hall]}
  \BibitemShut {NoStop}%
\bibitem [{\citenamefont {Zhu}\ \emph {et~al.}(2026)\citenamefont {Zhu},
  \citenamefont {Zhou}, \citenamefont {Wang}, \citenamefont {Wei},\ and\
  \citenamefont {Ruan}}]{7ywb-ml2q}%
  \BibitemOpen
  \bibfield  {author} {\bibinfo {author} {\bibfnamefont {T.}~\bibnamefont
  {Zhu}}, \bibinfo {author} {\bibfnamefont {D.}~\bibnamefont {Zhou}}, \bibinfo
  {author} {\bibfnamefont {H.}~\bibnamefont {Wang}}, \bibinfo {author}
  {\bibfnamefont {S.-H.}\ \bibnamefont {Wei}},\ and\ \bibinfo {author}
  {\bibfnamefont {J.}~\bibnamefont {Ruan}},\ }\bibinfo {title} {Floquet
  odd-parity collinear magnets},\ \href {https://doi.org/10.1103/7ywb-ml2q}
  {\bibfield  {journal} {\bibinfo  {journal} {Phys. Rev. Lett.}\ ,\ } (\bibinfo
  {year} {2026})}\BibitemShut {NoStop}%
\bibitem [{\citenamefont {Liu}\ \emph {et~al.}(2026)\citenamefont {Liu},
  \citenamefont {Zhuang}, \citenamefont {Zhu}, \citenamefont {Wu},\ and\
  \citenamefont {Yan}}]{wnqs-3djt}%
  \BibitemOpen
  \bibfield  {author} {\bibinfo {author} {\bibfnamefont {D.}~\bibnamefont
  {Liu}}, \bibinfo {author} {\bibfnamefont {Z.-Y.}\ \bibnamefont {Zhuang}},
  \bibinfo {author} {\bibfnamefont {D.}~\bibnamefont {Zhu}}, \bibinfo {author}
  {\bibfnamefont {Z.}~\bibnamefont {Wu}},\ and\ \bibinfo {author}
  {\bibfnamefont {Z.}~\bibnamefont {Yan}},\ }\bibinfo {title} {Light-induced
  odd-parity altermagnets on dimerized lattices},\ \href
  {https://doi.org/10.1103/wnqs-3djt} {\bibfield  {journal} {\bibinfo
  {journal} {Phys. Rev. B}\ }\textbf {\bibinfo {volume} {113}},\ \bibinfo
  {pages} {L060409} (\bibinfo {year} {2026})}\BibitemShut {NoStop}%
\bibitem [{\citenamefont {Zhuang}\ \emph {et~al.}(2025)\citenamefont {Zhuang},
  \citenamefont {Zhu}, \citenamefont {Liu}, \citenamefont {Wu},\ and\
  \citenamefont {Yan}}]{zhuang2025oddparityaltermagnetismoriginatedorbital}%
  \BibitemOpen
  \bibfield  {author} {\bibinfo {author} {\bibfnamefont {Z.-Y.}\ \bibnamefont
  {Zhuang}}, \bibinfo {author} {\bibfnamefont {D.}~\bibnamefont {Zhu}},
  \bibinfo {author} {\bibfnamefont {D.}~\bibnamefont {Liu}}, \bibinfo {author}
  {\bibfnamefont {Z.}~\bibnamefont {Wu}},\ and\ \bibinfo {author}
  {\bibfnamefont {Z.}~\bibnamefont {Yan}},\ }\href
  {https://arxiv.org/abs/2508.18361} {\bibinfo {title} {Odd-parity
  altermagnetism originated from orbital orders}} (\bibinfo {year} {2025}),\
  \Eprint {https://arxiv.org/abs/2508.18361} {arXiv:2508.18361
  [cond-mat.mes-hall]} \BibitemShut {NoStop}%
\bibitem [{\citenamefont {Zhu}\ \emph {et~al.}(2026)\citenamefont {Zhu},
  \citenamefont {Liu}, \citenamefont {Zhuang}, \citenamefont {Wu},\ and\
  \citenamefont {Yan}}]{zhu2026lightinducedevenparityunidirectionalspin}%
  \BibitemOpen
  \bibfield  {author} {\bibinfo {author} {\bibfnamefont {D.}~\bibnamefont
  {Zhu}}, \bibinfo {author} {\bibfnamefont {D.}~\bibnamefont {Liu}}, \bibinfo
  {author} {\bibfnamefont {Z.-Y.}\ \bibnamefont {Zhuang}}, \bibinfo {author}
  {\bibfnamefont {Z.}~\bibnamefont {Wu}},\ and\ \bibinfo {author}
  {\bibfnamefont {Z.}~\bibnamefont {Yan}},\ }\href
  {https://arxiv.org/abs/2601.03358} {\bibinfo {title} {Light-induced
  even-parity unidirectional spin splitting in coplanar antiferromagnets}}
  (\bibinfo {year} {2026}),\ \Eprint {https://arxiv.org/abs/2601.03358}
  {arXiv:2601.03358 [cond-mat.mtrl-sci]} \BibitemShut {NoStop}%
\bibitem [{\citenamefont {Fu}\ \emph {et~al.}(2026)\citenamefont {Fu},
  \citenamefont {Mondal}, \citenamefont {Liu}, \citenamefont {Tanaka},\ and\
  \citenamefont {Cayao}}]{lkf9-jgv6}%
  \BibitemOpen
  \bibfield  {author} {\bibinfo {author} {\bibfnamefont {P.-H.}\ \bibnamefont
  {Fu}}, \bibinfo {author} {\bibfnamefont {S.}~\bibnamefont {Mondal}}, \bibinfo
  {author} {\bibfnamefont {J.-F.}\ \bibnamefont {Liu}}, \bibinfo {author}
  {\bibfnamefont {Y.}~\bibnamefont {Tanaka}},\ and\ \bibinfo {author}
  {\bibfnamefont {J.}~\bibnamefont {Cayao}},\ }\bibinfo {title} {Floquet
  Engineering Spin Triplet States in Unconventional Magnets},\ \href
  {https://doi.org/10.1103/lkf9-jgv6} {\bibfield  {journal} {\bibinfo
  {journal} {Phys. Rev. Lett.}\ }\textbf {\bibinfo {volume} {136}},\ \bibinfo
  {pages} {066703} (\bibinfo {year} {2026})}\BibitemShut {NoStop}%
\bibitem [{\citenamefont {Li}\ \emph {et~al.}(2026)\citenamefont {Li},
  \citenamefont {Li}, \citenamefont {Guan},\ and\ \citenamefont
  {Meng}}]{li2026robusttunablefloquetaltermagnets}%
  \BibitemOpen
  \bibfield  {author} {\bibinfo {author} {\bibfnamefont {Z.}~\bibnamefont
  {Li}}, \bibinfo {author} {\bibfnamefont {L.}~\bibnamefont {Li}}, \bibinfo
  {author} {\bibfnamefont {M.}~\bibnamefont {Guan}},\ and\ \bibinfo {author}
  {\bibfnamefont {S.}~\bibnamefont {Meng}},\ }\href
  {https://arxiv.org/abs/2512.06416} {\bibinfo {title} {Robust and tunable
  floquet altermagnets in sliding a-type antiferromagnetic bilayers}} (\bibinfo
  {year} {2026}),\ \Eprint {https://arxiv.org/abs/2512.06416} {arXiv:2512.06416
  [cond-mat.mtrl-sci]} \BibitemShut {NoStop}%
\bibitem [{\citenamefont {Lin}\ and\ \citenamefont
  {Vila}(2026)}]{lin2026oddparityaltermagnetismsublatticecurrents}%
  \BibitemOpen
  \bibfield  {author} {\bibinfo {author} {\bibfnamefont {Y.-P.}\ \bibnamefont
  {Lin}}\ and\ \bibinfo {author} {\bibfnamefont {M.}~\bibnamefont {Vila}},\
  }\href {https://arxiv.org/abs/2503.09602} {\bibinfo {title} {Odd-parity
  altermagnetism through sublattice currents: From haldane-hubbard model to
  general bipartite lattices}} (\bibinfo {year} {2026}),\ \Eprint
  {https://arxiv.org/abs/2503.09602} {arXiv:2503.09602 [cond-mat.str-el]}
  \BibitemShut {NoStop}%
\bibitem [{\citenamefont {Bernevig}\ \emph {et~al.}(2005)\citenamefont
  {Bernevig}, \citenamefont {Hughes},\ and\ \citenamefont
  {Zhang}}]{PhysRevLett.95.066601}%
  \BibitemOpen
  \bibfield  {author} {\bibinfo {author} {\bibfnamefont {B.~A.}\ \bibnamefont
  {Bernevig}}, \bibinfo {author} {\bibfnamefont {T.~L.}\ \bibnamefont
  {Hughes}},\ and\ \bibinfo {author} {\bibfnamefont {S.-C.}\ \bibnamefont
  {Zhang}},\ }\bibinfo {title} {Orbitronics: The Intrinsic Orbital Current in
  $p$-Doped Silicon},\ \href {https://doi.org/10.1103/PhysRevLett.95.066601}
  {\bibfield  {journal} {\bibinfo  {journal} {Phys. Rev. Lett.}\ }\textbf
  {\bibinfo {volume} {95}},\ \bibinfo {pages} {066601} (\bibinfo {year}
  {2005})}\BibitemShut {NoStop}%
\bibitem [{\citenamefont {Go}\ \emph {et~al.}(2021)\citenamefont {Go},
  \citenamefont {Jo}, \citenamefont {Lee}, \citenamefont {Kläui},\ and\
  \citenamefont {Mokrousov}}]{Go2021Orbitronics}%
  \BibitemOpen
  \bibfield  {author} {\bibinfo {author} {\bibfnamefont {D.}~\bibnamefont
  {Go}}, \bibinfo {author} {\bibfnamefont {D.}~\bibnamefont {Jo}}, \bibinfo
  {author} {\bibfnamefont {H.-W.}\ \bibnamefont {Lee}}, \bibinfo {author}
  {\bibfnamefont {M.}~\bibnamefont {Kläui}},\ and\ \bibinfo {author}
  {\bibfnamefont {Y.}~\bibnamefont {Mokrousov}},\ }\bibinfo {title}
  {Orbitronics: Orbital currents in solids},\ \href
  {https://doi.org/10.1209/0295-5075/ac2653} {\bibfield  {journal} {\bibinfo
  {journal} {Europhysics Letters}\ }\textbf {\bibinfo {volume} {135}},\
  \bibinfo {pages} {37001} (\bibinfo {year} {2021})}\BibitemShut {NoStop}%
\bibitem [{\citenamefont {Go}\ \emph {et~al.}(2018)\citenamefont {Go},
  \citenamefont {Jo}, \citenamefont {Kim},\ and\ \citenamefont
  {Lee}}]{PhysRevLett.121.086602}%
  \BibitemOpen
  \bibfield  {author} {\bibinfo {author} {\bibfnamefont {D.}~\bibnamefont
  {Go}}, \bibinfo {author} {\bibfnamefont {D.}~\bibnamefont {Jo}}, \bibinfo
  {author} {\bibfnamefont {C.}~\bibnamefont {Kim}},\ and\ \bibinfo {author}
  {\bibfnamefont {H.-W.}\ \bibnamefont {Lee}},\ }\bibinfo {title} {Intrinsic
  Spin and Orbital Hall Effects from Orbital Texture},\ \href
  {https://doi.org/10.1103/PhysRevLett.121.086602} {\bibfield  {journal}
  {\bibinfo  {journal} {Phys. Rev. Lett.}\ }\textbf {\bibinfo {volume} {121}},\
  \bibinfo {pages} {086602} (\bibinfo {year} {2018})}\BibitemShut {NoStop}%
\bibitem [{\citenamefont {Jo}\ \emph {et~al.}(2018)\citenamefont {Jo},
  \citenamefont {Go},\ and\ \citenamefont {Lee}}]{PhysRevB.98.214405}%
  \BibitemOpen
  \bibfield  {author} {\bibinfo {author} {\bibfnamefont {D.}~\bibnamefont
  {Jo}}, \bibinfo {author} {\bibfnamefont {D.}~\bibnamefont {Go}},\ and\
  \bibinfo {author} {\bibfnamefont {H.-W.}\ \bibnamefont {Lee}},\ }\bibinfo
  {title} {Gigantic intrinsic orbital Hall effects in weakly spin-orbit coupled
  metals},\ \href {https://doi.org/10.1103/PhysRevB.98.214405} {\bibfield
  {journal} {\bibinfo  {journal} {Phys. Rev. B}\ }\textbf {\bibinfo {volume}
  {98}},\ \bibinfo {pages} {214405} (\bibinfo {year} {2018})}\BibitemShut
  {NoStop}%
\bibitem [{\citenamefont {Choi}\ \emph {et~al.}(2023)\citenamefont {Choi},
  \citenamefont {Jo}, \citenamefont {Ko}, \citenamefont {Go}, \citenamefont
  {Kim}, \citenamefont {Park}, \citenamefont {Kim}, \citenamefont {Min},
  \citenamefont {Choi},\ and\ \citenamefont {Lee}}]{Choi2023Observation}%
  \BibitemOpen
  \bibfield  {author} {\bibinfo {author} {\bibfnamefont {Y.-G.}\ \bibnamefont
  {Choi}}, \bibinfo {author} {\bibfnamefont {D.}~\bibnamefont {Jo}}, \bibinfo
  {author} {\bibfnamefont {K.-H.}\ \bibnamefont {Ko}}, \bibinfo {author}
  {\bibfnamefont {D.}~\bibnamefont {Go}}, \bibinfo {author} {\bibfnamefont
  {K.-H.}\ \bibnamefont {Kim}}, \bibinfo {author} {\bibfnamefont {H.~G.}\
  \bibnamefont {Park}}, \bibinfo {author} {\bibfnamefont {C.}~\bibnamefont
  {Kim}}, \bibinfo {author} {\bibfnamefont {B.-C.}\ \bibnamefont {Min}},
  \bibinfo {author} {\bibfnamefont {G.-M.}\ \bibnamefont {Choi}},\ and\
  \bibinfo {author} {\bibfnamefont {H.-W.}\ \bibnamefont {Lee}},\ }\bibinfo
  {title} {Observation of the orbital Hall effect in a light metal Ti},\ \href
  {https://doi.org/10.1038/s41586-023-06101-9} {\bibfield  {journal} {\bibinfo
  {journal} {Nature}\ }\textbf {\bibinfo {volume} {619}},\ \bibinfo {pages}
  {52} (\bibinfo {year} {2023})}\BibitemShut {NoStop}%
\bibitem [{\citenamefont {Lyalin}\ \emph {et~al.}(2023)\citenamefont {Lyalin},
  \citenamefont {Alikhah}, \citenamefont {Berritta}, \citenamefont {Oppeneer},\
  and\ \citenamefont {Kawakami}}]{PhysRevLett.131.156702}%
  \BibitemOpen
  \bibfield  {author} {\bibinfo {author} {\bibfnamefont {I.}~\bibnamefont
  {Lyalin}}, \bibinfo {author} {\bibfnamefont {S.}~\bibnamefont {Alikhah}},
  \bibinfo {author} {\bibfnamefont {M.}~\bibnamefont {Berritta}}, \bibinfo
  {author} {\bibfnamefont {P.~M.}\ \bibnamefont {Oppeneer}},\ and\ \bibinfo
  {author} {\bibfnamefont {R.~K.}\ \bibnamefont {Kawakami}},\ }\bibinfo {title}
  {Magneto-Optical Detection of the Orbital Hall Effect in Chromium},\ \href
  {https://doi.org/10.1103/PhysRevLett.131.156702} {\bibfield  {journal}
  {\bibinfo  {journal} {Phys. Rev. Lett.}\ }\textbf {\bibinfo {volume} {131}},\
  \bibinfo {pages} {156702} (\bibinfo {year} {2023})}\BibitemShut {NoStop}%
\bibitem [{\citenamefont {Sala}\ \emph {et~al.}(2023)\citenamefont {Sala},
  \citenamefont {Wang}, \citenamefont {Legrand},\ and\ \citenamefont
  {Gambardella}}]{PhysRevLett.131.156703}%
  \BibitemOpen
  \bibfield  {author} {\bibinfo {author} {\bibfnamefont {G.}~\bibnamefont
  {Sala}}, \bibinfo {author} {\bibfnamefont {H.}~\bibnamefont {Wang}}, \bibinfo
  {author} {\bibfnamefont {W.}~\bibnamefont {Legrand}},\ and\ \bibinfo {author}
  {\bibfnamefont {P.}~\bibnamefont {Gambardella}},\ }\bibinfo {title} {Orbital
  Hanle Magnetoresistance in a $3d$ Transition Metal},\ \href
  {https://doi.org/10.1103/PhysRevLett.131.156703} {\bibfield  {journal}
  {\bibinfo  {journal} {Phys. Rev. Lett.}\ }\textbf {\bibinfo {volume} {131}},\
  \bibinfo {pages} {156703} (\bibinfo {year} {2023})}\BibitemShut {NoStop}%
\bibitem [{\citenamefont {Ding}\ \emph {et~al.}(2020)\citenamefont {Ding},
  \citenamefont {Ross}, \citenamefont {Go}, \citenamefont {Baldrati},
  \citenamefont {Ren}, \citenamefont {Freimuth}, \citenamefont {Becker},
  \citenamefont {Kammerbauer}, \citenamefont {Yang}, \citenamefont {Jakob},
  \citenamefont {Mokrousov},\ and\ \citenamefont
  {Kl\"aui}}]{PhysRevLett.125.177201}%
  \BibitemOpen
  \bibfield  {author} {\bibinfo {author} {\bibfnamefont {S.}~\bibnamefont
  {Ding}}, \bibinfo {author} {\bibfnamefont {A.}~\bibnamefont {Ross}}, \bibinfo
  {author} {\bibfnamefont {D.}~\bibnamefont {Go}}, \bibinfo {author}
  {\bibfnamefont {L.}~\bibnamefont {Baldrati}}, \bibinfo {author}
  {\bibfnamefont {Z.}~\bibnamefont {Ren}}, \bibinfo {author} {\bibfnamefont
  {F.}~\bibnamefont {Freimuth}}, \bibinfo {author} {\bibfnamefont
  {S.}~\bibnamefont {Becker}}, \bibinfo {author} {\bibfnamefont
  {F.}~\bibnamefont {Kammerbauer}}, \bibinfo {author} {\bibfnamefont
  {J.}~\bibnamefont {Yang}}, \bibinfo {author} {\bibfnamefont {G.}~\bibnamefont
  {Jakob}}, \bibinfo {author} {\bibfnamefont {Y.}~\bibnamefont {Mokrousov}},\
  and\ \bibinfo {author} {\bibfnamefont {M.}~\bibnamefont {Kl\"aui}},\
  }\bibinfo {title} {Harnessing Orbital-to-Spin Conversion of Interfacial
  Orbital Currents for Efficient Spin-Orbit Torques},\ \href
  {https://doi.org/10.1103/PhysRevLett.125.177201} {\bibfield  {journal}
  {\bibinfo  {journal} {Phys. Rev. Lett.}\ }\textbf {\bibinfo {volume} {125}},\
  \bibinfo {pages} {177201} (\bibinfo {year} {2020})}\BibitemShut {NoStop}%
\bibitem [{\citenamefont {Sala}\ and\ \citenamefont
  {Gambardella}(2022)}]{PhysRevResearch.4.033037}%
  \BibitemOpen
  \bibfield  {author} {\bibinfo {author} {\bibfnamefont {G.}~\bibnamefont
  {Sala}}\ and\ \bibinfo {author} {\bibfnamefont {P.}~\bibnamefont
  {Gambardella}},\ }\bibinfo {title} {Giant orbital Hall effect and
  orbital-to-spin conversion in $3d$, $5d$, and $4f$ metallic
  heterostructures},\ \href {https://doi.org/10.1103/PhysRevResearch.4.033037}
  {\bibfield  {journal} {\bibinfo  {journal} {Phys. Rev. Res.}\ }\textbf
  {\bibinfo {volume} {4}},\ \bibinfo {pages} {033037} (\bibinfo {year}
  {2022})}\BibitemShut {NoStop}%
\bibitem [{\citenamefont {Liu}\ \emph {et~al.}(2025)\citenamefont {Liu},
  \citenamefont {Cullen}, \citenamefont {Arovas},\ and\ \citenamefont
  {Culcer}}]{PhysRevLett.134.036304}%
  \BibitemOpen
  \bibfield  {author} {\bibinfo {author} {\bibfnamefont {H.}~\bibnamefont
  {Liu}}, \bibinfo {author} {\bibfnamefont {J.~H.}\ \bibnamefont {Cullen}},
  \bibinfo {author} {\bibfnamefont {D.~P.}\ \bibnamefont {Arovas}},\ and\
  \bibinfo {author} {\bibfnamefont {D.}~\bibnamefont {Culcer}},\ }\bibinfo
  {title} {Quantum Correction to the Orbital Hall Effect},\ \href
  {https://doi.org/10.1103/PhysRevLett.134.036304} {\bibfield  {journal}
  {\bibinfo  {journal} {Phys. Rev. Lett.}\ }\textbf {\bibinfo {volume} {134}},\
  \bibinfo {pages} {036304} (\bibinfo {year} {2025})}\BibitemShut {NoStop}%
\bibitem [{\citenamefont {Kontani}\ \emph {et~al.}(2008)\citenamefont
  {Kontani}, \citenamefont {Tanaka}, \citenamefont {Hirashima}, \citenamefont
  {Yamada},\ and\ \citenamefont {Inoue}}]{PhysRevLett.100.096601}%
  \BibitemOpen
  \bibfield  {author} {\bibinfo {author} {\bibfnamefont {H.}~\bibnamefont
  {Kontani}}, \bibinfo {author} {\bibfnamefont {T.}~\bibnamefont {Tanaka}},
  \bibinfo {author} {\bibfnamefont {D.~S.}\ \bibnamefont {Hirashima}}, \bibinfo
  {author} {\bibfnamefont {K.}~\bibnamefont {Yamada}},\ and\ \bibinfo {author}
  {\bibfnamefont {J.}~\bibnamefont {Inoue}},\ }\bibinfo {title} {Giant
  Intrinsic Spin and Orbital Hall Effects in
  ${\mathrm{Sr}}_{2}M{\mathrm{O}}_{4}$ ($M=\mathrm{Ru}$, Rh, Mo)},\ \href
  {https://doi.org/10.1103/PhysRevLett.100.096601} {\bibfield  {journal}
  {\bibinfo  {journal} {Phys. Rev. Lett.}\ }\textbf {\bibinfo {volume} {100}},\
  \bibinfo {pages} {096601} (\bibinfo {year} {2008})}\BibitemShut {NoStop}%
\bibitem [{\citenamefont {Veneri}\ \emph {et~al.}(2025)\citenamefont {Veneri},
  \citenamefont {Rappoport},\ and\ \citenamefont
  {Ferreira}}]{PhysRevLett.134.136201}%
  \BibitemOpen
  \bibfield  {author} {\bibinfo {author} {\bibfnamefont {A.}~\bibnamefont
  {Veneri}}, \bibinfo {author} {\bibfnamefont {T.~G.}\ \bibnamefont
  {Rappoport}},\ and\ \bibinfo {author} {\bibfnamefont {A.}~\bibnamefont
  {Ferreira}},\ }\bibinfo {title} {Extrinsic Orbital Hall Effect: Orbital Skew
  Scattering and Crossover between Diffusive and Intrinsic Orbital Transport},\
  \href {https://doi.org/10.1103/PhysRevLett.134.136201} {\bibfield  {journal}
  {\bibinfo  {journal} {Phys. Rev. Lett.}\ }\textbf {\bibinfo {volume} {134}},\
  \bibinfo {pages} {136201} (\bibinfo {year} {2025})}\BibitemShut {NoStop}%
\bibitem [{\citenamefont {Kumar}\ and\ \citenamefont
  {Kumar}(2023)}]{Kumar2023Ultrafast}%
  \BibitemOpen
  \bibfield  {author} {\bibinfo {author} {\bibfnamefont {S.}~\bibnamefont
  {Kumar}}\ and\ \bibinfo {author} {\bibfnamefont {S.}~\bibnamefont {Kumar}},\
  }\bibinfo {title} {Ultrafast THz probing of nonlocal orbital current in
  transverse multilayer metallic heterostructures},\ \href
  {https://doi.org/10.1038/s41467-023-43956-y} {\bibfield  {journal} {\bibinfo
  {journal} {Nature Communications}\ }\textbf {\bibinfo {volume} {14}}
  (\bibinfo {year} {2023})}\BibitemShut {NoStop}%
\bibitem [{\citenamefont {Mishra}\ \emph {et~al.}(2024)\citenamefont {Mishra},
  \citenamefont {Lourembam}, \citenamefont {Lin},\ and\ \citenamefont
  {Singh}}]{Mishra2024Active}%
  \BibitemOpen
  \bibfield  {author} {\bibinfo {author} {\bibfnamefont {S.~S.}\ \bibnamefont
  {Mishra}}, \bibinfo {author} {\bibfnamefont {J.}~\bibnamefont {Lourembam}},
  \bibinfo {author} {\bibfnamefont {D.~J.~X.}\ \bibnamefont {Lin}},\ and\
  \bibinfo {author} {\bibfnamefont {R.}~\bibnamefont {Singh}},\ }\bibinfo
  {title} {Active ballistic orbital transport in Ni/Pt heterostructure},\ \href
  {https://doi.org/10.1038/s41467-024-48891-0} {\bibfield  {journal} {\bibinfo
  {journal} {Nature Communications}\ }\textbf {\bibinfo {volume} {15}}
  (\bibinfo {year} {2024})}\BibitemShut {NoStop}%
\bibitem [{\citenamefont {Beaulieu}\ \emph {et~al.}(2020)\citenamefont
  {Beaulieu}, \citenamefont {Schusser}, \citenamefont {Dong}, \citenamefont
  {Sch\"uler}, \citenamefont {Pincelli}, \citenamefont {Dendzik}, \citenamefont
  {Maklar}, \citenamefont {Neef}, \citenamefont {Ebert}, \citenamefont
  {Hricovini}, \citenamefont {Wolf}, \citenamefont {Braun}, \citenamefont
  {Rettig}, \citenamefont {Min\'ar},\ and\ \citenamefont
  {Ernstorfer}}]{PhysRevLett.125.216404}%
  \BibitemOpen
  \bibfield  {author} {\bibinfo {author} {\bibfnamefont {S.}~\bibnamefont
  {Beaulieu}}, \bibinfo {author} {\bibfnamefont {J.}~\bibnamefont {Schusser}},
  \bibinfo {author} {\bibfnamefont {S.}~\bibnamefont {Dong}}, \bibinfo {author}
  {\bibfnamefont {M.}~\bibnamefont {Sch\"uler}}, \bibinfo {author}
  {\bibfnamefont {T.}~\bibnamefont {Pincelli}}, \bibinfo {author}
  {\bibfnamefont {M.}~\bibnamefont {Dendzik}}, \bibinfo {author} {\bibfnamefont
  {J.}~\bibnamefont {Maklar}}, \bibinfo {author} {\bibfnamefont
  {A.}~\bibnamefont {Neef}}, \bibinfo {author} {\bibfnamefont {H.}~\bibnamefont
  {Ebert}}, \bibinfo {author} {\bibfnamefont {K.}~\bibnamefont {Hricovini}},
  \bibinfo {author} {\bibfnamefont {M.}~\bibnamefont {Wolf}}, \bibinfo {author}
  {\bibfnamefont {J.}~\bibnamefont {Braun}}, \bibinfo {author} {\bibfnamefont
  {L.}~\bibnamefont {Rettig}}, \bibinfo {author} {\bibfnamefont
  {J.}~\bibnamefont {Min\'ar}},\ and\ \bibinfo {author} {\bibfnamefont
  {R.}~\bibnamefont {Ernstorfer}},\ }\bibinfo {title} {Revealing Hidden Orbital
  Pseudospin Texture with Time-Reversal Dichroism in Photoelectron Angular
  Distributions},\ \href {https://doi.org/10.1103/PhysRevLett.125.216404}
  {\bibfield  {journal} {\bibinfo  {journal} {Phys. Rev. Lett.}\ }\textbf
  {\bibinfo {volume} {125}},\ \bibinfo {pages} {216404} (\bibinfo {year}
  {2020})}\BibitemShut {NoStop}%
\bibitem [{\citenamefont {Canonico}\ \emph {et~al.}(2020)\citenamefont
  {Canonico}, \citenamefont {Cysne}, \citenamefont {Molina-Sanchez},
  \citenamefont {Muniz},\ and\ \citenamefont
  {Rappoport}}]{PhysRevB.101.161409}%
  \BibitemOpen
  \bibfield  {author} {\bibinfo {author} {\bibfnamefont {L.~M.}\ \bibnamefont
  {Canonico}}, \bibinfo {author} {\bibfnamefont {T.~P.}\ \bibnamefont {Cysne}},
  \bibinfo {author} {\bibfnamefont {A.}~\bibnamefont {Molina-Sanchez}},
  \bibinfo {author} {\bibfnamefont {R.~B.}\ \bibnamefont {Muniz}},\ and\
  \bibinfo {author} {\bibfnamefont {T.~G.}\ \bibnamefont {Rappoport}},\
  }\bibinfo {title} {Orbital Hall insulating phase in transition metal
  dichalcogenide monolayers},\ \href
  {https://doi.org/10.1103/PhysRevB.101.161409} {\bibfield  {journal} {\bibinfo
   {journal} {Phys. Rev. B}\ }\textbf {\bibinfo {volume} {101}},\ \bibinfo
  {pages} {161409} (\bibinfo {year} {2020})}\BibitemShut {NoStop}%
\bibitem [{\citenamefont {Crasto~de Lima}\ \emph {et~al.}(2019)\citenamefont
  {Crasto~de Lima}, \citenamefont {Ferreira},\ and\ \citenamefont
  {Miwa}}]{doi:10.1021/acs.nanolett.9b02802}%
  \BibitemOpen
  \bibfield  {author} {\bibinfo {author} {\bibfnamefont {F.}~\bibnamefont
  {Crasto~de Lima}}, \bibinfo {author} {\bibfnamefont {G.~J.}\ \bibnamefont
  {Ferreira}},\ and\ \bibinfo {author} {\bibfnamefont {R.~H.}\ \bibnamefont
  {Miwa}},\ }\bibinfo {title} {Orbital Pseudospin-Momentum Locking in
  Two-Dimensional Chiral Borophene},\ \href
  {https://doi.org/10.1021/acs.nanolett.9b02802} {\bibfield  {journal}
  {\bibinfo  {journal} {Nano Letters}\ }\textbf {\bibinfo {volume} {19}},\
  \bibinfo {pages} {6564} (\bibinfo {year} {2019})}\BibitemShut {NoStop}%
\bibitem [{\citenamefont {Chen}\ \emph {et~al.}(2020)\citenamefont {Chen},
  \citenamefont {Ruan}, \citenamefont {Wu}, \citenamefont {Tang}, \citenamefont
  {Ryu}, \citenamefont {Tsai}, \citenamefont {Lee}, \citenamefont {Kahn},
  \citenamefont {Liou}, \citenamefont {Jia}, \citenamefont {Albertini},
  \citenamefont {Xiong}, \citenamefont {Jia}, \citenamefont {Liu},
  \citenamefont {Sobota}, \citenamefont {Liu}, \citenamefont {Moore},
  \citenamefont {Shen}, \citenamefont {Louie}, \citenamefont {Mo},\ and\
  \citenamefont {Crommie}}]{Chen2020Strong}%
  \BibitemOpen
  \bibfield  {author} {\bibinfo {author} {\bibfnamefont {Y.}~\bibnamefont
  {Chen}}, \bibinfo {author} {\bibfnamefont {W.}~\bibnamefont {Ruan}}, \bibinfo
  {author} {\bibfnamefont {M.}~\bibnamefont {Wu}}, \bibinfo {author}
  {\bibfnamefont {S.}~\bibnamefont {Tang}}, \bibinfo {author} {\bibfnamefont
  {H.}~\bibnamefont {Ryu}}, \bibinfo {author} {\bibfnamefont {H.-Z.}\
  \bibnamefont {Tsai}}, \bibinfo {author} {\bibfnamefont {R.~L.}\ \bibnamefont
  {Lee}}, \bibinfo {author} {\bibfnamefont {S.}~\bibnamefont {Kahn}}, \bibinfo
  {author} {\bibfnamefont {F.}~\bibnamefont {Liou}}, \bibinfo {author}
  {\bibfnamefont {C.}~\bibnamefont {Jia}}, \bibinfo {author} {\bibfnamefont
  {O.~R.}\ \bibnamefont {Albertini}}, \bibinfo {author} {\bibfnamefont
  {H.}~\bibnamefont {Xiong}}, \bibinfo {author} {\bibfnamefont
  {T.}~\bibnamefont {Jia}}, \bibinfo {author} {\bibfnamefont {Z.}~\bibnamefont
  {Liu}}, \bibinfo {author} {\bibfnamefont {J.~A.}\ \bibnamefont {Sobota}},
  \bibinfo {author} {\bibfnamefont {A.~Y.}\ \bibnamefont {Liu}}, \bibinfo
  {author} {\bibfnamefont {J.~E.}\ \bibnamefont {Moore}}, \bibinfo {author}
  {\bibfnamefont {Z.-X.}\ \bibnamefont {Shen}}, \bibinfo {author}
  {\bibfnamefont {S.~G.}\ \bibnamefont {Louie}}, \bibinfo {author}
  {\bibfnamefont {S.-K.}\ \bibnamefont {Mo}},\ and\ \bibinfo {author}
  {\bibfnamefont {M.~F.}\ \bibnamefont {Crommie}},\ }\bibinfo {title} {Strong
  correlations and orbital texture in single-layer 1T-TaSe$_2$},\ \href
  {https://doi.org/10.1038/s41567-019-0744-9} {\bibfield  {journal} {\bibinfo
  {journal} {Nature Physics}\ }\textbf {\bibinfo {volume} {16}},\ \bibinfo
  {pages} {218} (\bibinfo {year} {2020})}\BibitemShut {NoStop}%
\bibitem [{\citenamefont {Liu}\ and\ \citenamefont
  {Culcer}(2024)}]{PhysRevLett.132.186302}%
  \BibitemOpen
  \bibfield  {author} {\bibinfo {author} {\bibfnamefont {H.}~\bibnamefont
  {Liu}}\ and\ \bibinfo {author} {\bibfnamefont {D.}~\bibnamefont {Culcer}},\
  }\bibinfo {title} {Dominance of Extrinsic Scattering Mechanisms in the
  Orbital Hall Effect: Graphene, Transition Metal Dichalcogenides, and
  Topological Antiferromagnets},\ \href
  {https://doi.org/10.1103/PhysRevLett.132.186302} {\bibfield  {journal}
  {\bibinfo  {journal} {Phys. Rev. Lett.}\ }\textbf {\bibinfo {volume} {132}},\
  \bibinfo {pages} {186302} (\bibinfo {year} {2024})}\BibitemShut {NoStop}%
\bibitem [{\citenamefont {Hanke}\ \emph {et~al.}(2016)\citenamefont {Hanke},
  \citenamefont {Freimuth}, \citenamefont {Nandy}, \citenamefont {Zhang},
  \citenamefont {Bl\"ugel},\ and\ \citenamefont
  {Mokrousov}}]{PhysRevB.94.121114}%
  \BibitemOpen
  \bibfield  {author} {\bibinfo {author} {\bibfnamefont {J.-P.}\ \bibnamefont
  {Hanke}}, \bibinfo {author} {\bibfnamefont {F.}~\bibnamefont {Freimuth}},
  \bibinfo {author} {\bibfnamefont {A.~K.}\ \bibnamefont {Nandy}}, \bibinfo
  {author} {\bibfnamefont {H.}~\bibnamefont {Zhang}}, \bibinfo {author}
  {\bibfnamefont {S.}~\bibnamefont {Bl\"ugel}},\ and\ \bibinfo {author}
  {\bibfnamefont {Y.}~\bibnamefont {Mokrousov}},\ }\bibinfo {title} {Role of
  Berry phase theory for describing orbital magnetism: From magnetic
  heterostructures to topological orbital ferromagnets},\ \href
  {https://doi.org/10.1103/PhysRevB.94.121114} {\bibfield  {journal} {\bibinfo
  {journal} {Phys. Rev. B}\ }\textbf {\bibinfo {volume} {94}},\ \bibinfo
  {pages} {121114} (\bibinfo {year} {2016})}\BibitemShut {NoStop}%
\bibitem [{\citenamefont {Jiang}\ \emph {et~al.}(2024)\citenamefont {Jiang},
  \citenamefont {Song}, \citenamefont {Zhu}, \citenamefont {Fang},
  \citenamefont {Weng}, \citenamefont {Liu}, \citenamefont {Yang},\ and\
  \citenamefont {Fang}}]{PhysRevX.14.031039}%
  \BibitemOpen
  \bibfield  {author} {\bibinfo {author} {\bibfnamefont {Y.}~\bibnamefont
  {Jiang}}, \bibinfo {author} {\bibfnamefont {Z.}~\bibnamefont {Song}},
  \bibinfo {author} {\bibfnamefont {T.}~\bibnamefont {Zhu}}, \bibinfo {author}
  {\bibfnamefont {Z.}~\bibnamefont {Fang}}, \bibinfo {author} {\bibfnamefont
  {H.}~\bibnamefont {Weng}}, \bibinfo {author} {\bibfnamefont {Z.-X.}\
  \bibnamefont {Liu}}, \bibinfo {author} {\bibfnamefont {J.}~\bibnamefont
  {Yang}},\ and\ \bibinfo {author} {\bibfnamefont {C.}~\bibnamefont {Fang}},\
  }\bibinfo {title} {Enumeration of Spin-Space Groups: Toward a Complete
  Description of Symmetries of Magnetic Orders},\ \href
  {https://doi.org/10.1103/PhysRevX.14.031039} {\bibfield  {journal} {\bibinfo
  {journal} {Phys. Rev. X}\ }\textbf {\bibinfo {volume} {14}},\ \bibinfo
  {pages} {031039} (\bibinfo {year} {2024})}\BibitemShut {NoStop}%
\bibitem [{\citenamefont {Chen}\ \emph {et~al.}(2024)\citenamefont {Chen},
  \citenamefont {Ren}, \citenamefont {Zhu}, \citenamefont {Yu}, \citenamefont
  {Zhang}, \citenamefont {Liu}, \citenamefont {Li}, \citenamefont {Liu},
  \citenamefont {Li},\ and\ \citenamefont {Liu}}]{PhysRevX.14.031038}%
  \BibitemOpen
  \bibfield  {author} {\bibinfo {author} {\bibfnamefont {X.}~\bibnamefont
  {Chen}}, \bibinfo {author} {\bibfnamefont {J.}~\bibnamefont {Ren}}, \bibinfo
  {author} {\bibfnamefont {Y.}~\bibnamefont {Zhu}}, \bibinfo {author}
  {\bibfnamefont {Y.}~\bibnamefont {Yu}}, \bibinfo {author} {\bibfnamefont
  {A.}~\bibnamefont {Zhang}}, \bibinfo {author} {\bibfnamefont
  {P.}~\bibnamefont {Liu}}, \bibinfo {author} {\bibfnamefont {J.}~\bibnamefont
  {Li}}, \bibinfo {author} {\bibfnamefont {Y.}~\bibnamefont {Liu}}, \bibinfo
  {author} {\bibfnamefont {C.}~\bibnamefont {Li}},\ and\ \bibinfo {author}
  {\bibfnamefont {Q.}~\bibnamefont {Liu}},\ }\bibinfo {title} {Enumeration and
  Representation Theory of Spin Space Groups},\ \href
  {https://doi.org/10.1103/PhysRevX.14.031038} {\bibfield  {journal} {\bibinfo
  {journal} {Phys. Rev. X}\ }\textbf {\bibinfo {volume} {14}},\ \bibinfo
  {pages} {031038} (\bibinfo {year} {2024})}\BibitemShut {NoStop}%
\bibitem [{\citenamefont {Xiao}\ \emph {et~al.}(2024)\citenamefont {Xiao},
  \citenamefont {Zhao}, \citenamefont {Li}, \citenamefont {Shindou},\ and\
  \citenamefont {Song}}]{PhysRevX.14.031037}%
  \BibitemOpen
  \bibfield  {author} {\bibinfo {author} {\bibfnamefont {Z.}~\bibnamefont
  {Xiao}}, \bibinfo {author} {\bibfnamefont {J.}~\bibnamefont {Zhao}}, \bibinfo
  {author} {\bibfnamefont {Y.}~\bibnamefont {Li}}, \bibinfo {author}
  {\bibfnamefont {R.}~\bibnamefont {Shindou}},\ and\ \bibinfo {author}
  {\bibfnamefont {Z.-D.}\ \bibnamefont {Song}},\ }\bibinfo {title} {Spin Space
  Groups: Full Classification and Applications},\ \href
  {https://doi.org/10.1103/PhysRevX.14.031037} {\bibfield  {journal} {\bibinfo
  {journal} {Phys. Rev. X}\ }\textbf {\bibinfo {volume} {14}},\ \bibinfo
  {pages} {031037} (\bibinfo {year} {2024})}\BibitemShut {NoStop}%
\bibitem [{\citenamefont {LaMountain}\ \emph {et~al.}(2018)\citenamefont
  {LaMountain}, \citenamefont {Bergeron}, \citenamefont {Balla}, \citenamefont
  {Stanev}, \citenamefont {Hersam},\ and\ \citenamefont
  {Stern}}]{PhysRevB.97.045307}%
  \BibitemOpen
  \bibfield  {author} {\bibinfo {author} {\bibfnamefont {T.}~\bibnamefont
  {LaMountain}}, \bibinfo {author} {\bibfnamefont {H.}~\bibnamefont
  {Bergeron}}, \bibinfo {author} {\bibfnamefont {I.}~\bibnamefont {Balla}},
  \bibinfo {author} {\bibfnamefont {T.~K.}\ \bibnamefont {Stanev}}, \bibinfo
  {author} {\bibfnamefont {M.~C.}\ \bibnamefont {Hersam}},\ and\ \bibinfo
  {author} {\bibfnamefont {N.~P.}\ \bibnamefont {Stern}},\ }\bibinfo {title}
  {Valley-selective optical Stark effect probed by Kerr rotation},\ \href
  {https://doi.org/10.1103/PhysRevB.97.045307} {\bibfield  {journal} {\bibinfo
  {journal} {Phys. Rev. B}\ }\textbf {\bibinfo {volume} {97}},\ \bibinfo
  {pages} {045307} (\bibinfo {year} {2018})}\BibitemShut {NoStop}%
\bibitem [{\citenamefont {Sie}\ \emph {et~al.}(2015)\citenamefont {Sie},
  \citenamefont {McIver}, \citenamefont {Lee}, \citenamefont {Fu},
  \citenamefont {Kong},\ and\ \citenamefont {Gedik}}]{Sie2015Valleyselective}%
  \BibitemOpen
  \bibfield  {author} {\bibinfo {author} {\bibfnamefont {E.~J.}\ \bibnamefont
  {Sie}}, \bibinfo {author} {\bibfnamefont {J.~W.}\ \bibnamefont {McIver}},
  \bibinfo {author} {\bibfnamefont {Y.-H.}\ \bibnamefont {Lee}}, \bibinfo
  {author} {\bibfnamefont {L.}~\bibnamefont {Fu}}, \bibinfo {author}
  {\bibfnamefont {J.}~\bibnamefont {Kong}},\ and\ \bibinfo {author}
  {\bibfnamefont {N.}~\bibnamefont {Gedik}},\ }\bibinfo {title}
  {Valley-selective optical Stark effect in monolayer WS$_2$},\ \href
  {https://doi.org/10.1038/nmat4156} {\bibfield  {journal} {\bibinfo  {journal}
  {Nature Materials}\ }\textbf {\bibinfo {volume} {14}},\ \bibinfo {pages}
  {290} (\bibinfo {year} {2015})}\BibitemShut {NoStop}%
\bibitem [{\citenamefont {Sie}\ \emph {et~al.}(2017)\citenamefont {Sie},
  \citenamefont {Lui}, \citenamefont {Lee}, \citenamefont {Fu}, \citenamefont
  {Kong},\ and\ \citenamefont {Gedik}}]{doi:10.1126/science.aal2241}%
  \BibitemOpen
  \bibfield  {author} {\bibinfo {author} {\bibfnamefont {E.~J.}\ \bibnamefont
  {Sie}}, \bibinfo {author} {\bibfnamefont {C.~H.}\ \bibnamefont {Lui}},
  \bibinfo {author} {\bibfnamefont {Y.-H.}\ \bibnamefont {Lee}}, \bibinfo
  {author} {\bibfnamefont {L.}~\bibnamefont {Fu}}, \bibinfo {author}
  {\bibfnamefont {J.}~\bibnamefont {Kong}},\ and\ \bibinfo {author}
  {\bibfnamefont {N.}~\bibnamefont {Gedik}},\ }\bibinfo {title} {Large,
  valley-exclusive Bloch-Siegert shift in monolayer WS$_2$},\ \href
  {https://doi.org/10.1126/science.aal2241} {\bibfield  {journal} {\bibinfo
  {journal} {Science}\ }\textbf {\bibinfo {volume} {355}},\ \bibinfo {pages}
  {1066} (\bibinfo {year} {2017})}\BibitemShut {NoStop}%
\bibitem [{\citenamefont {De~Giovannini}\ \emph {et~al.}(2016)\citenamefont
  {De~Giovannini}, \citenamefont {Hübener},\ and\ \citenamefont
  {Rubio}}]{DeGiovannini2016Monitoring}%
  \BibitemOpen
  \bibfield  {author} {\bibinfo {author} {\bibfnamefont {U.}~\bibnamefont
  {De~Giovannini}}, \bibinfo {author} {\bibfnamefont {H.}~\bibnamefont
  {Hübener}},\ and\ \bibinfo {author} {\bibfnamefont {A.}~\bibnamefont
  {Rubio}},\ }\bibinfo {title} {Monitoring Electron-Photon Dressing in
  WSe$_2$},\ \href {https://doi.org/10.1021/acs.nanolett.6b04419} {\bibfield
  {journal} {\bibinfo  {journal} {Nano Letters}\ }\textbf {\bibinfo {volume}
  {16}},\ \bibinfo {pages} {7993} (\bibinfo {year} {2016})}\BibitemShut
  {NoStop}%
\bibitem [{\citenamefont {Chen}\ \emph {et~al.}(2024)\citenamefont {Chen},
  \citenamefont {Li}, \citenamefont {Bai}, \citenamefont {Mao}, \citenamefont
  {Zeer}, \citenamefont {Go}, \citenamefont {Dai}, \citenamefont {Huang},
  \citenamefont {Mokrousov},\ and\ \citenamefont
  {Niu}}]{doi:10.1021/acs.nanolett.3c05129}%
  \BibitemOpen
  \bibfield  {author} {\bibinfo {author} {\bibfnamefont {Z.}~\bibnamefont
  {Chen}}, \bibinfo {author} {\bibfnamefont {R.}~\bibnamefont {Li}}, \bibinfo
  {author} {\bibfnamefont {Y.}~\bibnamefont {Bai}}, \bibinfo {author}
  {\bibfnamefont {N.}~\bibnamefont {Mao}}, \bibinfo {author} {\bibfnamefont
  {M.}~\bibnamefont {Zeer}}, \bibinfo {author} {\bibfnamefont {D.}~\bibnamefont
  {Go}}, \bibinfo {author} {\bibfnamefont {Y.}~\bibnamefont {Dai}}, \bibinfo
  {author} {\bibfnamefont {B.}~\bibnamefont {Huang}}, \bibinfo {author}
  {\bibfnamefont {Y.}~\bibnamefont {Mokrousov}},\ and\ \bibinfo {author}
  {\bibfnamefont {C.}~\bibnamefont {Niu}},\ }\bibinfo {title}
  {Topology-Engineered Orbital Hall Effect in Two-Dimensional Ferromagnets},\
  \href {https://doi.org/10.1021/acs.nanolett.3c05129} {\bibfield  {journal}
  {\bibinfo  {journal} {Nano Letters}\ }\textbf {\bibinfo {volume} {24}},\
  \bibinfo {pages} {4826} (\bibinfo {year} {2024})}\BibitemShut {NoStop}%
\bibitem [{\citenamefont {Guo}\ \emph {et~al.}(2025)\citenamefont {Guo},
  \citenamefont {Li},\ and\ \citenamefont {Wang}}]{PhysRevB.111.L140404}%
  \BibitemOpen
  \bibfield  {author} {\bibinfo {author} {\bibfnamefont {S.-D.}\ \bibnamefont
  {Guo}}, \bibinfo {author} {\bibfnamefont {P.}~\bibnamefont {Li}},\ and\
  \bibinfo {author} {\bibfnamefont {G.}~\bibnamefont {Wang}},\ }\bibinfo
  {title} {First-principles calculations study of valley polarization in
  antiferromagnetic bilayer systems},\ \href
  {https://doi.org/10.1103/PhysRevB.111.L140404} {\bibfield  {journal}
  {\bibinfo  {journal} {Phys. Rev. B}\ }\textbf {\bibinfo {volume} {111}},\
  \bibinfo {pages} {L140404} (\bibinfo {year} {2025})}\BibitemShut {NoStop}%
\bibitem [{\citenamefont {Wu}\ \emph {et~al.}(2025)\citenamefont {Wu},
  \citenamefont {Sun}, \citenamefont {Dong}, \citenamefont {Wu},\ and\
  \citenamefont {Li}}]{https://doi.org/10.1002/adfm.202501506}%
  \BibitemOpen
  \bibfield  {author} {\bibinfo {author} {\bibfnamefont {C.}~\bibnamefont
  {Wu}}, \bibinfo {author} {\bibfnamefont {H.}~\bibnamefont {Sun}}, \bibinfo
  {author} {\bibfnamefont {P.}~\bibnamefont {Dong}}, \bibinfo {author}
  {\bibfnamefont {Y.-Z.}\ \bibnamefont {Wu}},\ and\ \bibinfo {author}
  {\bibfnamefont {P.}~\bibnamefont {Li}},\ }\bibinfo {title} {Coexisting
  Triferroic and Multiple Types of Valley Polarization by Structural Phase
  Transition in 2D Materials},\ \href
  {https://doi.org/https://doi.org/10.1002/adfm.202501506} {\bibfield
  {journal} {\bibinfo  {journal} {Advanced Functional Materials}\ }\textbf
  {\bibinfo {volume} {35}},\ \bibinfo {pages} {2501506} (\bibinfo {year}
  {2025})}\BibitemShut {NoStop}%
\bibitem [{\citenamefont {Zhang}\ \emph {et~al.}(2025)\citenamefont {Zhang},
  \citenamefont {Wang}, \citenamefont {Xu}, \citenamefont {Dai},\ and\
  \citenamefont {Ma}}]{D5MH00242G}%
  \BibitemOpen
  \bibfield  {author} {\bibinfo {author} {\bibfnamefont {T.}~\bibnamefont
  {Zhang}}, \bibinfo {author} {\bibfnamefont {M.}~\bibnamefont {Wang}},
  \bibinfo {author} {\bibfnamefont {X.}~\bibnamefont {Xu}}, \bibinfo {author}
  {\bibfnamefont {Y.}~\bibnamefont {Dai}},\ and\ \bibinfo {author}
  {\bibfnamefont {Y.}~\bibnamefont {Ma}},\ }\bibinfo {title} {Gate-controllable
  quadri-layertronics in a 2D multiferroic antiferromagnet},\ \href
  {https://doi.org/10.1039/D5MH00242G} {\bibfield  {journal} {\bibinfo
  {journal} {Mater. Horiz.}\ }\textbf {\bibinfo {volume} {12}},\ \bibinfo
  {pages} {6919} (\bibinfo {year} {2025})}\BibitemShut {NoStop}%
\bibitem [{Sup()}]{SuppMat}%
  \BibitemOpen
  \href@noop {} {\bibinfo {title} {See {S}upplemental {M}aterial at
  \url{http://link.aps.org/supplemental/xxx}, which includes
  refs.\cite{Bukov04032015,PhysRevB.54.11169,KRESSE199615,PhysRevLett.77.3865,PhysRevLett.100.136406,PhysRevB.50.17953,PhysRevB.59.1758,PhysRevB.13.5188,PhysRevB.74.125106,10.1063/1.3382344,WANG2021108033,MOSTOFI2008685,WU2018405,10.1063/1.1329672,doi:10.1126/science.abb7023,10.1038/s41467-021-22324-8,PhysRevMaterials.7.064002},
  for more details about {TB} model; first-principles calculation method; full
  crystal structures; full band structures; phase diagram of {TB} trilayer and
  {VS}i$_2${N}$_4$ trilayer under {RCPL}; phase diagram of {VS}i$_2${N}$_4$
  bilayer under {LCPL}}}\BibitemShut {NoStop}%
\bibitem [{\citenamefont {Phong}\ \emph {et~al.}(2019)\citenamefont {Phong},
  \citenamefont {Addison}, \citenamefont {Ahn}, \citenamefont {Min},
  \citenamefont {Agarwal},\ and\ \citenamefont
  {Mele}}]{PhysRevLett.123.236403}%
  \BibitemOpen
  \bibfield  {author} {\bibinfo {author} {\bibfnamefont {V.~o.~T.}\
  \bibnamefont {Phong}}, \bibinfo {author} {\bibfnamefont {Z.}~\bibnamefont
  {Addison}}, \bibinfo {author} {\bibfnamefont {S.}~\bibnamefont {Ahn}},
  \bibinfo {author} {\bibfnamefont {H.}~\bibnamefont {Min}}, \bibinfo {author}
  {\bibfnamefont {R.}~\bibnamefont {Agarwal}},\ and\ \bibinfo {author}
  {\bibfnamefont {E.~J.}\ \bibnamefont {Mele}},\ }\bibinfo {title} {Optically
  Controlled Orbitronics on a Triangular Lattice},\ \href
  {https://doi.org/10.1103/PhysRevLett.123.236403} {\bibfield  {journal}
  {\bibinfo  {journal} {Phys. Rev. Lett.}\ }\textbf {\bibinfo {volume} {123}},\
  \bibinfo {pages} {236403} (\bibinfo {year} {2019})}\BibitemShut {NoStop}%
\bibitem [{\citenamefont {Shi}\ and\ \citenamefont
  {Zhou}(2021)}]{PhysRevB.104.155146}%
  \BibitemOpen
  \bibfield  {author} {\bibinfo {author} {\bibfnamefont {Y.}~\bibnamefont
  {Shi}}\ and\ \bibinfo {author} {\bibfnamefont {J.}~\bibnamefont {Zhou}},\
  }\bibinfo {title} {Coherence control of directional nonlinear photocurrent in
  spatially symmetric systems},\ \href
  {https://doi.org/10.1103/PhysRevB.104.155146} {\bibfield  {journal} {\bibinfo
   {journal} {Phys. Rev. B}\ }\textbf {\bibinfo {volume} {104}},\ \bibinfo
  {pages} {155146} (\bibinfo {year} {2021})}\BibitemShut {NoStop}%
\bibitem [{\citenamefont {Chen}\ \emph {et~al.}(2020)\citenamefont {Chen},
  \citenamefont {Chen}, \citenamefont {Gao}, \citenamefont {Zhou},\ and\
  \citenamefont {Xu}}]{PhysRevLett.124.036803}%
  \BibitemOpen
  \bibfield  {author} {\bibinfo {author} {\bibfnamefont {R.}~\bibnamefont
  {Chen}}, \bibinfo {author} {\bibfnamefont {C.-Z.}\ \bibnamefont {Chen}},
  \bibinfo {author} {\bibfnamefont {J.-H.}\ \bibnamefont {Gao}}, \bibinfo
  {author} {\bibfnamefont {B.}~\bibnamefont {Zhou}},\ and\ \bibinfo {author}
  {\bibfnamefont {D.-H.}\ \bibnamefont {Xu}},\ }\bibinfo {title} {Higher-Order
  Topological Insulators in Quasicrystals},\ \href
  {https://doi.org/10.1103/PhysRevLett.124.036803} {\bibfield  {journal}
  {\bibinfo  {journal} {Phys. Rev. Lett.}\ }\textbf {\bibinfo {volume} {124}},\
  \bibinfo {pages} {036803} (\bibinfo {year} {2020})}\BibitemShut {NoStop}%
\bibitem [{\citenamefont {Li}\ \emph {et~al.}(2024)\citenamefont {Li},
  \citenamefont {Liu},\ and\ \citenamefont {Liu}}]{PhysRevB.109.L201109}%
  \BibitemOpen
  \bibfield  {author} {\bibinfo {author} {\bibfnamefont {Y.-X.}\ \bibnamefont
  {Li}}, \bibinfo {author} {\bibfnamefont {Y.}~\bibnamefont {Liu}},\ and\
  \bibinfo {author} {\bibfnamefont {C.-C.}\ \bibnamefont {Liu}},\ }\bibinfo
  {title} {Creation and manipulation of higher-order topological states by
  altermagnets},\ \href {https://doi.org/10.1103/PhysRevB.109.L201109}
  {\bibfield  {journal} {\bibinfo  {journal} {Phys. Rev. B}\ }\textbf {\bibinfo
  {volume} {109}},\ \bibinfo {pages} {L201109} (\bibinfo {year}
  {2024})}\BibitemShut {NoStop}%
\bibitem [{\citenamefont {Choi}\ \emph {et~al.}(2020)\citenamefont {Choi},
  \citenamefont {Xie}, \citenamefont {Chen}, \citenamefont {Park},
  \citenamefont {Song}, \citenamefont {Yoon}, \citenamefont {Kim},
  \citenamefont {Taniguchi}, \citenamefont {Watanabe}, \citenamefont {Kim},
  \citenamefont {Fong}, \citenamefont {Ali}, \citenamefont {Law},\ and\
  \citenamefont {Lee}}]{Choi_2020_HigherOrderTopologyWTe2}%
  \BibitemOpen
  \bibfield  {author} {\bibinfo {author} {\bibfnamefont {Y.-B.}\ \bibnamefont
  {Choi}}, \bibinfo {author} {\bibfnamefont {Y.}~\bibnamefont {Xie}}, \bibinfo
  {author} {\bibfnamefont {C.-Z.}\ \bibnamefont {Chen}}, \bibinfo {author}
  {\bibfnamefont {J.}~\bibnamefont {Park}}, \bibinfo {author} {\bibfnamefont
  {S.-B.}\ \bibnamefont {Song}}, \bibinfo {author} {\bibfnamefont
  {J.}~\bibnamefont {Yoon}}, \bibinfo {author} {\bibfnamefont {B.~J.}\
  \bibnamefont {Kim}}, \bibinfo {author} {\bibfnamefont {T.}~\bibnamefont
  {Taniguchi}}, \bibinfo {author} {\bibfnamefont {K.}~\bibnamefont {Watanabe}},
  \bibinfo {author} {\bibfnamefont {J.}~\bibnamefont {Kim}}, \bibinfo {author}
  {\bibfnamefont {K.~C.}\ \bibnamefont {Fong}}, \bibinfo {author}
  {\bibfnamefont {M.~N.}\ \bibnamefont {Ali}}, \bibinfo {author} {\bibfnamefont
  {K.~T.}\ \bibnamefont {Law}},\ and\ \bibinfo {author} {\bibfnamefont {G.-H.}\
  \bibnamefont {Lee}},\ }\bibinfo {title} {Evidence of higher-order topology in
  multilayer WTe$_2$ from Josephson coupling through anisotropic hinge
  states},\ \href {https://doi.org/10.1038/s41563-020-0721-9} {\bibfield
  {journal} {\bibinfo  {journal} {Nature Materials}\ }\textbf {\bibinfo
  {volume} {19}},\ \bibinfo {pages} {974} (\bibinfo {year} {2020})}\BibitemShut
  {NoStop}%
\bibitem [{\citenamefont {Li}\ and\ \citenamefont
  {Liu}(2023)}]{PhysRevB.108.205410}%
  \BibitemOpen
  \bibfield  {author} {\bibinfo {author} {\bibfnamefont {Y.-X.}\ \bibnamefont
  {Li}}\ and\ \bibinfo {author} {\bibfnamefont {C.-C.}\ \bibnamefont {Liu}},\
  }\bibinfo {title} {Majorana corner modes and tunable patterns in an
  altermagnet heterostructure},\ \href
  {https://doi.org/10.1103/PhysRevB.108.205410} {\bibfield  {journal} {\bibinfo
   {journal} {Phys. Rev. B}\ }\textbf {\bibinfo {volume} {108}},\ \bibinfo
  {pages} {205410} (\bibinfo {year} {2023})}\BibitemShut {NoStop}%
\bibitem [{\citenamefont {Milfeld}\ and\ \citenamefont
  {Wyatt}(1983)}]{PhysRevA.27.72}%
  \BibitemOpen
  \bibfield  {author} {\bibinfo {author} {\bibfnamefont {K.~F.}\ \bibnamefont
  {Milfeld}}\ and\ \bibinfo {author} {\bibfnamefont {R.~E.}\ \bibnamefont
  {Wyatt}},\ }\bibinfo {title} {Study, extension, and application of Floquet
  theory for quantum molecular systems in an oscillating field},\ \href
  {https://doi.org/10.1103/PhysRevA.27.72} {\bibfield  {journal} {\bibinfo
  {journal} {Phys. Rev. A}\ }\textbf {\bibinfo {volume} {27}},\ \bibinfo
  {pages} {72} (\bibinfo {year} {1983})}\BibitemShut {NoStop}%
\bibitem [{\citenamefont {G\'omez-Le\'on}\ and\ \citenamefont
  {Platero}(2013)}]{PhysRevLett.110.200403}%
  \BibitemOpen
  \bibfield  {author} {\bibinfo {author} {\bibfnamefont {A.}~\bibnamefont
  {G\'omez-Le\'on}}\ and\ \bibinfo {author} {\bibfnamefont {G.}~\bibnamefont
  {Platero}},\ }\bibinfo {title} {Floquet-Bloch Theory and Topology in
  Periodically Driven Lattices},\ \href
  {https://doi.org/10.1103/PhysRevLett.110.200403} {\bibfield  {journal}
  {\bibinfo  {journal} {Phys. Rev. Lett.}\ }\textbf {\bibinfo {volume} {110}},\
  \bibinfo {pages} {200403} (\bibinfo {year} {2013})}\BibitemShut {NoStop}%
\bibitem [{\citenamefont {Zhou}\ \emph {et~al.}(2023)\citenamefont {Zhou},
  \citenamefont {Bao}, \citenamefont {Fan}, \citenamefont {Zhou}, \citenamefont
  {Gao}, \citenamefont {Zhong}, \citenamefont {Lin}, \citenamefont {Liu},
  \citenamefont {Yu}, \citenamefont {Tang}, \citenamefont {Meng}, \citenamefont
  {Duan},\ and\ \citenamefont
  {Zhou}}]{Zhou_2023_PseudospinFloquetBlackPhosphorus}%
  \BibitemOpen
  \bibfield  {author} {\bibinfo {author} {\bibfnamefont {S.}~\bibnamefont
  {Zhou}}, \bibinfo {author} {\bibfnamefont {C.}~\bibnamefont {Bao}}, \bibinfo
  {author} {\bibfnamefont {B.}~\bibnamefont {Fan}}, \bibinfo {author}
  {\bibfnamefont {H.}~\bibnamefont {Zhou}}, \bibinfo {author} {\bibfnamefont
  {Q.}~\bibnamefont {Gao}}, \bibinfo {author} {\bibfnamefont {H.}~\bibnamefont
  {Zhong}}, \bibinfo {author} {\bibfnamefont {T.}~\bibnamefont {Lin}}, \bibinfo
  {author} {\bibfnamefont {H.}~\bibnamefont {Liu}}, \bibinfo {author}
  {\bibfnamefont {P.}~\bibnamefont {Yu}}, \bibinfo {author} {\bibfnamefont
  {P.}~\bibnamefont {Tang}}, \bibinfo {author} {\bibfnamefont {S.}~\bibnamefont
  {Meng}}, \bibinfo {author} {\bibfnamefont {W.}~\bibnamefont {Duan}},\ and\
  \bibinfo {author} {\bibfnamefont {S.}~\bibnamefont {Zhou}},\ }\bibinfo
  {title} {Pseudospin-selective Floquet band engineering in black phosphorus},\
  \href {https://doi.org/10.1038/s41586-022-05610-3} {\bibfield  {journal}
  {\bibinfo  {journal} {Nature}\ }\textbf {\bibinfo {volume} {614}},\ \bibinfo
  {pages} {75} (\bibinfo {year} {2023})}\BibitemShut {NoStop}%
\bibitem [{\citenamefont {Sun}\ \emph {et~al.}(2025)\citenamefont {Sun},
  \citenamefont {Du},\ and\ \citenamefont {Zhao}}]{PhysRevB.111.155404}%
  \BibitemOpen
  \bibfield  {author} {\bibinfo {author} {\bibfnamefont {L.}~\bibnamefont
  {Sun}}, \bibinfo {author} {\bibfnamefont {Y.}~\bibnamefont {Du}},\ and\
  \bibinfo {author} {\bibfnamefont {M.}~\bibnamefont {Zhao}},\ }\bibinfo
  {title} {Giant Berry curvature dipole in monolayer ${\mathrm{MoS}}_{2}$ in a
  periodic electric field},\ \href
  {https://doi.org/10.1103/PhysRevB.111.155404} {\bibfield  {journal} {\bibinfo
   {journal} {Phys. Rev. B}\ }\textbf {\bibinfo {volume} {111}},\ \bibinfo
  {pages} {155404} (\bibinfo {year} {2025})}\BibitemShut {NoStop}%
\bibitem [{\citenamefont {Hong}\ \emph {et~al.}(2020)\citenamefont {Hong},
  \citenamefont {Liu}, \citenamefont {Wang}, \citenamefont {Zhou},
  \citenamefont {Ma}, \citenamefont {Xu}, \citenamefont {Feng}, \citenamefont
  {Chen}, \citenamefont {Chen}, \citenamefont {Sun}, \citenamefont {Chen},
  \citenamefont {Cheng},\ and\ \citenamefont
  {Ren}}]{doi:10.1126/science.abb7023}%
  \BibitemOpen
  \bibfield  {author} {\bibinfo {author} {\bibfnamefont {Y.-L.}\ \bibnamefont
  {Hong}}, \bibinfo {author} {\bibfnamefont {Z.}~\bibnamefont {Liu}}, \bibinfo
  {author} {\bibfnamefont {L.}~\bibnamefont {Wang}}, \bibinfo {author}
  {\bibfnamefont {T.}~\bibnamefont {Zhou}}, \bibinfo {author} {\bibfnamefont
  {W.}~\bibnamefont {Ma}}, \bibinfo {author} {\bibfnamefont {C.}~\bibnamefont
  {Xu}}, \bibinfo {author} {\bibfnamefont {S.}~\bibnamefont {Feng}}, \bibinfo
  {author} {\bibfnamefont {L.}~\bibnamefont {Chen}}, \bibinfo {author}
  {\bibfnamefont {M.-L.}\ \bibnamefont {Chen}}, \bibinfo {author}
  {\bibfnamefont {D.-M.}\ \bibnamefont {Sun}}, \bibinfo {author} {\bibfnamefont
  {X.-Q.}\ \bibnamefont {Chen}}, \bibinfo {author} {\bibfnamefont {H.-M.}\
  \bibnamefont {Cheng}},\ and\ \bibinfo {author} {\bibfnamefont
  {W.}~\bibnamefont {Ren}},\ }\bibinfo {title} {Chemical vapor deposition of
  layered two-dimensional MoSi$_2$N$_4$ materials},\ \href
  {https://doi.org/10.1126/science.abb7023} {\bibfield  {journal} {\bibinfo
  {journal} {Science}\ }\textbf {\bibinfo {volume} {369}},\ \bibinfo {pages}
  {670} (\bibinfo {year} {2020})}\BibitemShut {NoStop}%
\bibitem [{\citenamefont {Wang}\ \emph {et~al.}(2021)\citenamefont {Wang},
  \citenamefont {Shi}, \citenamefont {Liu}, \citenamefont {Zhang},
  \citenamefont {Hong}, \citenamefont {Li}, \citenamefont {Gao}, \citenamefont
  {Chen}, \citenamefont {Ren}, \citenamefont {Cheng}, \citenamefont {Li},\ and\
  \citenamefont {Chen}}]{Wang_2021_MA2Z4TopologicalMagneticSuperconducting}%
  \BibitemOpen
  \bibfield  {author} {\bibinfo {author} {\bibfnamefont {L.}~\bibnamefont
  {Wang}}, \bibinfo {author} {\bibfnamefont {Y.}~\bibnamefont {Shi}}, \bibinfo
  {author} {\bibfnamefont {M.}~\bibnamefont {Liu}}, \bibinfo {author}
  {\bibfnamefont {A.}~\bibnamefont {Zhang}}, \bibinfo {author} {\bibfnamefont
  {Y.-L.}\ \bibnamefont {Hong}}, \bibinfo {author} {\bibfnamefont
  {R.}~\bibnamefont {Li}}, \bibinfo {author} {\bibfnamefont {Q.}~\bibnamefont
  {Gao}}, \bibinfo {author} {\bibfnamefont {M.}~\bibnamefont {Chen}}, \bibinfo
  {author} {\bibfnamefont {W.}~\bibnamefont {Ren}}, \bibinfo {author}
  {\bibfnamefont {H.-M.}\ \bibnamefont {Cheng}}, \bibinfo {author}
  {\bibfnamefont {Y.}~\bibnamefont {Li}},\ and\ \bibinfo {author}
  {\bibfnamefont {X.-Q.}\ \bibnamefont {Chen}},\ }\bibinfo {title}
  {Intercalated architecture of MA$_2$Z$_4$ family layered van der Waals
  materials with emerging topological, magnetic and superconducting
  properties},\ \href@noop {} {\bibfield  {journal} {\bibinfo  {journal}
  {Nature Communications}\ }\textbf {\bibinfo {volume} {12}} (\bibinfo {year}
  {2021})}\BibitemShut {NoStop}%
\bibitem [{\citenamefont {Liu}\ \emph {et~al.}(2024)\citenamefont {Liu},
  \citenamefont {Ma}, \citenamefont {Li},\ and\ \citenamefont
  {Zhao}}]{Liu2024Tailoring}%
  \BibitemOpen
  \bibfield  {author} {\bibinfo {author} {\bibfnamefont {K.}~\bibnamefont
  {Liu}}, \bibinfo {author} {\bibfnamefont {X.}~\bibnamefont {Ma}}, \bibinfo
  {author} {\bibfnamefont {Y.}~\bibnamefont {Li}},\ and\ \bibinfo {author}
  {\bibfnamefont {M.}~\bibnamefont {Zhao}},\ }\bibinfo {title} {Tailoring the
  quantum anomalous layer Hall effect in multiferroic bilayers through
  sliding},\ \href {https://doi.org/10.1038/s41524-024-01306-6} {\bibfield
  {journal} {\bibinfo  {journal} {npj Computational Materials}\ }\textbf
  {\bibinfo {volume} {10}} (\bibinfo {year} {2024})}\BibitemShut {NoStop}%
\bibitem [{\citenamefont {Xun}\ \emph {et~al.}(2024)\citenamefont {Xun},
  \citenamefont {Wu}, \citenamefont {Sun}, \citenamefont {Zhang}, \citenamefont
  {Wu},\ and\ \citenamefont {Li}}]{doi:10.1021/acs.nanolett.4c00597}%
  \BibitemOpen
  \bibfield  {author} {\bibinfo {author} {\bibfnamefont {W.}~\bibnamefont
  {Xun}}, \bibinfo {author} {\bibfnamefont {C.}~\bibnamefont {Wu}}, \bibinfo
  {author} {\bibfnamefont {H.}~\bibnamefont {Sun}}, \bibinfo {author}
  {\bibfnamefont {W.}~\bibnamefont {Zhang}}, \bibinfo {author} {\bibfnamefont
  {Y.-Z.}\ \bibnamefont {Wu}},\ and\ \bibinfo {author} {\bibfnamefont
  {P.}~\bibnamefont {Li}},\ }\bibinfo {title} {Coexisting Magnetism,
  Ferroelectric, and Ferrovalley Multiferroic in Stacking-Dependent
  Two-Dimensional Materials},\ \href
  {https://doi.org/10.1021/acs.nanolett.4c00597} {\bibfield  {journal}
  {\bibinfo  {journal} {Nano Letters}\ }\textbf {\bibinfo {volume} {24}},\
  \bibinfo {pages} {3541} (\bibinfo {year} {2024})}\BibitemShut {NoStop}%
\bibitem [{\citenamefont {Zhang}\ \emph {et~al.}(2024)\citenamefont {Zhang},
  \citenamefont {Xu}, \citenamefont {Guo}, \citenamefont {Dai},\ and\
  \citenamefont {Ma}}]{doi:10.1021/acs.nanolett.3c04597}%
  \BibitemOpen
  \bibfield  {author} {\bibinfo {author} {\bibfnamefont {T.}~\bibnamefont
  {Zhang}}, \bibinfo {author} {\bibfnamefont {X.}~\bibnamefont {Xu}}, \bibinfo
  {author} {\bibfnamefont {J.}~\bibnamefont {Guo}}, \bibinfo {author}
  {\bibfnamefont {Y.}~\bibnamefont {Dai}},\ and\ \bibinfo {author}
  {\bibfnamefont {Y.}~\bibnamefont {Ma}},\ }\bibinfo {title} {Layer-Polarized
  Anomalous Hall Effects from Inversion-Symmetric Single-Layer Lattices},\
  \href {https://doi.org/10.1021/acs.nanolett.3c04597} {\bibfield  {journal}
  {\bibinfo  {journal} {Nano Letters}\ }\textbf {\bibinfo {volume} {24}},\
  \bibinfo {pages} {1009} (\bibinfo {year} {2024})}\BibitemShut {NoStop}%
\bibitem [{\citenamefont {Zhang}\ \emph {et~al.}(2023)\citenamefont {Zhang},
  \citenamefont {Xu}, \citenamefont {Huang}, \citenamefont {Dai}, \citenamefont
  {Kou},\ and\ \citenamefont {Ma}}]{D2MH00906D}%
  \BibitemOpen
  \bibfield  {author} {\bibinfo {author} {\bibfnamefont {T.}~\bibnamefont
  {Zhang}}, \bibinfo {author} {\bibfnamefont {X.}~\bibnamefont {Xu}}, \bibinfo
  {author} {\bibfnamefont {B.}~\bibnamefont {Huang}}, \bibinfo {author}
  {\bibfnamefont {Y.}~\bibnamefont {Dai}}, \bibinfo {author} {\bibfnamefont
  {L.}~\bibnamefont {Kou}},\ and\ \bibinfo {author} {\bibfnamefont
  {Y.}~\bibnamefont {Ma}},\ }\bibinfo {title} {Layer-polarized anomalous Hall
  effects in valleytronic van der Waals bilayers},\ \href
  {https://doi.org/10.1039/D2MH00906D} {\bibfield  {journal} {\bibinfo
  {journal} {Mater. Horiz.}\ }\textbf {\bibinfo {volume} {10}},\ \bibinfo
  {pages} {483} (\bibinfo {year} {2023})}\BibitemShut {NoStop}%
\bibitem [{\citenamefont {Cysne}\ \emph {et~al.}(2021)\citenamefont {Cysne},
  \citenamefont {Costa}, \citenamefont {Canonico}, \citenamefont {Nardelli},
  \citenamefont {Muniz},\ and\ \citenamefont
  {Rappoport}}]{PhysRevLett.126.056601}%
  \BibitemOpen
  \bibfield  {author} {\bibinfo {author} {\bibfnamefont {T.~P.}\ \bibnamefont
  {Cysne}}, \bibinfo {author} {\bibfnamefont {M.}~\bibnamefont {Costa}},
  \bibinfo {author} {\bibfnamefont {L.~M.}\ \bibnamefont {Canonico}}, \bibinfo
  {author} {\bibfnamefont {M.~B.}\ \bibnamefont {Nardelli}}, \bibinfo {author}
  {\bibfnamefont {R.~B.}\ \bibnamefont {Muniz}},\ and\ \bibinfo {author}
  {\bibfnamefont {T.~G.}\ \bibnamefont {Rappoport}},\ }\bibinfo {title}
  {Disentangling Orbital and Valley Hall Effects in Bilayers of Transition
  Metal Dichalcogenides},\ \href
  {https://doi.org/10.1103/PhysRevLett.126.056601} {\bibfield  {journal}
  {\bibinfo  {journal} {Phys. Rev. Lett.}\ }\textbf {\bibinfo {volume} {126}},\
  \bibinfo {pages} {056601} (\bibinfo {year} {2021})}\BibitemShut {NoStop}%
\bibitem [{\citenamefont {Li}\ \emph {et~al.}(2021)\citenamefont {Li},
  \citenamefont {Chen}, \citenamefont {Jiang},\ and\ \citenamefont
  {Xie}}]{PhysRevLett.127.236402}%
  \BibitemOpen
  \bibfield  {author} {\bibinfo {author} {\bibfnamefont {H.}~\bibnamefont
  {Li}}, \bibinfo {author} {\bibfnamefont {C.-Z.}\ \bibnamefont {Chen}},
  \bibinfo {author} {\bibfnamefont {H.}~\bibnamefont {Jiang}},\ and\ \bibinfo
  {author} {\bibfnamefont {X.~C.}\ \bibnamefont {Xie}},\ }\bibinfo {title}
  {Coexistence of Quantum Hall and Quantum Anomalous Hall Phases in Disordered
  ${\mathrm{MnBi}}_{2}{\mathrm{Te}}_{4}$},\ \href
  {https://doi.org/10.1103/PhysRevLett.127.236402} {\bibfield  {journal}
  {\bibinfo  {journal} {Phys. Rev. Lett.}\ }\textbf {\bibinfo {volume} {127}},\
  \bibinfo {pages} {236402} (\bibinfo {year} {2021})}\BibitemShut {NoStop}%
\bibitem [{\citenamefont {Kechedzhi}\ \emph {et~al.}(2007)\citenamefont
  {Kechedzhi}, \citenamefont {Fal'ko}, \citenamefont {McCann},\ and\
  \citenamefont {Altshuler}}]{PhysRevLett.98.176806}%
  \BibitemOpen
  \bibfield  {author} {\bibinfo {author} {\bibfnamefont {K.}~\bibnamefont
  {Kechedzhi}}, \bibinfo {author} {\bibfnamefont {V.~I.}\ \bibnamefont
  {Fal'ko}}, \bibinfo {author} {\bibfnamefont {E.}~\bibnamefont {McCann}},\
  and\ \bibinfo {author} {\bibfnamefont {B.~L.}\ \bibnamefont {Altshuler}},\
  }\bibinfo {title} {Influence of Trigonal Warping on Interference Effects in
  Bilayer Graphene},\ \href {https://doi.org/10.1103/PhysRevLett.98.176806}
  {\bibfield  {journal} {\bibinfo  {journal} {Phys. Rev. Lett.}\ }\textbf
  {\bibinfo {volume} {98}},\ \bibinfo {pages} {176806} (\bibinfo {year}
  {2007})}\BibitemShut {NoStop}%
\bibitem [{\citenamefont {Rakyta}\ \emph {et~al.}(2010)\citenamefont {Rakyta},
  \citenamefont {Korm\'anyos},\ and\ \citenamefont
  {Cserti}}]{PhysRevB.82.113405}%
  \BibitemOpen
  \bibfield  {author} {\bibinfo {author} {\bibfnamefont {P.}~\bibnamefont
  {Rakyta}}, \bibinfo {author} {\bibfnamefont {A.}~\bibnamefont
  {Korm\'anyos}},\ and\ \bibinfo {author} {\bibfnamefont {J.}~\bibnamefont
  {Cserti}},\ }\bibinfo {title} {Trigonal warping and anisotropic band
  splitting in monolayer graphene due to Rashba spin-orbit coupling},\ \href
  {https://doi.org/10.1103/PhysRevB.82.113405} {\bibfield  {journal} {\bibinfo
  {journal} {Phys. Rev. B}\ }\textbf {\bibinfo {volume} {82}},\ \bibinfo
  {pages} {113405} (\bibinfo {year} {2010})}\BibitemShut {NoStop}%
\bibitem [{\citenamefont {Zeng}\ \emph {et~al.}(2017)\citenamefont {Zeng},
  \citenamefont {Ren}, \citenamefont {Zhang},\ and\ \citenamefont
  {Qiao}}]{PhysRevB.95.045424}%
  \BibitemOpen
  \bibfield  {author} {\bibinfo {author} {\bibfnamefont {J.}~\bibnamefont
  {Zeng}}, \bibinfo {author} {\bibfnamefont {Y.}~\bibnamefont {Ren}}, \bibinfo
  {author} {\bibfnamefont {K.}~\bibnamefont {Zhang}},\ and\ \bibinfo {author}
  {\bibfnamefont {Z.}~\bibnamefont {Qiao}},\ }\bibinfo {title} {Topological
  phase transition from trigonal warping in van der Waals multilayers},\ \href
  {https://doi.org/10.1103/PhysRevB.95.045424} {\bibfield  {journal} {\bibinfo
  {journal} {Phys. Rev. B}\ }\textbf {\bibinfo {volume} {95}},\ \bibinfo
  {pages} {045424} (\bibinfo {year} {2017})}\BibitemShut {NoStop}%
\bibitem [{\citenamefont {Wu}\ \emph {et~al.}(2021)\citenamefont {Wu},
  \citenamefont {Zhu},\ and\ \citenamefont {Yu}}]{PhysRevB.104.195427}%
  \BibitemOpen
  \bibfield  {author} {\bibinfo {author} {\bibfnamefont {Y.-L.}\ \bibnamefont
  {Wu}}, \bibinfo {author} {\bibfnamefont {G.-H.}\ \bibnamefont {Zhu}},\ and\
  \bibinfo {author} {\bibfnamefont {X.-Q.}\ \bibnamefont {Yu}},\ }\bibinfo
  {title} {Nonlinear anomalous Nernst effect in strained graphene induced by
  trigonal warping},\ \href {https://doi.org/10.1103/PhysRevB.104.195427}
  {\bibfield  {journal} {\bibinfo  {journal} {Phys. Rev. B}\ }\textbf {\bibinfo
  {volume} {104}},\ \bibinfo {pages} {195427} (\bibinfo {year}
  {2021})}\BibitemShut {NoStop}%
\bibitem [{\citenamefont {Kresse}\ and\ \citenamefont
  {Furthm\"uller}(1996)}]{PhysRevB.54.11169}%
  \BibitemOpen
  \bibfield  {author} {\bibinfo {author} {\bibfnamefont {G.}~\bibnamefont
  {Kresse}}\ and\ \bibinfo {author} {\bibfnamefont {J.}~\bibnamefont
  {Furthm\"uller}},\ }\bibinfo {title} {Efficient iterative schemes for ab
  initio total-energy calculations using a plane-wave basis set},\ \href
  {https://doi.org/10.1103/PhysRevB.54.11169} {\bibfield  {journal} {\bibinfo
  {journal} {Phys. Rev. B}\ }\textbf {\bibinfo {volume} {54}},\ \bibinfo
  {pages} {11169} (\bibinfo {year} {1996})}\BibitemShut {NoStop}%
\bibitem [{\citenamefont {Kresse}\ and\ \citenamefont
  {Furthmüller}(1996)}]{KRESSE199615}%
  \BibitemOpen
  \bibfield  {author} {\bibinfo {author} {\bibfnamefont {G.}~\bibnamefont
  {Kresse}}\ and\ \bibinfo {author} {\bibfnamefont {J.}~\bibnamefont
  {Furthmüller}},\ }\bibinfo {title} {Efficiency of ab-initio total energy
  calculations for metals and semiconductors using a plane-wave basis set},\
  \href {https://doi.org/https://doi.org/10.1016/0927-0256(96)00008-0}
  {\bibfield  {journal} {\bibinfo  {journal} {Computational Materials Science}\
  }\textbf {\bibinfo {volume} {6}},\ \bibinfo {pages} {15} (\bibinfo {year}
  {1996})}\BibitemShut {NoStop}%
\bibitem [{\citenamefont {Perdew}\ \emph {et~al.}(1996)\citenamefont {Perdew},
  \citenamefont {Burke},\ and\ \citenamefont
  {Ernzerhof}}]{PhysRevLett.77.3865}%
  \BibitemOpen
  \bibfield  {author} {\bibinfo {author} {\bibfnamefont {J.~P.}\ \bibnamefont
  {Perdew}}, \bibinfo {author} {\bibfnamefont {K.}~\bibnamefont {Burke}},\ and\
  \bibinfo {author} {\bibfnamefont {M.}~\bibnamefont {Ernzerhof}},\ }\bibinfo
  {title} {Generalized Gradient Approximation Made Simple},\ \href
  {https://doi.org/10.1103/PhysRevLett.77.3865} {\bibfield  {journal} {\bibinfo
   {journal} {Phys. Rev. Lett.}\ }\textbf {\bibinfo {volume} {77}},\ \bibinfo
  {pages} {3865} (\bibinfo {year} {1996})}\BibitemShut {NoStop}%
\bibitem [{\citenamefont {Perdew}\ \emph {et~al.}(2008)\citenamefont {Perdew},
  \citenamefont {Ruzsinszky}, \citenamefont {Csonka}, \citenamefont {Vydrov},
  \citenamefont {Scuseria}, \citenamefont {Constantin}, \citenamefont {Zhou},\
  and\ \citenamefont {Burke}}]{PhysRevLett.100.136406}%
  \BibitemOpen
  \bibfield  {author} {\bibinfo {author} {\bibfnamefont {J.~P.}\ \bibnamefont
  {Perdew}}, \bibinfo {author} {\bibfnamefont {A.}~\bibnamefont {Ruzsinszky}},
  \bibinfo {author} {\bibfnamefont {G.~I.}\ \bibnamefont {Csonka}}, \bibinfo
  {author} {\bibfnamefont {O.~A.}\ \bibnamefont {Vydrov}}, \bibinfo {author}
  {\bibfnamefont {G.~E.}\ \bibnamefont {Scuseria}}, \bibinfo {author}
  {\bibfnamefont {L.~A.}\ \bibnamefont {Constantin}}, \bibinfo {author}
  {\bibfnamefont {X.}~\bibnamefont {Zhou}},\ and\ \bibinfo {author}
  {\bibfnamefont {K.}~\bibnamefont {Burke}},\ }\bibinfo {title} {Restoring the
  Density-Gradient Expansion for Exchange in Solids and Surfaces},\ \href
  {https://doi.org/10.1103/PhysRevLett.100.136406} {\bibfield  {journal}
  {\bibinfo  {journal} {Phys. Rev. Lett.}\ }\textbf {\bibinfo {volume} {100}},\
  \bibinfo {pages} {136406} (\bibinfo {year} {2008})}\BibitemShut {NoStop}%
\bibitem [{\citenamefont {Bl\"ochl}(1994)}]{PhysRevB.50.17953}%
  \BibitemOpen
  \bibfield  {author} {\bibinfo {author} {\bibfnamefont {P.~E.}\ \bibnamefont
  {Bl\"ochl}},\ }\bibinfo {title} {Projector augmented-wave method},\ \href
  {https://doi.org/10.1103/PhysRevB.50.17953} {\bibfield  {journal} {\bibinfo
  {journal} {Phys. Rev. B}\ }\textbf {\bibinfo {volume} {50}},\ \bibinfo
  {pages} {17953} (\bibinfo {year} {1994})}\BibitemShut {NoStop}%
\bibitem [{\citenamefont {Kresse}\ and\ \citenamefont
  {Joubert}(1999)}]{PhysRevB.59.1758}%
  \BibitemOpen
  \bibfield  {author} {\bibinfo {author} {\bibfnamefont {G.}~\bibnamefont
  {Kresse}}\ and\ \bibinfo {author} {\bibfnamefont {D.}~\bibnamefont
  {Joubert}},\ }\bibinfo {title} {From ultrasoft pseudopotentials to the
  projector augmented-wave method},\ \href
  {https://doi.org/10.1103/PhysRevB.59.1758} {\bibfield  {journal} {\bibinfo
  {journal} {Phys. Rev. B}\ }\textbf {\bibinfo {volume} {59}},\ \bibinfo
  {pages} {1758} (\bibinfo {year} {1999})}\BibitemShut {NoStop}%
\bibitem [{\citenamefont {Monkhorst}\ and\ \citenamefont
  {Pack}(1976)}]{PhysRevB.13.5188}%
  \BibitemOpen
  \bibfield  {author} {\bibinfo {author} {\bibfnamefont {H.~J.}\ \bibnamefont
  {Monkhorst}}\ and\ \bibinfo {author} {\bibfnamefont {J.~D.}\ \bibnamefont
  {Pack}},\ }\bibinfo {title} {Special points for Brillouin-zone
  integrations},\ \href {https://doi.org/10.1103/PhysRevB.13.5188} {\bibfield
  {journal} {\bibinfo  {journal} {Phys. Rev. B}\ }\textbf {\bibinfo {volume}
  {13}},\ \bibinfo {pages} {5188} (\bibinfo {year} {1976})}\BibitemShut
  {NoStop}%
\bibitem [{\citenamefont {Aryasetiawan}\ \emph {et~al.}(2006)\citenamefont
  {Aryasetiawan}, \citenamefont {Karlsson}, \citenamefont {Jepsen},\ and\
  \citenamefont {Sch\"onberger}}]{PhysRevB.74.125106}%
  \BibitemOpen
  \bibfield  {author} {\bibinfo {author} {\bibfnamefont {F.}~\bibnamefont
  {Aryasetiawan}}, \bibinfo {author} {\bibfnamefont {K.}~\bibnamefont
  {Karlsson}}, \bibinfo {author} {\bibfnamefont {O.}~\bibnamefont {Jepsen}},\
  and\ \bibinfo {author} {\bibfnamefont {U.}~\bibnamefont {Sch\"onberger}},\
  }\bibinfo {title} {Calculations of Hubbard $U$ from first-principles},\ \href
  {https://doi.org/10.1103/PhysRevB.74.125106} {\bibfield  {journal} {\bibinfo
  {journal} {Phys. Rev. B}\ }\textbf {\bibinfo {volume} {74}},\ \bibinfo
  {pages} {125106} (\bibinfo {year} {2006})}\BibitemShut {NoStop}%
\bibitem [{\citenamefont {Grimme}\ \emph {et~al.}(2010)\citenamefont {Grimme},
  \citenamefont {Antony}, \citenamefont {Ehrlich},\ and\ \citenamefont
  {Krieg}}]{10.1063/1.3382344}%
  \BibitemOpen
  \bibfield  {author} {\bibinfo {author} {\bibfnamefont {S.}~\bibnamefont
  {Grimme}}, \bibinfo {author} {\bibfnamefont {J.}~\bibnamefont {Antony}},
  \bibinfo {author} {\bibfnamefont {S.}~\bibnamefont {Ehrlich}},\ and\ \bibinfo
  {author} {\bibfnamefont {H.}~\bibnamefont {Krieg}},\ }\bibinfo {title} {A
  consistent and accurate ab initio parametrization of density functional
  dispersion correction (DFT-D) for the 94 elements H-Pu},\ \href@noop {}
  {\bibfield  {journal} {\bibinfo  {journal} {The Journal of Chemical Physics}\
  }\textbf {\bibinfo {volume} {132}},\ \bibinfo {pages} {154104} (\bibinfo
  {year} {2010})}\BibitemShut {NoStop}%
\bibitem [{\citenamefont {Wang}\ \emph {et~al.}(2021)\citenamefont {Wang},
  \citenamefont {Xu}, \citenamefont {Liu}, \citenamefont {Tang},\ and\
  \citenamefont {Geng}}]{WANG2021108033}%
  \BibitemOpen
  \bibfield  {author} {\bibinfo {author} {\bibfnamefont {V.}~\bibnamefont
  {Wang}}, \bibinfo {author} {\bibfnamefont {N.}~\bibnamefont {Xu}}, \bibinfo
  {author} {\bibfnamefont {J.-C.}\ \bibnamefont {Liu}}, \bibinfo {author}
  {\bibfnamefont {G.}~\bibnamefont {Tang}},\ and\ \bibinfo {author}
  {\bibfnamefont {W.-T.}\ \bibnamefont {Geng}},\ }\bibinfo {title} {VASPKIT: A
  user-friendly interface facilitating high-throughput computing and analysis
  using VASP code},\ \href
  {https://doi.org/https://doi.org/10.1016/j.cpc.2021.108033} {\bibfield
  {journal} {\bibinfo  {journal} {Computer Physics Communications}\ }\textbf
  {\bibinfo {volume} {267}},\ \bibinfo {pages} {108033} (\bibinfo {year}
  {2021})}\BibitemShut {NoStop}%
\bibitem [{\citenamefont {Mostofi}\ \emph {et~al.}(2008)\citenamefont
  {Mostofi}, \citenamefont {Yates}, \citenamefont {Lee}, \citenamefont {Souza},
  \citenamefont {Vanderbilt},\ and\ \citenamefont {Marzari}}]{MOSTOFI2008685}%
  \BibitemOpen
  \bibfield  {author} {\bibinfo {author} {\bibfnamefont {A.~A.}\ \bibnamefont
  {Mostofi}}, \bibinfo {author} {\bibfnamefont {J.~R.}\ \bibnamefont {Yates}},
  \bibinfo {author} {\bibfnamefont {Y.-S.}\ \bibnamefont {Lee}}, \bibinfo
  {author} {\bibfnamefont {I.}~\bibnamefont {Souza}}, \bibinfo {author}
  {\bibfnamefont {D.}~\bibnamefont {Vanderbilt}},\ and\ \bibinfo {author}
  {\bibfnamefont {N.}~\bibnamefont {Marzari}},\ }\bibinfo {title} {wannier90: A
  tool for obtaining maximally-localised Wannier functions},\ \href
  {https://doi.org/https://doi.org/10.1016/j.cpc.2007.11.016} {\bibfield
  {journal} {\bibinfo  {journal} {Computer Physics Communications}\ }\textbf
  {\bibinfo {volume} {178}},\ \bibinfo {pages} {685} (\bibinfo {year}
  {2008})}\BibitemShut {NoStop}%
\bibitem [{\citenamefont {Wu}\ \emph {et~al.}(2018)\citenamefont {Wu},
  \citenamefont {Zhang}, \citenamefont {Song}, \citenamefont {Troyer},\ and\
  \citenamefont {Soluyanov}}]{WU2018405}%
  \BibitemOpen
  \bibfield  {author} {\bibinfo {author} {\bibfnamefont {Q.}~\bibnamefont
  {Wu}}, \bibinfo {author} {\bibfnamefont {S.}~\bibnamefont {Zhang}}, \bibinfo
  {author} {\bibfnamefont {H.-F.}\ \bibnamefont {Song}}, \bibinfo {author}
  {\bibfnamefont {M.}~\bibnamefont {Troyer}},\ and\ \bibinfo {author}
  {\bibfnamefont {A.~A.}\ \bibnamefont {Soluyanov}},\ }\bibinfo {title}
  {WannierTools: An open-source software package for novel topological
  materials},\ \href
  {https://doi.org/https://doi.org/10.1016/j.cpc.2017.09.033} {\bibfield
  {journal} {\bibinfo  {journal} {Computer Physics Communications}\ }\textbf
  {\bibinfo {volume} {224}},\ \bibinfo {pages} {405} (\bibinfo {year}
  {2018})}\BibitemShut {NoStop}%
\bibitem [{\citenamefont {Henkelman}\ \emph {et~al.}(2000)\citenamefont
  {Henkelman}, \citenamefont {Uberuaga},\ and\ \citenamefont
  {Jónsson}}]{10.1063/1.1329672}%
  \BibitemOpen
  \bibfield  {author} {\bibinfo {author} {\bibfnamefont {G.}~\bibnamefont
  {Henkelman}}, \bibinfo {author} {\bibfnamefont {B.~P.}\ \bibnamefont
  {Uberuaga}},\ and\ \bibinfo {author} {\bibfnamefont {H.}~\bibnamefont
  {Jónsson}},\ }\bibinfo {title} {A climbing image nudged elastic band method
  for finding saddle points and minimum energy paths},\ \href@noop {}
  {\bibfield  {journal} {\bibinfo  {journal} {The Journal of Chemical Physics}\
  }\textbf {\bibinfo {volume} {113}},\ \bibinfo {pages} {9901} (\bibinfo {year}
  {2000})}\BibitemShut {NoStop}%
\bibitem [{\citenamefont {Wang}\ \emph {et~al.}(2021)\citenamefont {Wang},
  \citenamefont {Shi}, \citenamefont {Liu}, \citenamefont {Zhang},
  \citenamefont {Hong}, \citenamefont {Li}, \citenamefont {Qiang},
  \citenamefont {Chen}, \citenamefont {Ren}, \citenamefont {Cheng},
  \citenamefont {Li},\ and\ \citenamefont
  {Xing-Qiu}}]{10.1038/s41467-021-22324-8}%
  \BibitemOpen
  \bibfield  {author} {\bibinfo {author} {\bibfnamefont {L.}~\bibnamefont
  {Wang}}, \bibinfo {author} {\bibfnamefont {Y.}~\bibnamefont {Shi}}, \bibinfo
  {author} {\bibfnamefont {M.}~\bibnamefont {Liu}}, \bibinfo {author}
  {\bibfnamefont {A.}~\bibnamefont {Zhang}}, \bibinfo {author} {\bibfnamefont
  {Y.-L.}\ \bibnamefont {Hong}}, \bibinfo {author} {\bibfnamefont
  {R.}~\bibnamefont {Li}}, \bibinfo {author} {\bibfnamefont {G.}~\bibnamefont
  {Qiang}}, \bibinfo {author} {\bibfnamefont {M.}~\bibnamefont {Chen}},
  \bibinfo {author} {\bibfnamefont {W.}~\bibnamefont {Ren}}, \bibinfo {author}
  {\bibfnamefont {H.-M.}\ \bibnamefont {Cheng}}, \bibinfo {author}
  {\bibfnamefont {Y.}~\bibnamefont {Li}},\ and\ \bibinfo {author}
  {\bibfnamefont {C.}~\bibnamefont {Xing-Qiu}},\ }\bibinfo {title}
  {Intercalated architecture of MA2Z4 family layered van der Waals materials
  with emerging topological, magnetic and superconducting properties},\ \href
  {https://doi.org/10.1038/s41467-021-22324-8} {\bibfield  {journal} {\bibinfo
  {journal} {Nat Commun}\ }\textbf {\bibinfo {volume} {12}},\ \bibinfo {pages}
  {2361} (\bibinfo {year} {2021})}\BibitemShut {NoStop}%
\bibitem [{\citenamefont {Li}\ \emph {et~al.}(2023)\citenamefont {Li},
  \citenamefont {Yang}, \citenamefont {Jiang}, \citenamefont {Wu},\ and\
  \citenamefont {Xun}}]{PhysRevMaterials.7.064002}%
  \BibitemOpen
  \bibfield  {author} {\bibinfo {author} {\bibfnamefont {P.}~\bibnamefont
  {Li}}, \bibinfo {author} {\bibfnamefont {X.}~\bibnamefont {Yang}}, \bibinfo
  {author} {\bibfnamefont {Q.-S.}\ \bibnamefont {Jiang}}, \bibinfo {author}
  {\bibfnamefont {Y.-Z.}\ \bibnamefont {Wu}},\ and\ \bibinfo {author}
  {\bibfnamefont {W.}~\bibnamefont {Xun}},\ }\bibinfo {title} {Built-in
  electric field and strain tunable valley-related multiple topological phase
  transitions in
  $\mathrm{VSi}X{\mathrm{N}}_{4}(X=\mathrm{C},\mathrm{Si},\mathrm{Ge},\mathrm{Sn},\mathrm{Pb})$
  monolayers},\ \href {https://doi.org/10.1103/PhysRevMaterials.7.064002}
  {\bibfield  {journal} {\bibinfo  {journal} {Phys. Rev. Mater.}\ }\textbf
  {\bibinfo {volume} {7}},\ \bibinfo {pages} {064002} (\bibinfo {year}
  {2023})}\BibitemShut {NoStop}%
\end{thebibliography}%
\bibliographystyle{apsrev4-2}

\end{document}